\documentclass[11pt,3p,review,authoryear]{elsarticle}

\journal{International Journal of Forecasting}
\biboptions{longnamesfirst}
\usepackage[utf8]{inputenc}
\usepackage{arydshln}
\usepackage{slashbox}
\usepackage{color}
\usepackage[]{bm}
\usepackage[]{amsmath}
\usepackage{rotating}
\usepackage[]{verbatim}
\usepackage{amsthm}
\usepackage{enumerate}
\usepackage[]{bm}
\usepackage{bbm}
\usepackage{soul}
\usepackage[titletoc]{appendix}
\usepackage[flushleft]{threeparttable}
\usepackage{float}
\usepackage[titletoc]{appendix}
\usepackage{setspace}
\usepackage{epstopdf} 
\usepackage{graphicx}
\usepackage{lscape}
\usepackage{amssymb}
\usepackage[]{geometry}
\usepackage[inline]{enumitem}
\usepackage{array}
\usepackage{subcaption}
\usepackage[normalem]{ulem}
\usepackage{xr}
\newtheorem{proposition}{Proposition}
\newtheorem{corollary}{Corollary}
\newtheorem{assumption}{Assumption}
\newtheorem{theorem}{Theorem}

\newtheorem{lemma}{Lemma}

\usepackage{longtable}[=v4.13]
\usepackage{booktabs}
\usepackage{graphicx}
\usepackage{multirow}
\usepackage{adjustbox}

\begin{document}

\begin{frontmatter}

\title{Equal Predictive Ability Tests Based on Panel Data \\
with Applications to OECD and IMF Forecasts}

\author[ledi]{Oguzhan Akgun}
\address[ledi]{LEDi, University of Burgundy, France}

\author[pa]{Alain Pirotte\corref{cor}}
\cortext[cor]{Corresponding author}
\ead{Alain.Pirotte@u-paris2.fr; 21 rue Valette, 75005, Paris, France}
\address[pa]{University of Paris Pantheon-Assas, CRED}

\author[cass]{Giovanni Urga}
\address[cass]{Bayes Business School (formerly Cass), London, United Kingdom and Bergamo University, Italy}

\author[smu]{Zhenlin Yang}
\address[smu]{School of Economics, Singapore Management University, Singapore}

\begin{abstract}
We propose two types of equal predictive ability (EPA) tests with panels to compare the predictions made by two forecasters.
The first type, namely $S$-statistics, focuses on \textit{the overall EPA hypothesis} which states that the EPA holds on average over all panel units and over time.
The second, called $C$-statistics, focuses on \textit{the clustered EPA hypothesis} where the EPA holds jointly for a fixed number of clusters of panel units.
The asymptotic properties of the proposed tests are evaluated under weak and strong cross-sectional dependence.
An extensive Monte Carlo simulation shows that the proposed tests have very good finite sample properties even with little information about the cross-sectional dependence in the data.
The proposed framework is applied to compare the economic growth forecasts of the OECD and the IMF, and to evaluate the performance of the consumer price inflation forecasts of the IMF.
\end{abstract}


\end{frontmatter}

\section{Introduction}
\label{sec:intro}

Formal tests of the null hypothesis of no difference in forecast accuracy using two time series of forecast errors have been widely considered in the literature: see \cite{vuong89}, \citet[][hereafter DM]{diebold95}, \cite{west96}, \cite{clark01}, \cite{clark15}, \cite{giacomini06}, \cite{clark07}, {\cite{mariano12}}, among others.
On the contrary, the literature on such tests using panel data is scarce with {a few} exceptions: \citet{keane90}, \cite{davies95} and \citet[][QTZ, hereafter]{qu22}.

The main aim of this paper is to develop testing procedures for equal predictive ability (EPA) hypotheses based on panel data, taking into account the cross-sectional dependence (CD) and the temporal dependence in the data set.
{Let $\widehat{y}_{l,it}$ be the forecast of agent $l=1,2$ made for the target value $y_{it}$ of time $t=1,2,\dots,T$ for unit (e.g., country, firm) $i=1,2,\dots,n$.
We propose tests for comparing the predictive ability of two forecasting agents, using $n$ time series of loss differentials, $\Delta L_{it} = L(y_{it};\widehat{y}_{1,it})-L(y_{it};\widehat{y}_{1,it})$, of length $T$, where $L\left(\cdot\right)$ is a generic loss function \citep[for different loss functions in use, see][]{gneiting11}.}
Our setting differs from that of {\citet{keane90}} and \cite{davies95} in that we consider forecasts made by two forecasters on multiple economic units over time, whereas they consider forecasts made by multiple forecasters on a single economic unit over time.
Our paper fills an important gap in the literature by allowing for multiple target values for each point in time {with respect to these two papers}.

We develop two types of tests corresponding to two EPA hypotheses.
The first type focuses on \textit{the overall EPA hypothesis} which states that the EPA holds on average over all units and over time. 
The statistics of this type, which we call $S$-statistics, are useful when a researcher is not interested in the differences of predictive ability for a specific unit {or clusters of units} but in the overall differences.
The statistics of the second type, namely $C$-statistics, focus on \textit{the clustered EPA hypothesis} which states that the EPA holds jointly for a fixed number of clusters of units in the panel.
QTZ consider the same set of null hypotheses.
Our work differs from that of QTZ in the way the test statistics are constructed, most notably the statistics for testing the clustered EPA hypothesis.
While both works set up a joint null hypothesis as ``all cluster means are zero" (see, \eqref{eq5} in our paper and (11) in QTZ), we formulate multivariate test statistics in line with the joint hypotheses (the $C$-statistics in Section \ref{sec:jointtests}) with limiting null distribution being $\chi_G^2$, where $G$ is the number of clusters.
On the other hand, QTZ formulate a univariate test statistic (the $J_n^D$ given in (12) in QTZ) based on the ``average of the cluster averages of the loss differentials."
If cluster sizes are equal, the numerator of $J_n^D$ is equivalent to these of $J_{n,T}^{DM}$ of QTZ and $S$-statistics in our paper, tests for the overall EPA.
As a result, in general $J_n^D$ fails to detect departure from the clustered EPA hypothesis if the overall EPA hypothesis holds.
See the detailed discussion given at the end of Section \ref{sec:jointtests}.

The applied literature in comparing the accuracy of two or more forecasts suggests that the forecast errors of units, such as countries, are affected by common shocks such as the global financial crisis.
For instance, \citet{pain14} show that the economic growth projections of the OECD for the period 2007-2012 are systematically upward biased.
A similar tendency exists for other forecasters, such as the IMF.
Moreover, these effects are carried into the loss differentials such that they follow a similar pattern, as we highlight later in this paper.
The results of \cite{pain14} indicate also that the effect of these common shocks is heterogeneous across economies and some country clusters exists.

Following these insights, we build our testing framework around the loss differentials which follow an approximate factor model where some common factors affect all units in the panel with heterogeneous loadings.
In addition, the error terms are allowed to be cross-sectionally weakly correlated.
We therefore simultaneously allow the loss differentials to contain weak cross-sectional dependence (WCD) arising from, e.g., spatial error correlation, and strong cross-sectional dependence (SCD) due to the existence of common factors, using the terminology of \cite{chudik11}.
To develop our tests under WCD, we use non-parametric methods of variance estimation, based on geographic or economic distances between panel units \citep[][]{kelejian07,kim13}.
In addition, we propose a novel partial sample variance estimator for large panels which deals with the case of unknown distances while being robust to arbitrary WCD and temporal dependence.
To deal with SCD, we use the principal components estimator (PCE) built for large dimensional approximate factor models \citep{bai02,bai03}.
Following {the insights of} \cite{driscoll98}, we also propose tests for the case of unknown number of common factors contrary to the ones based on the PCE.
This last approach
is flexible in the sense that it is robust under different types of CD.
Moreover, they do not rely on a linear factor model contrary to the PCE based tests.
They are also very easy to calculate as in the case of overall EPA testing, the proposed test is identical to applying the \cite{diebold95} test to cross-sectional averages of the loss differentials.

We analyze the asymptotic properties of the proposed test statistics using joint limits.
Under mild conditions, the overall EPA test statistics and the clustered EPA test statistics are shown to converge in distribution to standard normal and chi-square with $G$ degrees of freedom under the null of interest, respectively.
The finite sample properties of the tests are examined via Monte Carlo simulations.
The results show in general that our tests have very good finite sample performance.
Specifically, we find that \cite{driscoll98} based overall and clustered EPA tests are robust to arbitrary CD in the loss differentials.
However, they have lower power than the tests developed specifically for the case of WCD if loss differentials are only weakly cross-sectionally correlated.
Moreover, if the linear approximate factor model assumption is satisfied by the loss differentials, they have lower power compared to the tests based on PCE.

The proposed tests are used in two applications.
In the first one, we compare the economic growth forecast errors of the OECD and the IMF using data for 29 countries over the period between 1998 and 2016.
In the second application, the quality of the IMF consumer price inflation (CPI) forecasts is challenged by comparing them with random walk forecasts.
This data set contains 127 countries over the period 1991-2019.
In both applications, we find strong evidence of SCD in loss differentials of forecast errors.
{The results of the first application show that neither overall EPA hypothesis nor clustered EPA hypothesis can be rejected for the economic growth forecasts of the OECD and the IMF.
In contrast, in the second application, we find significant evidence against the overall and clustered EPA hypotheses on the comparison of the IMF and random walk forecasts, especially in the pre-crisis period.}

The remainder of the paper is as follows:
The testing framework and the hypotheses of interest are presented in Section \ref{sec:modeltest}.
The panel EPA test statistics and their asymptotic properties are reported in Section \ref{sec:tests}.
Section \ref{sec:monte} is on the small sample properties of the proposed tests.
{Section \ref{sec:imp} gives guidelines to practitioners willing to test the EPA hypotheses using panels.}
In Section \ref{sec:app} the economic growth forecast errors of the OECD and the IMF are compared using the proposed tests.
Section \ref{sec:conc} concludes.
Three appendices contain additional justification of our testing framework, proofs of results and {the details of the} second application comparing the IMF CPI forecasts with random walk forecasts.

\textbf{Notation.} Let $\mathbf{A} = [a_{ij}]$ be an $n \times n$ matrix.
The column and row norms of $\mathbf{A}$ are $||\mathbf{A}||_1 = \max_{1\leq j\leq n}\sum_{i=1}^{n}|a_{ij}|$ and $||\mathbf{A}||_{\infty} = \max_{1\leq i\leq n}\sum_{j=1}^{n}|a_{ij}|$, respectively.
The Euclidean norm of an $n \times m$ matrix $\mathbf{B}$ is $||\mathbf{B}|| = [\mathrm{Tr}(\mathbf{B}'\mathbf{B})]^{1/2}$. $M$ is a finite positive constant.
$\overset{d}{\rightarrow}$ and $\overset{p}{\rightarrow}$ denote convergence in distribution and probability, respectively.
Joint passage to infinity of $T$ and $n$ is denoted by $(T,n) \rightarrow \infty$.

\section{Testing Framework and the Hypotheses}\label{sec:modeltest}

We are interested in comparing the errors of forecasts made by two forecasters on an economic variable observed for units $i=1,2,
\dots, n$ at time $t=1,2, \dots, T$.
The loss differentials, $\Delta L_{it} = L(y_{it};\widehat{y}_{1,it})-L(y_{it};\widehat{y}_{1,it})$, are {potentially cross-sectionally correlated.
The CD may be of strong or weak types or may even be generated by two distinct components of these two different types.
To simplify the distinction, we assume the following structure on the loss differentials:}
\begin{equation}\label{model}
\begin{split}
\Delta L_{it} &= \mu_i + \boldsymbol{\lambda }_{i}^{\prime }\mathbf{f}_{t}+\varepsilon_{it}, \\
\varepsilon_{it} &= \sum_{j = 1}^n r_{ij} {\epsilon}_{jt},
\end{split}
\end{equation}
where $\mathbf{f}_{t} = (f_{1t},f_{2t},\dots,f_{mt})'$ is an $m\times 1$ vector of unobservable common factors and $\boldsymbol{\lambda }_{i}= (\lambda_{1i},\lambda_{2i},\dots,\lambda_{mi})'$ is an $m\times 1$ vector of {fixed} factor loadings.
The coefficients $r_{ij}$ are fixed, unknown elements of an $n\times n$ matrix $\mathbf{R}_{n}$.
{We impose $r_{ii}=1$ for normalization of the variances.}
$\mathbf{f}_{t}$ and ${\epsilon}_{it}$ are assumed to be zero mean weakly stationary time series allowed to be autocorrelated
through time.
In addition, ${\epsilon}_{it}$ is cross-sectionally uncorrelated.
The unit specific means satisfy $|\mu_i|<\infty$.

We assume that the error terms $\varepsilon_{it}$ carry WCD, meaning that the variance of their cross-sectional average vanishes asymptotically:
\begin{equation}\label{eq:wcdcond}
\lim_{n \to \infty} \mathrm{Var}\left( n^{-1} \sum_{i=1}^n \varepsilon_{it} \right) = \lim_{n \to \infty} \left( n^{-2} \sum_{i,j=1}^n \mathbf{r}_{i.}' \boldsymbol{\gamma}_{n,0} \mathbf{r}_{j.}\right) = 0,
\end{equation}
where $\mathbf{r}_{i.}=(r_{i1},r_{i2},\dots,r_{in})^{\prime}$, $\boldsymbol{\gamma}_{n,0} = \mathrm{diag}(\gamma_{11,0},\dots,\gamma_{nn,0})$ and $\gamma_{ii,0} = \mathrm{E}({\epsilon}^2_{it})$.
For this to be true, it requires the row and column sum norms of $\mathbf{R}_{n}$ to be bounded; see Assumption \ref{ass_r1} below.
The common component in \eqref{model} is assumed to induce SCD such that the variance of its cross-sectional average is bounded away from zero:
\begin{equation}\label{eq:scdcond}
\mathrm{Var}\left( n^{-1} \sum_{i=1}^n \boldsymbol{\lambda }_{i}^{\prime }\mathbf{f}_{t} \right) = n^{-2} \sum_{i,j=1}^n \boldsymbol{\lambda }_{i}^{\prime } \boldsymbol{\Gamma}_{0} \boldsymbol{\lambda }_{j} > 0, \text{ for all } n,
\end{equation}
where $\boldsymbol{\Gamma}_{0} = \mathrm{E} (\mathbf{f}_{t} \mathbf{f}_{t}^{\prime })$.
The conditions under which this holds true are given below, in Assumption \ref{ass_comcomp}.
The WCD can be modelled by a spatial process, such as spatial autoregression, spatial moving average as well as their higher order versions, or a factor model with a possibly infinite number of weak common factors \citep[see, for instance,][]{chudik11}.

{In this paper,} we follow closely the methodological implications of DM.
Firstly, we adopt the view that the forecast errors may be related to forecasts made using some econometric models or simply by expert knowledge.
As in DM, our approach is built on a model-free environment, meaning that it is agnostic about the process generating the forecasts.
This is contrary to the extensions of DM in papers such as \cite{west96} and \cite{giacomini06} which focus, among other things, on problems associated with testing using forecast errors generated by nested models.
Secondly, following DM, our assumptions concern directly the loss differentials.
Forecast errors can be non-stationary and induce CD.
What is important is the properties which are carried into the loss differentials, not the properties of the forecast errors specifically.
As an example, in Appendix \ref{appendixloss} we show how {the specification in \eqref{model}} can be obtained for two different loss functions, namely absolute and quadratic loss, starting with a pure factor model for the forecast errors.
However, our approach is not necessarily limited to these loss functions.
As in {the} DM approach, any other (symmetric or asymmetric) function can be used.
Eventually, our methodology relies on the empirical analysis of time series and CD properties of the loss differentials, using CD tests \citep[][]{pesaran15} and information criteria \citep[][]{bai02}.

\textbf{The null hypotheses of EPA and the alternatives.} The first hypothesis of interest is
\begin{equation}\label{eq4}
H^a_{0}:\bar{\mu}=0,
\end{equation}%
where $\bar{\mu} = \lim_{n\to\infty} \bar{\mu}_n$ with $\bar{\mu}_n = n^{-1} \sum^n_{i = 1} \mu_i$.
This hypothesis{, which was also considered by QTZ} states that the forecasts generated by the two agents are equally accurate on average over all $i$ and $t$.
If a researcher is not interested in the difference in predictive power for a particular unit but the average difference over units, this hypothesis should be considered.
It is particularly plausible to consider this in a micro forecasting study where the units are random draws from a population.
{Throughout the paper, this hypothesis is called \textit{the overall EPA hypothesis}.}
{The alternative hypothesis in this case is that overall EPA does not hold:
\begin{equation}\label{eq41}
H^a_{1}:\bar{\mu} \neq 0.
\end{equation}
Of course it is also possible to consider one-sided alternatives.

In a macro forecasting study, differences between clusters of units can have a specific economic importance and may be of interest from a policy perspective.
For instance, a question of interest is whether the forecasts made by agents are more accurate for a particular cluster of countries in the sample.
In this case, the null hypothesis can be formulated such that the predictive equality holds for $G$ clusters of units: 
\begin{equation}\label{eq5}
H^{ca}_{0}: \bar{\mu}_{g} = 0\text{, \ for all } g=1,2,\dots,G,
\end{equation}
where $\bar{\mu}_{g} = \lim_{n_g\to\infty}\bar{\mu}_{g,n_g}$ with $\bar{\mu}_{g,n_g} = n_g^{-1} \sum_{i \in G_g} \mu_i$ and $G_g$ is the set of indexes of $n_g$ cross-sectional units which belong to cluster $g$. 
In this paper, this hypothesis is referred to as \textit{the clustered EPA hypothesis}.
The alternative hypothesis associated is the following:
\begin{equation}\label{eq51}
H^{ca}_{1}: \bar{\mu}_{g} \neq 0\text{, \ for at least one } g=1,2,\dots,G.
\end{equation}
QTZ consider the null hypothesis $H^{ca}_{0}$ without stating an alternative hypothesis.
Their test against $H^{ca}_{0}$ is inconsistent in particular cases, as for instance, their clustered EPA test fails to reject $H^{ca}_{0}$ if $H^a_{0}$ and $H^{ca}_{1}$ hold simultaneously.
This issue is discussed theoretically below in Section \ref{sec:jointtests}.
The difference between the two hypotheses is {therefore} important and the choice depends on the specific interest.
The overall EPA may hold even if the two forecasters have different predictive ability for different clusters.
This occurs if the average loss differentials of different country clusters are different from zero {(that is $H^{ca}_{1}$ holds)} but they add up to zero when pooled {(that is $H^a_{0}$ holds)}.
If these clusters have an economic meaning, it is important to test the clustered EPA hypothesis.

In what follows, we propose test statistics for these two null hypotheses, under different assumptions on the CD structure in the loss differentials.
These are No CD, WCD or SCD.
The main difference between the proposed test statistics is the methods of calculation of the asymptotic variance depending on this structure.
A summary of these test statistics is given in Table \ref{tab:tests}.
Most of our test statistics do not require the approximate factor model given in \eqref{model} with the exceptions being $\underline{S}_{nT}^{(3)}$ and $\underline{C}_{nT}^{(3)}$.
Nevertheless, we retain the factor model to simplify the discussion of different types of CD.

\begin{table}[htbp!]
\caption{Test Statistics Proposed}
\label{tab:tests}\centering
    {\begin{tabular}{ccc}
    \toprule
          & Overall EPA Tests & Clustered EPA Tests \\
          & $S$-statistics & $C$-statistics \\
    \midrule
    {No cross-sectional dependence} & $S_{nT}^{(1)}$ & $C_{nT}^{(1)}$ \\
    \midrule
    {Weak cross-sectional dependence} & & \\
    \textit{Distance based tests} & $S_{nT}^{(2)}$ & $C_{nT}^{(2)}$ \\
    \textit{Tests with unknown distance} & $\underline{S}_{nT}^{(2)}$ & $\underline{C}_{nT}^{(2)}$ \\
    \midrule
    {Strong cross-sectional dependence} & & \\
    \textit{No restriction on CD} & $S_{nT}^{(3)}$, $\tilde{S}_{nT}^{(3)}$ & $C_{nT}^{(3)}$ \\
    \textit{Factor model based tests} & $\underline{S}_{nT}^{(3)}$ & $\underline{C}_{nT}^{(3)}$ \\
    \bottomrule
    \end{tabular}%
    }
\end{table}

\section{The Test Statistics and Their Asymptotic Properties}\label{sec:tests}

\subsection{Tests for overall equal predictive ability: $S$-statistics}\label{sec:overalltests}

Consider the sample mean loss differential over time and units: 
\begin{equation*}
\Delta \bar{L}_{nT}=\frac{1}{nT}\sum_{i=1}^{n}\sum_{t=1}^{T}\Delta L_{it}.
\end{equation*}
We provide testing procedures for the overall EPA implied in \eqref{eq4} based on $\Delta\bar{L}_{nT}$ under different assumptions about the structure of CD in the loss differentials.
Let $k_T(\cdot)$ be a kernel function and $d_T$ a sequence of {positive,} non-random bandwidth parameters.
In all cases, the limiting null distribution of the test statistics is obtained under Assumptions \ref{ass_epsilon} and \ref{ass_tkernel}.
\begin{assumption}\label{ass_epsilon}
${\epsilon}_{it}$ follows the linear process ${\epsilon}_{it} = \sum_{h=0}^{\infty} c_{ih} \psi_{i,t-h}$ for each $i$, with $\psi_{it} \sim iid(0,1)$ over $i$ and $t$, $\mathrm{E} |\psi_{it}|^4 < \infty$, $\max_{1\leq i\leq n} \sum_{h=0}^{\infty} h | c_{ih} | < M < \infty $ and $\sum_{h=0}^{\infty} c_{ih}>0$.
\end{assumption}
\begin{assumption}\label{ass_tkernel}
\begin{enumerate*}[label=(\alph*)]
\item\label{ass_tkernel1} $k_T(x):\mathbb{R} \rightarrow [-1,1]$ is continuous at zero,
$k_T(0)=1$,
$k_T(x) = k_T(-x)$ $\forall x \in \mathbb{R}$,
\item\label{ass_tkernel2} $\lim_{T \rightarrow \infty} d_T^{-1} \sum_{h=1}^{T-1} |k_{T} \left({h/d_T} \right) | < \infty$,
\item\label{ass_tsbw} $d_T \rightarrow \infty$ such that $d^2_T/T \rightarrow 0$ as $T \rightarrow \infty$.
\end{enumerate*}
\end{assumption}
Assumption \ref{ass_epsilon} is sufficient to obtain a CLT for the sample mean of ${\epsilon}_{it}$ for each $i$.
Assumption \ref{ass_tkernel}\ref{ass_tkernel1} is standard in time series literature.
Assumption \ref{ass_tkernel}\ref{ass_tkernel2} is a high level assumption that is used for consistent estimation of long-run variance for each $i$. The conditions under which this assumption holds are given by \cite{jansson02}.
Consistency also requires Assumption \ref{ass_tkernel}\ref{ass_tsbw} which controls the expansion of $d_T$ relative to $T$.
\medskip

\textbf{Overall EPA tests under cross-sectional independence.}
We first consider the simplest case where $\boldsymbol{\lambda }_{i}^{\prime }\mathbf{f}_{t} = 0$ for all $i,t$ and $r_{ij} = 0$ for all $i \neq j$.
In order to test the null hypothesis $H^a_{0}$, we propose to use the following statistic:
\begin{equation}\label{eq13}
S_{nT}^{(1)} =\frac{\sqrt{nT} \Delta \bar{L}_{nT}}{\hat{{\sigma}}_{1,nT}%
},
\end{equation}
where ${\hat{\sigma}}_{1,nT}^{2}= (nT)^{-1}\sum_{i=1}^{n} \sum_{t,s=1}^{T} k_{T}\left( {d_{ts}/d_T}\right) \Delta \tilde{L}_{it}\Delta \tilde{L}_{is}$, with $\Delta \tilde{L}_{it}=\Delta L_{it}-\Delta \bar{L}_{i,T}$, where $\Delta \bar{L}_{i,T} = T^{-1}\sum_{t=1}^{T} \Delta {L}_{it}$ and $d_{ts} = |t-s|$.

\begin{proposition}\label{res1}
Suppose $\Delta L_{it}$ follows the model in \eqref{model} with $\boldsymbol{\lambda }_{i}^{\prime }\mathbf{f}_{t} = 0$ for all $i,t$, $r_{ij} = 0$ for all $i \neq j$, and Assumptions \ref{ass_epsilon}-\ref{ass_tkernel} hold.
Then, under $H^a_{0}$ and as $(T,n) \rightarrow \infty$, $S_{nT}^{(1)} \overset{D}{\rightarrow} N(0,1)$.
\end{proposition}
To prove this result, it is sufficient to show that $\sqrt{nT}(\Delta \bar{L}_{nT}-\bar{\mu}_{n})/\sigma_{1,nT} \overset{d}{\rightarrow} N(0,1)$ where $\sigma^2_{1,nT} = n^{-1}\sum_{i=1}^{n} \bar{\gamma}_{i,T}$ with $\bar{\gamma}_{i,T} = T^{-1} \sum_{t,s=1}^{T} {\gamma}_{i,d_{ts}}$, ${\gamma}_{i,d_{ts}} = \mathrm{E}(\epsilon_{it} \epsilon_{is})$, and $\hat{\sigma}_{1,nT}^{2} - \sigma^2_{1,nT} \overset{p}{\rightarrow} 0$. {It is easy to prove that the test statistic is divergent, hence consistent, under $H^a_{1}$.
To see this, note that in this case $\Delta \bar{L}_{nT} \overset{p}{\rightarrow} \bar{\mu} \neq 0$, $\hat{\sigma}_{1,nT}^{2} \overset{p}{\rightarrow} \sigma^2_{1} > 0$ where $\sigma^2_{1} = \lim_{n,T\to\infty} \sigma^2_{1,nT}$.
It follows that $P[|S_{nT}^{(1)}|>c]{\rightarrow}1$ for any constant $c \in \mathbb{R}$ as $(T,n) \rightarrow \infty$.
}
\medskip

\textbf{Overall EPA tests under WCD.}
Suppose $\boldsymbol{\lambda }_{i}^{\prime }\mathbf{f}_{t} = 0$ for all $i,t$ but $r_{ij}$ is not necessarily zero for all $i \neq j$.
In this case, the loss differentials $\Delta L_{it}$ are no longer independent across $i$.
Define $d_{ij}=d_{ji}\geq 0$ as the distance between units $i$ and $j$.
We make Assumptions \ref{ass_r1} and \ref{ass_r2} on the coefficients $r_{ij}$.

\begin{assumption}\label{ass_r1}
For all $n \in \mathbb{Z}^+$, $||\mathbf{R}_n||_{1}<\infty$ and $||\mathbf{R}_n||_{\infty}<\infty$.
\end{assumption}
\begin{assumption}\label{ass_r2}
$\sum_{j=1}^{n} |\mathbf{r}_{i.}'\mathbf{r}_{j.}|d^{\rho}_{ij}<\infty$ for some $\rho \geq 1$.
\end{assumption}

Assumption \ref{ass_r1} is standard in spatial econometrics literature and implies the WCD defined in \eqref{eq:wcdcond}.
The role of Assumption \ref{ass_r2} is to restrict the spatial correlation among panel units in relation to the distances between them.
As noted by \cite{kelejian07}, the corresponding condition in the time series context is fading memory over time.
Under this assumption, as the distance between two panel units increase, the correlation between them decreases.
This in turn sets a basis for using the spatial kernel function $k_{S}(\cdot)$ which gives smaller weights to the covariance between units which are more distant from each other.

To deal with the WCD when $T=1$, \cite{kelejian07} proposed a spatial heteroskedasticity and autocorrelation consistent estimator of the variance-covariance matrix.
This estimator can be seen as a spatial version of the kernel estimators for time series such that it uses a spatial kernel based on the distance between units.
The estimator is generalized by \cite{kim13} to panel data regression, by combining the spatial and time kernels.
Define $k_{q} = \lim_{x\to0} [1-k_S(x)]/|x|^{q}$ and let $\rho_s = \max \{ q : k_{q} < \infty \}$ be the Parzen exponent of $k_S(\cdot)$ \citep{andrews91}{, and $d_n$ a sequence of positive, non-random bandwidth parameters.}
Assumption \ref{ass_skernel} is placed on the spatial kernel function.

\begin{assumption}\label{ass_skernel}
\begin{enumerate*}[label=(\alph*)]
\item\label{ass_skernel1}
$k_S(x):\mathbb{R} \rightarrow [-1,1]$ is continuous at zero with $\rho_s \geq 1$,
$k_S(0)=1$,
$k_S(x) = k_S(-x)$ $\forall x \in \mathbb{R}$,
\item\label{ass_skernel3} $\max_{1 \leq i \leq n} \lim_{n \rightarrow \infty} d_n^{-1} \sum_{j=1}^{n} |k_{S} \left({d_{ij}/d_n} \right) | < \infty$,
\item\label{ass_csbw} $d_n \rightarrow \infty$ such that $d_n/n \rightarrow 0$ as $n \rightarrow \infty$.
\end{enumerate*}
\end{assumption}
The condition in \ref{ass_skernel}\ref{ass_skernel1} is satisfied by all kernel functions used in practice, such as Bartlett, Parzen, Tukey–Hanning and quadratic spectral \citep[see][]{andrews91}.
Conditions similar to that in \ref{ass_skernel}\ref{ass_skernel3} are used by \cite{kelejian07}, \cite{moscone12} and \cite{kim13}.
All these studies allow solely for kernels which truncate, i.e. those which equal zero after a certain value of the bandwidth parameter.
The first two papers place assumptions on the relative expansion of $l_n = \max_{1 \leq i \leq n} l_{i,n}$ where $l_{i,n} = \sum^n_{j=1} 1\{d_{ij}\leq d_n\}$.
Assumption \ref{ass_skernel}\ref{ass_tsbw} controls the expansion of $d_n$ relative to $n$.
This condition, as well as \ref{ass_skernel}\ref{ass_skernel3}, is not necessary for the consistent estimation of the variance-covariance matrices in our study.
Nevertheless, these conditions are retained to compare our results with the existing literature.

We propose the following test statistic in this case of WCD:
\begin{equation}\label{eq15}
S_{nT}^{(2)}=\frac{\sqrt{nT}\Delta \bar{L}_{nT}}{\hat{\sigma}_{2,nT}},  
\end{equation}
where
\begin{equation}\label{eq16}
\hat{\sigma}_{2,nT}^{2}=\frac{1}{nT}\sum_{i,j=1}^{n} \sum_{t,s=1}^{T} k_{S}\left(\frac{d_{ij}}{d_n}\right) k_{T}\left(\frac{d_{ts}}{d_{T}}\right) \Delta \tilde{L}_{it}\Delta \tilde{L}_{js}.
\end{equation}%
The theoretical properties of the estimator $\hat{\sigma}_{2,nT}^{2}$ are explored by \cite{kim13}.
\citet{moscone12} use a similar estimator with the difference being that they set $k_{T}\left( \cdot \right) =1$.

The disadvantage of this variance estimator is that for its implementation the distances between all pairs of units, $d_{ij}$ have to be known to the researcher.
Furthermore, the cut-off distance $d_{n}$ has to be chosen.
In practice there may be many possible distance metrics and economic theory does not always help to choose between them.
When a distance metric is not available we can use a partial sample estimator given by
\begin{equation}\label{eq162}
\underline{\hat{\sigma}}_{2,nT}^{2}=\frac{1}{\underline{n} T}\sum_{i,j=1}^{\underline{n}} \sum_{t,s=1}^{T} k_{T}\left( \frac{d_{ts}}{d_{T}}\right)  \Delta \tilde{L}_{it}\Delta \tilde{L}_{js},
\end{equation}
where $\underline{n}$, an increasing function of $n$, is the number of observations used to calculate the variance.
It is strictly smaller than $n$.
Similar variance estimators are used by \citet{bai06} and \citet{moscone15}.
The first study focuses on the factor models whereas the second one deals with panel regression models with small $T$.
Our variance estimator generalizes that of the \citet{moscone15} by allowing for a large $T$ with the help of a time kernel.
The corresponding test statistic is given by
\begin{equation*}
\underline{S}_{nT}^{(2)}=\frac{\sqrt{nT}\Delta \bar{L}_{nT}}{\underline{\hat{\sigma}}_{2,nT}}.
\end{equation*}

\begin{proposition}\label{res2}
Suppose $\Delta L_{it}$ follows the model in \eqref{model} with $\boldsymbol{\lambda }_{i}^{\prime }\mathbf{f}_{t} = 0$ for all $i,t$
and Assumptions \ref{ass_epsilon}-\ref{ass_skernel} hold. Then, under $H^a_{0}$ and as $(T,n) \rightarrow \infty$,
\begin{enumerate*}[label=(\roman*)]
\item $S_{nT}^{(2)} \overset{d}{\rightarrow} N(0,1)$ if $d_n \rightarrow \infty$,
\item and $\underline{S}_{nT}^{(2)} \overset{d}{\rightarrow} N(0,1)$ if $\underline{n} \rightarrow \infty$ such that $\underline{n}/n \rightarrow \kappa \in [0,1]$.
\end{enumerate*}
\end{proposition}

Under Assumptions \ref{ass_epsilon}-\ref{ass_skernel}, $\sqrt{nT}(\Delta \bar{L}_{nT}-\bar{\mu}_{n})/\sigma_{2,nT} \overset{d}{\rightarrow} N(0,1)$ as $(T,n) \rightarrow \infty$, where $\sigma_{2,nT}^{2} = n^{-1} \sum_{i,j=1}^{n}\mathbf{r}_{i.}' \bar{\boldsymbol{\gamma}}_{nT} \mathbf{r}_{j.}$ with $\bar{\boldsymbol{\gamma}}_{nT} = T^{-1} \sum_{t,s=1}^{T} \boldsymbol{\gamma}_{n,d_{ts}}$ and $\boldsymbol{\gamma}_{n,d_{ts}} = \mathrm{diag}(\gamma_{1,d_{ts}},\gamma_{2,d_{ts}},\dots,\gamma_{n,d_{ts}})$.
Hence, the stated results follow by the consistency of the variance estimators.
{Similar arguments made in the case of no CD lead to the consistency of the test statistics in the case which the alternative $H^a_{1}$ holds.}

{The test statistic $\underline{S}_{nT}^{(2)}$ is an asymptotic one which, in practice, can be hard to implement.
Despite its advantage of not relying on a distance metric, it needs special care as there is not clear indications on which cross-sectional units to use in the calculation.
A solution in practice is to fix $\underline{n}$, calculate the test statistic for a large number of subsamples of cross-sectional units of size $\underline{n}$, and to take the smallest test statistic.
This provides a conservative test statistic which would be correctly sized but potentially with low power.
Nevertheless, below, we consider another test statistic, ${S}_{nT}^{(3)}$, which is robust to WCD and does not pose similar problems.
}
\medskip

\textbf{Overall EPA tests under SCD.} This is the most general case with no specific restriction imposed on the CD of the loss differentials. To obtain a CLT for the means, we make Assumption \ref{ass_comcomp} on the common factors, their loadings and the error terms.

\begin{assumption}\label{ass_comcomp}
\begin{enumerate*}[label=(\alph*)]
\item\label{ass_comcomp_factors} $\mathbf{f}_{t}$ is independent of $\varepsilon_{it}$ and follows the linear process $\mathbf{f}_{t} = \sum_{h=0}^{\infty} \mathbf{C}_{h} \Psi_{t-h}$, with $\Psi_{t} \sim iid(0,\mathbf{I}_m)$, $\mathrm{E} ||\Psi_{t}||^4 < \infty$, $\sum_{h=0}^{\infty} h || \mathbf{C}_{h} || < M < \infty $ and $\sum_{h=0}^{\infty} \mathbf{C}_{h}$ is full rank.
\item\label{ass_comcomp_loadingssecondorder} Factor loadings $\boldsymbol{\lambda}_{i}$ are fixed parameters such that $||\boldsymbol{\lambda}_{i}|| < \infty$, $n^{-1} \sum_{i=1}^{n} {\boldsymbol{\lambda }}_{i} {\boldsymbol{\lambda }}'_{i} {\rightarrow} \boldsymbol{\Sigma}_{\lambda}>0$ for an $m \times m$ matrix $\boldsymbol{\Sigma}_{\lambda}$.
\item\label{ass_comcomp_loadingsfirstorder} There exists at least one common factor $f_{kt}$, $k \in \{1,\dots,m\}$, for which the loadings satisfy $|n^{-1}\sum_{i=1}^{n}\lambda_{ki}| > 0$ for all $n \in \mathbb{Z}^+$.
\end{enumerate*}
\end{assumption}

%

It is easy to see that, ${\boldsymbol{\lambda }}_{i}'\mathbf{f}_{t}$ satisfying Assumption \ref{ass_comcomp} lead to SCD as defined in \eqref{eq:scdcond}.
We require Assumption \ref{ass_comcomp}\ref{ass_comcomp_factors} for the consistent estimation of the long-run variance of the common factors.
Assumption \ref{ass_comcomp}\ref{ass_comcomp_loadingssecondorder} is standard for factor models.
Assumption \ref{ass_comcomp}\ref{ass_comcomp_loadingsfirstorder} ensures that at least one of the factors contribute to the asymptotic variance of the cross-sectional averages.
Although nonstandard, this assumption is not restrictive as it does not affect the validity of our testing procedures.

In this case of SCD, the variance estimator given in (\ref{eq16}) can be modified by setting $k_{S}(\cdot)=1$.
This variance estimator does not require any knowledge of a distance measure between the units.
Moreover, it assigns weights equal to one for all covariances from the same time period, hence {it is} robust to SCD as well as WCD.
The test statistic takes the form: 
\begin{equation}\label{eq17}
S_{nT}^{(3)}=\frac{\sqrt{T}\Delta \bar{L}_{nT}}{\hat{\sigma}_{3,nT}},
\end{equation}
where 
\begin{equation}\label{eq18}
\hat{\sigma}_{3,nT}^{2}=\frac{1}{n^2T}\sum_{i,j=1}^{n} \sum_{t,s=1}^{T} k_{T}\left( \frac{d_{ts}}{d_{T}}\right) \Delta \tilde{L}%
_{it}\Delta \tilde{L}_{js}.
\end{equation}%
The variance estimator \eqref{eq18} is valid when $T$ is large, regardless of $n$ being finite or infinite \citep[see][]{driscoll98}.
{The special case of this test statistic where $k_{T}$ is the Bartlett kernel corresponds to the test statistic $J_{n,T}^{DM}$ of QTZ.
As the authors note, an important advantage of this test statistic is that it does not rely on the linear approximate factor model in \eqref{model}.
It is robust even under DGPs such as $\Delta L_{it} = \mu_i + g(\boldsymbol{\lambda }_{i},\mathbf{f}_{t})+\varepsilon_{it}$ where $g(\cdot,\cdot)$ is a non-parametric function.
Furthermore, the test statistic $S_{nT}^{(3)}$ is very easy to calculate as it is identical to the DM test statistic for the cross-sectional averages of loss differentials.
To see this, it suffices to write $\Delta \bar{L}_{nT} = T^{-1} \sum_{t=1}^{T} \bar{L}_{n,t}$ and $\hat{\sigma}_{3,nT}^{2}=T^{-1} \sum_{t,s=1}^{T} k_{T}\left(d_{ts}/d_{T}\right) \Delta \tilde{L}_{n,t}\Delta \tilde{L}_{n,s}$, where $\bar{L}_{n,t} = n^{-1} \sum_{i=1}^{n} \Delta L_{it}$ and $\Delta \tilde{L}_{n,t} = \bar{L}_{n,t} - \Delta \bar{L}_{nT}$.}

An alternative way to estimate the covariance matrix is to exploit the factor structure of the DGP.
The PCE of large panels is investigated by \citet{stock02}, \citet{bai02}, \citet{bai03}, among others.
This is a flexible estimator which is robust to WCD and autocorrelation in the error terms.
It is obtained by minimizing the average squared residuals computed for $m$ common factors:
\begin{equation}\label{eq128}
V(m) = \frac{1}{nT} \sum_{i=1}^{n} \sum_{t=1}^{T} (\Delta \tilde{L}_{it} - \boldsymbol{\lambda }_{i}^{\prime }{\mathbf{f}}_{t})^2
\end{equation}%
subject to $\mathrm{Var}(\mathbf{f}_{t})=\mathbf{I}_{m}$ and $\boldsymbol{\Lambda}_n'\boldsymbol{\Lambda}_n$ being diagonal where $\boldsymbol{\Lambda}_n = (\boldsymbol{\lambda}_1,\boldsymbol{\lambda}_2,\dots,\boldsymbol{\lambda}_n)'$.
Then the solution for the estimates of the common factors, $\widehat{\mathbf{f}}_{t}$, are given by $\sqrt{T}$ times the first $m$
eigenvectors of the matrix $\sum_{i=1}^{n} \Delta \tilde{\mathbf{L}}_{i.} \Delta \tilde{\mathbf{L}}_{i.}^{\prime}$ with $\Delta \tilde{\mathbf{L}}_{i.}=(\Delta \tilde{L}_{i1},\Delta \tilde{L}_{i2},\dots ,\Delta \tilde{L}_{iT})^{\prime}$ and the factor loadings can be estimated as $\widehat{\boldsymbol{\lambda }}_{i}=\frac{1}{T} \sum_{t=1}^{T}\widehat{\mathbf{f}}_{t}\Delta \tilde{L}_{it}$.
{We make the following high level assumption on the asymptotic properties of the PC estimates.

\begin{assumption}\label{ass_commoncomprate}
$\delta_{nT}(\widehat{\boldsymbol{\lambda}}'_i \widehat{\mathbf{f}}_{t} - \boldsymbol{\lambda}'_i  {\mathbf{f}}_{t}) = O_p(1)$ where $\delta_{nT} =\min (\sqrt{T},\sqrt{n})$.
\end{assumption}
\noindent This is a standard result in the literature on the approximate factor models. The conditions under which it holds are given in \cite{bai03} (see their Theorem 3).}
Then the overall EPA hypothesis can be tested using
\begin{equation}\label{eq1732}
\underline{S}_{nT}^{(3)}=\frac{\sqrt{T}\Delta \bar{L}_{nT}}{\hat{\underline{\sigma}}_{3,nT}},
\end{equation}
where 
\begin{equation}\label{eq1722}
\hat{\underline{\sigma}}_{3,nT}^{2} = \frac{1}{n^2T} \sum_{i,j=1}^{n}\sum_{t,s=1}^{T} k_{T}\left( \frac{d_{ts}}{d_{T}}\right) \widehat{\boldsymbol{\lambda }}'_{i} \widehat{\mathbf{f}}_{t} \widehat{\mathbf{f}}'_{s} \widehat{\boldsymbol{\lambda }}_{j} + \frac{1}{n^2T} \sum_{i=1}^{n} \sum_{t,s=1}^{T} k_{T}\left( \frac{d_{ts}}{d_{T}}\right) \widehat{\varepsilon}_{it} \widehat{\varepsilon}_{is},
\end{equation}
with $\widehat{\varepsilon}_{it} = \Delta \tilde{L}_{it} - \widehat{\boldsymbol{\lambda }}'_{i} \widehat{\mathbf{f}}_{t}$.

\begin{proposition}\label{res3}
Suppose $\Delta L_{it}$ follows the model in \eqref{model}, Assumptions \ref{ass_epsilon}-\ref{ass_r1}, \ref{ass_comcomp} and \ref{ass_commoncomprate} hold. Then, under $H^a_{0}$ and as $(T,n) \rightarrow \infty$,
\begin{enumerate*}[label=(\roman*)]
\item $S_{nT}^{(3)} \overset{d}{\rightarrow} N(0,1)$,
\item and $\underline{S}_{nT}^{(3)} \overset{d}{\rightarrow} N(0,1)$.
\end{enumerate*}
%
\end{proposition}

Under Assumptions \ref{ass_epsilon}-\ref{ass_r1} and \ref{ass_comcomp}, $\sqrt{T}(\Delta \bar{L}_{nT}-\bar{\mu}_{n})/\sigma_{3,nT} \overset{d}{\rightarrow} N(0,1)$, where
\begin{equation}\label{eqscdvar}
\sigma^2_{3,nT} = \frac{1}{n^2} \sum_{i,j=1}^{n} \left( \boldsymbol{\lambda}'_i \bar{\boldsymbol{\Gamma}}_{T} \boldsymbol{\lambda}_j + \mathbf{r}_{i.}' \bar{\boldsymbol{\gamma}}_{nT} \mathbf{r}_{j.} \right),
\end{equation}
with $\bar{\boldsymbol{\Gamma}}_{T} = T^{-1} \sum_{t,s=1}^{T} \boldsymbol{\Gamma}_{d_{ts}}$.
The rate of convergence in the CLT is $T^{1/2}$ instead of the usual rate of $(nT)^{1/2}$ in the cases of no CD and WCD.
This follows from the SCD characterized in \eqref{eq:scdcond}.
{We can show the consistency of the test statistics using similar arguments to the ones made earlier, after Propositions \ref{res1} and \ref{res2}.
It can be easily shown that the variance estimator $\hat{\sigma}_{3,nT}^{2}$ is consistent under no CD as well as WCD.
Hence the test statistic $S_{nT}^{(3)}$ is consistent in these cases as well.
However, the rate of divergence is smaller than those of Propositions \ref{res1} and \ref{res2}.
Thus, the test is expected to have a lower power against the null $H^a_{0}$.
}

In \eqref{eqscdvar}, the first term in parentheses dominate{s} the second one.
This is because, under Assumptions \ref{ass_epsilon}, \ref{ass_r1} and \ref{ass_comcomp}, $n^{-2} \sum_{i,j=1}^{n} \boldsymbol{\lambda}'_i \bar{\boldsymbol{\Gamma}}_{T} \boldsymbol{\lambda}_j = O(1)$ but $n^{-2} \sum_{i,j=1}^{n} \mathbf{r}_{i.}' \bar{\boldsymbol{\gamma}}_{nT} \mathbf{r}_{j.} = O(1/n)$.
Hence, the latter is asymptotically negligible.
This means that under SCD, one can use a simpler variance estimator without the second term in \eqref{eq1722}.

{
When $T$ is fixed but $n$ is large, the test statistic $S_{nT}^{(3)}$ is not valid as it relies on the $T$-asymptotics.
However, under additional assumptions on the DGP of the loss differentials, we can obtain a test statistic for fixed $T$.
Specifically, assume that the loss differentials are based on optimal one-step ahead forecasts in the sense that they are serially uncorrelated.
Consider the following test statistic:
\begin{equation}\label{eq173}
\tilde{S}_{nT}^{(3)}=\frac{\sqrt{T}\Delta \bar{L}_{nT}}{{\tilde{\sigma}}_{3,nT}},
\end{equation}
where
\begin{equation}\label{eq182}
\tilde{\sigma}_{3,nT}^{2}=\frac{1}{T-1} \sum_{t=1}^{T} \Delta \tilde{L}_{n,t}^2.
\end{equation}
We have the following result:
\begin{corollary}
Suppose that the loss differentials $\Delta L_{it}$ follows the model in \eqref{model}, they are serially uncorrelated and one of the following holds:
\begin{enumerate*}[label=(\alph*)]
\item $\boldsymbol{\lambda }_{i}^{\prime }\mathbf{f}_{t} = 0$ for all $i,t$
and Assumptions \ref{ass_epsilon}-\ref{ass_skernel} hold;
\item Assumptions \ref{ass_epsilon}-\ref{ass_r1}, \ref{ass_comcomp} and \ref{ass_commoncomprate} hold and $\epsilon_{it} \sim N(0,v_i^2)$.
\end{enumerate*}
Then, under $H^a_{0}$ and as $n \rightarrow \infty$, $\tilde{S}_{nT}^{(3)}\overset{d}{\rightarrow} t(T-1)$.
\end{corollary}
Here $t(T-1)$ represents the Student's $t$-distribution with $T-1$ degrees of freedom.
The result shows that, under additional assumptions to those of Propositions \ref{res2} and \ref{res3}, we can use the Student's $t$ critical values after adjusting the degrees of freedom in the test statistic ${S}_{nT}^{(3)}$.
If the loss differentials carry SCD, we need normality of the error terms for the result to hold.
If they are weakly cross-sectionally correlated, the normality assumption is not required as under WCD assumption cross-sectional averages of the loss differentials admit a CLT, i.e. they are normally distributed for each $t=1,2,\dots,T$.
}


\subsection{Tests for clustered equal predictive ability: $C$-statistics}\label{sec:jointtests}

Define $G_g$, $g=1,\dots,G$, as the set of indexes of $n_g$ cross-sectional units which belong to cluster $g$ such that $G_{g} \cap G_{g'} = \emptyset$, $\forall g \neq g'$. In this subsection, we are interested in testing the null hypothesis $H^{ca}_{0}$ which can be written as
\begin{equation}\label{eq5_v2}
H^{ca}_{0}: \bar{\boldsymbol{\mu}} = \mathbf{0},
\end{equation}
where $\bar{\boldsymbol{\mu}} =(\bar{\mu}_{1},\bar{\mu}_{2},\dots,\bar{\mu}_{G})^{\prime }$.
Our tests are based on the empirical counterpart of this quantity: $\Delta\bar{\mathbf{L}}_{nT}=(\Delta \bar{L}_{1,n_1,T},\Delta \bar{L}_{2,n_2,T}, \ldots ,\Delta \bar{L}_{G,n_G,T})^{\prime }$ where $\Delta \bar{L}_{g,n_g,T} = (n_gT)^{-1} \sum_{i \in G_g} \sum_{t = 1}^T \Delta L_{it}$.
We assume that the sets of indexes $G_g$, $g=1,\dots,G$ are known.
Assumption \ref{ass_groupnumbers} is placed to control the asymptotic number of units per cluster.

\begin{assumption}\label{ass_groupnumbers}
{For all $G>1$, }as $n\rightarrow \infty$, $n_g/n \rightarrow \tau_g \in (0,1)$ for each $g=1,\dots,G$.
\end{assumption}

With this assumption we do not rule out the possibility of having $G = 1$ in which case we have $n_1 = n$. This particular case corresponds to the overall EPA tests of the previous subsection.
In what follows, these test statistics are generalized for $G > 1$ for each case of CD.

\textbf{Clustered EPA tests under cross-sectional independence.} In this case $\boldsymbol{\lambda }_{i}^{\prime }\mathbf{f}_{t} = 0$ for all $i,t$ and $r_{ij}
= 0$ for all $i \neq j$.
We propose the following statistic to test the hypothesis in \eqref{eq5_v2}:

\begin{equation}\label{eq2421}
C^{(1)}_{nT} = nT\Delta\bar{\mathbf{L}}'_{nT} \widehat{\boldsymbol{\Omega}}_{1,nT}^{-1} \Delta\bar{\mathbf{L}}_{nT},
\end{equation}
where
\begin{equation*}
\widehat{\boldsymbol{\Omega}}_{1,nT} = \frac{1}{T}\sum_{i=1}^{n} \frac{n}{n^2_{g_i}} \sum_{t,s=1}^{T} k_{T}\left( \frac{d_{ts}}{d_{T}}\right) \boldsymbol{\iota}_{g_i}\boldsymbol{\iota}_{g_i}^{\prime} \Delta \tilde{L}_{it}\Delta \tilde{L}_{is},
\end{equation*}
with $g_i \in \{1,2,\dots,G\}$ being a variable which states the cluster which $i$th unit belongs to and $\boldsymbol{\iota}_{g_i}$ being the $g_i$th column of $\mathbf{I}_{G}$. Notice that $\widehat{\boldsymbol{\Omega}}_{1,nT}$ is a diagonal matrix which contains an estimate of the average long-run variances of each cluster as diagonal elements {up to a factor which is asymptotically equal to $\tau^{-1}_{g}$}.

\begin{proposition}\label{res4}
Suppose $\Delta L_{it}$ follows the model in \eqref{model} with $\boldsymbol{\lambda }_{i}^{\prime }\mathbf{f}_{t} = 0$ for all $i,t$, $r_{ij} = 0$ for all $i \neq j$, and Assumptions \ref{ass_epsilon}, \ref{ass_tkernel} and \ref{ass_groupnumbers} hold. Then, under $H^{ca}_{0}$ and as $(T,n) \rightarrow \infty$, $C^{(1)}_{nT} \overset{d}{\rightarrow} \chi^2_G$.
\end{proposition}

Under Assumptions \ref{ass_epsilon}, \ref{ass_tkernel} and \ref{ass_groupnumbers}, we have $\sqrt{nT}\boldsymbol{\Omega}_{1,nT}^{-1/2}(\Delta\bar{\mathbf{L}}_{nT} - \bar{\boldsymbol{\mu}}_{n})\overset{d}{\rightarrow} N(\mathbf{0},\mathbf{I}_{G})$ as $(T,n) \rightarrow \infty$, where $\bar{\boldsymbol{\mu}}_{n} =(\bar{\mu}_{1,n_1},\bar{\mu}_{2,n_2},\dots,\bar{\mu}_{G,n_G})^{\prime }$ and $\boldsymbol{\Omega}_{1,nT} = \sum_{i=1}^{n} \frac{n}{n^2_{g_i}} \boldsymbol{\iota}_{g_i}\boldsymbol{\iota}_{g_i}^{\prime} \bar{\gamma}_{i,T}$.
This is a generalization of the CLT following Proposition \ref{res1} which is obtained when $G=1$.
{It is easy to see that the test is consistent under the alternative hypothesis $H^{ca}_{1}$.
This is because under the assumptions of the proposition, the matrix $\widehat{\boldsymbol{\Omega}}_{1,nT}$ tends to a positive definite matrix as $(T,n) \rightarrow \infty$ whereas at least one entry of the vector $\Delta\bar{\mathbf{L}}_{nT}$ tends to a non-zero constant provided that there exists a $g \in \{1,\dots,G \}$ for which $\bar{\mu}_{g} \neq 0$.
Furthermore, the test is consistent against $H^a_{0}$ as this hypothesis fails if any cluster has a non-zero mean.
However, we expect this test to have lower power against $H^a_{0}$ compared to $S^{(1)}_{nT}$.}

\textbf{Clustered EPA tests under WCD.} Suppose $\boldsymbol{\lambda }_{i}^{\prime }\mathbf{f}_{t} = 0$ for all $i,t$ but $r_{ij}$ is not necessarily zero for all $i \neq j$.
We can use the following statistic in order to test $H^{ca}_{0}$:
\begin{equation}\label{eq2422}
C^{(2)}_{nT} = nT\Delta\bar{\mathbf{L}}'_{nT} \widehat{\boldsymbol{\Omega}}_{2,nT}^{-1} \Delta\bar{\mathbf{L}}_{nT},
\end{equation}
where
\begin{equation*}
\widehat{\boldsymbol{\Omega}}_{2,nT} = \frac{1}{T}\sum_{i,j=1}^{n} \frac{n}{n_{g_i} n_{g_j}} k_{S} \left(\frac{d_{ij}}{d_n}\right) \sum_{t,s=1}^{T} k_{T}\left( \frac{d_{ts}}{d_{T}}\right)  \boldsymbol{\iota}_{g_i}\boldsymbol{\iota}_{g_j}^{\prime} \Delta \tilde{L}_{it}\Delta \tilde{L}_{js}.
\end{equation*}

As discussed in the previous section, the estimator has the disadvantage of relying on a known distance between each unit in the panel.
An alternative variance estimator for this case of cross-sectional clusters can be constructed as in \eqref{eq162}.
Such an estimator is
\begin{equation}\label{eq2322}
\underline{\widehat{\boldsymbol{\Omega}}}_{2,nT} = \frac{1}{\underline{n}T}\sum_{i=1}^{\underline{n}_{g_i}} \sum_{j=1}^{\underline{n}_{g_j}} \frac{{n}^2}{{n}_{g_i} {n}_{g_j}} \sum_{t,s=1}^{T} k_{T}\left( \frac{d_{ts}}{d_{T}}\right)  \boldsymbol{\iota}_{g_i}\boldsymbol{\iota}_{g_j}^{\prime} \Delta \tilde{L}_{it}\Delta \tilde{L}_{js},
\end{equation}
where $\underline{n}_{g}$ is the number of observations taken into account in the calculation of variance in cluster $g$ hence we have $\underline{n} = \sum_{g=1}^{G} \underline{n}_{g}$.
The corresponding test statistic is
\begin{equation}
\underline{C}^{(2)}_{nT} = nT\Delta\bar{\mathbf{L}}'_{nT} \underline{\widehat{\boldsymbol{\Omega}}}_{2,nT}^{-1} \Delta\bar{\mathbf{L}}_{nT}.
\end{equation}

\begin{theorem}\label{res5}
Suppose $\Delta L_{it}$ follows the model in \eqref{model} with $\boldsymbol{\lambda }_{i}^{\prime }\mathbf{f}_{t} = 0$ for all $i,t$
and Assumptions \ref{ass_epsilon}-\ref{ass_skernel} and \ref{ass_groupnumbers} hold. Then, under $H^{ca}_{0}$ and as $(T,n) \rightarrow \infty$, 
\begin{enumerate*}[label=(\roman*)]
\item $C_{nT}^{(2)} \overset{d}{\rightarrow}  \chi^2_G$ if $d_n \rightarrow \infty$,
\item and $\underline{C}_{nT}^{(2)} \overset{d}{\rightarrow} \chi^2_G$ if as $\underline{n}_{g}\rightarrow \infty,\forall g = 1,\dots,G$ such that $\underline{n}_{g}/{n}_{g} \rightarrow \kappa \in [0,1]$.
\end{enumerate*}
\end{theorem}

Under the assumptions of the theorem, we have $\sqrt{nT}\boldsymbol{\Omega}_{2,nT}^{-1/2}(\Delta\bar{\mathbf{L}}_{nT} - \bar{\boldsymbol{\mu}}_{n})\overset{d}{\rightarrow} N(\mathbf{0},\mathbf{I}_{G})$ as $(T,n) \rightarrow \infty$, where $\boldsymbol{\Omega}_{2,nT} = \sum_{i,j=1}^{n} \frac{n}{n_{g_i} n_{g_j}}  \boldsymbol{\iota}_{g_i}\boldsymbol{\iota}_{g_j}^{\prime} \mathbf{r}_{i.}' \bar{\boldsymbol{\gamma}}_{nT} \mathbf{r}_{j.}$. Then the stated results follow from the consistency of the variance estimators.
This theorem nests Propositions \ref{res1}, \ref{res2} and \ref{res4}.
The first is obtained when $\mathbf{R}_n = \mathbf{I}_n$ and $G=1$, the second is obtained with only $G=1$ and the last is when only $\mathbf{R}_n = \mathbf{I}_n$.
{The consistency of the tests follow from similar arguments to the ones above, those under Proposition \ref{res4}.}

\medskip

\textbf{Clustered EPA tests under SCD.} We now consider the case of no specific restrictions on the CD in the loss differentials. With SCD, the clustered EPA hypothesis $H^{ca}_{0}$ can be tested using:
\begin{equation}\label{eq2423}
C^{(3)}_{nT} = T\Delta\bar{\mathbf{L}}'_{nT} \widehat{\boldsymbol{\Omega}}_{3,nT}^{-1} \Delta\bar{\mathbf{L}}_{nT},
\end{equation}
\begin{equation*}\label{eq232}
\widehat{\boldsymbol{\Omega}}_{3,nT} = \frac{1}{T}\sum_{i,j=1}^{n} \frac{1}{n_{g_i} n_{g_j}} \sum_{t,s=1}^{T} k_{T}\left( \frac{d_{ts}}{d_{T}}\right)  \boldsymbol{\iota}_{g_i}\boldsymbol{\iota}_{g_j}^{\prime} \Delta \tilde{L}_{it}\Delta \tilde{L}_{js}.
\end{equation*}
As discussed previously, this type of variance estimator is robust to arbitrary CD and it has the advantage of not requiring known distances between units.
{Furthermore, like $S^{(3)}_{nT}$, it is very easy to calculate as it is identical to a Wald test applied to within-cluster cross-sectional averages where between cluster covariances as well as serial correlations are taken into account.}
However, its performance may be poor in cases of $n$ large relative to $T$.
Hence, once more the factor structure of the loss differentials can be used to form variance estimators.
A test statistic with such an estimate is
\begin{equation}\label{eq2424}
\underline{C}_{nT}^{(3)} = T\Delta\bar{\mathbf{L}}'_{nT} \underline{\widehat{\boldsymbol{\Omega}}}_{3,nT}^{-1} \Delta\bar{\mathbf{L}}_{nT},
\end{equation}
\begin{equation}
\underline{\widehat{\boldsymbol{\Omega}}}_{3,nT} = \frac{1}{T}\sum_{i,j=1}^{n} \frac{1}{n_{g_i} n_{g_j}} \sum_{t,s=1}^{T} k_{T}\left( \frac{d_{ts}}{d_{T}}\right) \boldsymbol{\iota}_{g_i}\boldsymbol{\iota}_{g_j}^{\prime} \widehat{\boldsymbol{\lambda }}'_{i} \widehat{\mathbf{f}}_{t} \widehat{\mathbf{f}}'_{s} \widehat{\boldsymbol{\lambda }}_{j} + \frac{1}{T}\sum_{i=1}^{n} \frac{1}{n^2_{g_i}} \sum_{t,s=1}^{T} k_{T}\left( \frac{d_{ts}}{d_{T}}\right) \boldsymbol{\iota}_{g_i}\boldsymbol{\iota}_{g_i}^{\prime} \widehat{\varepsilon}_{it} \widehat{\varepsilon}_{is}.
\end{equation}

\begin{theorem}\label{res6}
Suppose $\Delta L_{it}$ follows the model in \eqref{model}, Assumptions \ref{ass_epsilon}-\ref{ass_r1}, \ref{ass_comcomp}-\ref{ass_groupnumbers} hold. Then, under $H^{ca}_{0}$ and as $(T,n) \rightarrow \infty$,
\begin{enumerate*}[label=(\roman*)]
\item $C_{nT}^{(3)} \overset{d}{\rightarrow} \chi^2_G$,
\item and $\underline{C}_{nT}^{(3)} \overset{d}{\rightarrow} \chi^2_G$.
\end{enumerate*}
%
\end{theorem}

Under the assumptions of the theorem, we have $\sqrt{T}\boldsymbol{\Omega}_{3,nT}^{-1/2}(\Delta\bar{\mathbf{L}}_{nT} - \bar{\boldsymbol{\mu}}_{n})\overset{d}{\rightarrow} N(\mathbf{0},\mathbf{I}_{G})$, where
$
\boldsymbol{\Omega}_{3,nT} = \sum_{i,j=1}^{n} \frac{1}{n_{g_i} n_{g_j}} \boldsymbol{\iota}_{g_i}\boldsymbol{\iota}_{g_j}^{\prime} \left( \boldsymbol{\lambda}'_i \bar{\boldsymbol{\Gamma}}_{T} \boldsymbol{\lambda}_j + \mathbf{r}_{i.}' \bar{\boldsymbol{\gamma}}_{nT} \mathbf{r}_{j.} \right)
$.
Proposition \ref{res3} is a special case of this theorem with $G=1$.
It is easy to prove the consistency of the tests following arguments similar to the ones under Proposition \ref{res4} and Theorem \ref{res5}.
Similar to the discussion following Proposition \ref{res3}, $C_{nT}^{(3)}$ is expected to be consistent under WCD but have lower power than $C_{nT}^{(2)}$ in this case.

Furthermore, it is easy to see that the test is consistent against $H^a_{0}$, following similar arguments of the discussion of Proposition \ref{res4}.
However, it is important to note that none {of} the overall EPA test statistics is consistent under general alternatives of $H^{ca}_{0}$.
To see this, suppose that $G=2$ with $\bar{\mu}_{1,n_1} \to \mu^0 \neq 0$, $\bar{\mu}_{2,n_2} \to -\mu^0$ as $n\to\infty$.
Then $\bar{\mu}_n \to 0$ as $n\to\infty$, hence, the overall EPA hypothesis holds asymptotically while the clustered EPA hypothesis fails.
Therefore, the $S$-statistics follow their asymptotic null distributions under their respective assumptions, as stated in Propositions \ref{res1}-\ref{res3} and in Corollary 1.

\textbf{Comparison of $C$-statistics and $J_n^D$ of QTZ.} QTZ propose a test statistic for the clustered EPA hypothesis of the form
\begin{equation*}
J_n^D = \frac{\sqrt{G}\bar{D}}{\sqrt{(G-1)^{-1}\sum_{g=1}^G(D_g-\bar{D})^2}},
\end{equation*}
where $D_g = \frac{1}{\sqrt{n_gT}} \sum_{i \in G_g} \sum_{t = 1}^T \Delta L_{it}$ and $\bar{D} = G^{-1}\sum_{g=1}^G D_g$ and they suggest to reject the null $H^{ca}_{0}$ if $|J_n^D|>t_{G-1,1-\alpha/2}$ where $t_{G-1,1-\alpha}$ is the $1-\alpha$ quantile of the $t(G-1)$ distribution.
To see how $J_n^D$ is related to the test statistics that we propose, let us first suppose that $n_g=n/G$ for all $g=1,\dots,G$, that is, the cluster sizes are identical.
Then, the numerator of $J_n^D$ is equal to $\sqrt{nT}\Delta \bar{L}_{nT}$, the numerator of our $S$-statistics, and that of $J_{n,T}^{DM}$ of QTZ:
\begin{equation*}
\begin{split}
\sqrt{G}\bar{D} &= \sqrt{G} \left(  G^{-1}\sum_{g=1}^G \frac{1}{\sqrt{n_gT}} \sum_{i \in G_g} \sum_{t = 1}^T \Delta L_{it} \right) \\
&= \frac{1}{\sqrt{G}}\sum_{g=1}^G \sqrt{n_gT} \frac{1}{n_gT} \sum_{i \in G_g} \sum_{t = 1}^T \Delta L_{it} \\
&= \frac{1}{\sqrt{G}}\sum_{g=1}^G \sqrt{\frac{n}{G}T} \Delta \bar{L}_{g,n_g,T} \\
&= \sqrt{nT}\Delta \bar{L}_{nT}.
\end{split}
\end{equation*}
Hence, the test statistic is closer to the $S$-statistics rather than the $C$-statistics.

To have a clear idea on the asymptotic properties of $J_n^D$, let us suppose now that $G=2$ and the cluster sizes are asymptotically identical: $\tau_1=\tau_2=1/2$.
Suppose also that $\bar{\mu}_{1,n_1} \to \mu^0 \neq 0$, $\bar{\mu}_{2,n_2} \to -\mu^0$ as $n\to\infty$, therefore $\bar{\mu}_n \to 0$.
In this case $H^a_{0}$ holds and the asymptotic distribution of $\sqrt{G}\bar{D}$ is identical to that of $\sqrt{nT}\Delta \bar{L}_{nT}$ under WCD.
Hence, even if the clustered EPA hypothesis fails, the test cannot detect the deviations from it.


\subsection{Special cases and extensions}

\textbf{Tests for joint equal predictive ability.} In macroeconomic applications, the differences in the predictive ability for each cross-sectional unit can have a
specific economic importance. When this is the case, one may be interested in the following hypothesis:
\begin{equation*}\label{eq625}
H_{0}^j : \mathrm{E}(\Delta L_{it}) = \mu_i = 0 \text{, \ for all }i=1, 2, \dots,n.
\end{equation*}
This hypothesis, namely \textit{the joint EPA hypothesis}, states that the EPA hypothesis holds for each unit in the sample.
This hypothesis can be seen as a special case of the clustered EPA hypothesis $H^{ca}_{0}$ with the number of clusters being equal to the number of units in the panel, that is $G=n$.
However, in this case Assumption \ref{ass_groupnumbers} is violated as for each $g=1,\dots,G$, $n_g = 1$ and therefore $n_g/n \rightarrow 0$. Nevertheless, the hypothesis can still be tested using the above test statistics if $n$ is fixed, after suitable modification of the convergence rates.
For instance, a test statistic which is robust to arbitrary CD is
\begin{equation*}\label{eq24}
J_{nT}=T\Delta\bar{\mathbf{L}}_{nT}^{\prime }\widehat{%
\boldsymbol{\Omega}}_{nT}^{-1}\Delta\bar{\mathbf{L}}_{nT}%
\overset{D}{\rightarrow } \chi^2_n,
\end{equation*}
\begin{equation*}\label{eq26}
\widehat{\boldsymbol{\Omega}}_{nT}=\frac{1}{T}\sum_{i,j=1}^{n} \sum_{t,s=1}^{T}k_{T}\left( \frac{d_{ts}}{d_{T}}\right)
\boldsymbol{\iota}_{i}\boldsymbol{\iota}_{j}^{\prime }\Delta \tilde{L}_{it}\Delta \tilde{L}_{js},
\end{equation*}
with ${\Delta }\bar{\mathbf{L}}%
_{nT}=(\Delta \bar{L}_{1T},\Delta \bar{L}_{2T}, \ldots ,\Delta \bar{L}%
_{nT})^{\prime }$, and $\boldsymbol{\iota}_{i}$ being the $i$th column of $\mathbf{I}_{n}$.
Different cases of CD can be covered by modifications on the variance estimator.
These cases are discussed at length in a previous version of this paper \citep{akgun20}, where we also consider the case of large $n$.

\textbf{Tests for individual cross-sections.}
The overall test statistics $S_{nT}^{(1)}$, $S_{nT}^{(2)}$ and $\underline{S}_{nT}^{(2)}$, and clustered test statistics $C_{nT}^{(1)}$, $C_{nT}^{(2)}$ and $\underline{C}_{nT}^{(2)}$ are applicable to a single cross-section{, i.e. when $T=1$} {after minor modifications.
As an example, let us take $S_{nT}^{(1)}$.
The modified statistic will take the following form:
\begin{equation*}
S_{n}^{(1)} =\frac{\Delta \bar{L}_{n}}{\hat{{\sigma}}_{1,n}/\sqrt{n}%
},
\end{equation*}
where ${\hat{\sigma}}_{1,n}^{2}= n^{-1}\sum_{i=1}^{n} \Delta \tilde{L}_{it}\Delta \tilde{L}_{it}$, with $\Delta \tilde{L}_{it}=\Delta L_{it}-\Delta \bar{L}_{n,t}$, where $\Delta \bar{L}_{n,t} = n^{-1}\sum_{i=1}^{n} \Delta {L}_{it}$.
The modification is on the calculation of the variance, where we calculate the deviations from the overall mean, instead of the individual means.}
For a single cross-section, a CLT can be easily obtained as a basis for these statistics.
For instance, we can apply the CLT for independent but heterogeneous sequence \citep[see, e.g.][Theorem 5.10]{white01} to show the normality of tests with no CD, and the CLT for spatially correlated triangular arrays of \citet{kelejian98} to show the normality of tests with WCD, respectively.

\section{Monte Carlo Study}\label{sec:monte}

To investigate the finite sample properties of the test statistics given above, a set of Monte Carlo experiments are conducted.
2000 samples are generated from each DGP described below for the dimensions of $T\in \{10,20,30,50,100\}$ and $n\in \{10,20,30,50,100\}$. All tests are applied for the nominal size of 5\%.

\subsection{Design}

\label{ssec:design}

Two different DGPs are considered to explore the effect of WCD and SCD on the performance of the tests.
DGP1 contains only WCD {which is controlled using a stationary spatial AR(1) process which satisfies Assumption \ref{ass_r1}}.
{In this case, for each unit $i$, two independent forecast error series $(e_{1,it},e_{2,it})$ are generated using spatial AR(1) processes. Define first
\begin{equation} \label{eq27}
\zeta _{l,it}=\rho \sum_{j=1}^{n}w_{ij}\zeta _{l,jt}+u_{l,it},\quad \text{%
with},\quad u_{l,it}\sim iid.N(0,1),\quad l=1,2,
\end{equation}%
where $w_{ij}$ is the element of the spatial matrix $\mathbf{W}_{n}$ in row $i$ and column $j$ with $w_{ii} = 0$ for all $i=1,2,\dots,n$.
Then the forecast error series $e_{l,it}$, $l=1,2$ are generated as the $i$th element of the $n$-vector
\begin{equation}  \label{eq29}
\mathbf{e}_{l,n,t}= \frac{1}{\sqrt{\bar{s}_{2}}}\mathbf{S}_n\mathbf{u}_{l,n,t},
\end{equation}%
where $\mathbf{u}_{l,n,t}=(\mathbf{u}_{l,1t},\mathbf{u}_{l,2t},\dots ,\mathbf{u}_{l,nt})^{\prime }$, $\mathbf{S}_n=(\mathbf{I}_{n}-\rho \mathbf{W}_{n})^{-1}$ and $\bar{s}_{2} = n^{-1}\mathrm{tr}(\mathbf{S}_n\mathbf{S}_n^{\prime })$.
To explore the size of various tests, we use these two forecast error series $e_{l,it}$, $l=1,2$.
We set $\rho = 0.5$.\footnote{We also tried $\rho = 0$ and $\rho = 0.9$. The findings are similar to those obtained with $\rho = 0.5$ and the results omitted to save space.}
In this DGP a quadratic loss function is used.}

DGP2 contains SCD and WCD.
In this case, following \cite{giacomini06}, we generate the loss differential directly, so the tests do not rely on a specific loss function.
This is given by 
\begin{equation}  \label{eq301}
\Delta L_{it} = \xi(\mu_i + \lambda_{1i} f_{1t} + \lambda_{2i} f_{2t} +
\varepsilon_{it}).
\end{equation}
To investigate the size we set $\mu _{i}=0$ for each $%
i=1,2,\dots ,n$ and generate factor loadings as
$
\lambda _{1i},\lambda _{2i}\sim iid.N(1,0.2)
$.
The common factors are formed by
$
f_{1t},f_{2t}\sim iid.N(0,1)
$,
The error series $\varepsilon_{it}$ are generated in the same spirit as in (\ref{eq29}){, precisely we set $\varepsilon_{it} = e_{1,it}$}.
We finally set $\xi=\sqrt{1/3.4}$ to control for the variance of the loss differential series.

{We also undertake a robustness check to study the small sample properties of the tests $S_{n}^{(3)}$ and $\tilde{S}_{n}^{(3)}$ by considering a heavy-tailed error distribution.
We generate errors from a $t(6)$ distribution for half of the panel and from a standard normal for the other half.
Specifically, in \eqref{eq27} we set, for each $l=1,2$, $u_{l,it}\sim iid.t(6)$ if $i=1,\dots,N/2$ and $u_{l,it}\sim iid.N(0,1)$ otherwise.}

{We explore the power of various tests under two different alternative hypotheses.
For this purpose, in the case of DGP1, we generate a third series $e_{3i,t}$ as two different re-parametrization{s} of the series $e_{2,it}$.}
The first one corresponds to the homogeneous alternative and the second one to the heterogeneous alternative.
We generate $e_{3i,t}=\sqrt{1.2}e_{2,it}$ and report the results from testing the equality of
forecast accuracy of $e_{1,it}$ and $e_{3i,t}$.
In the heterogeneous scenario, we generate $e_{3i,t}=\sqrt{\theta_{i}}e_{2,it}$ where $\theta _{i} = 0.8$ for $i=1,\dots,n/2$ and $\theta _{i} = 1.2$ for $i=n/2+1,\dots,n$.
Similarly, in the case of DGP2, we set $\mu _{i}=1.2$ for each $i$ in the case of homogeneous alternative and $\mu _{i} = -0.2$ for $i=1,\dots,n/2$ and $\mu _{i} = 0.2$ for $i=n/2+1,\dots,n$
in the case of heterogeneous alternative.



As the error series and the common factors are serially uncorrelated for each unit, it is implicitly assumed that we are dealing with one-step ahead forecasts.
Hence, we set the time series kernel $k_{T}(\cdot)=1$ if $t=s$ and $k_{T}(\cdot)=0$ otherwise.
Spatial interactions between units are created with a row-normalized rook-type weight matrix.\footnote{The units are assumed to lie on a $p_1 \times p_2$ rectangular grid such that the first $p_1$ units are located in the first column of the grid, the second $p_1$ units are located in the second column and so on. We therefore have $n = p_1 p_2$. For each $n\in \{10,20,30,50,100\}$, we choose $p_1$ as 2, 4, 6, 10 and 50, respectively.}
The distance between two units is the given by the Euclidean distance.
In the computation of the spatial kernel $k_{S}(\cdot)$, we use these distances.
In addition, we use distances based on the wrong assumption that the units are located on a line.
We use Bartlett kernel for all experiments and following \cite{kelejian07}, we set the spatial kernel bandwidth to $d_n = \lceil n^{1/4} \rceil+1$ where $\lceil \cdot \rceil$ stands for the smallest integer bigger than its argument. Similarly, the overall and clustered EPA test statistics using partial sample variance estimators are calculated by setting $\underline{n} = \lceil n^{1/2} \rceil+1$ and $\underline{n}_{1} = \underline{n}_{2} = \lceil n^{1/2}/2 \rceil+1$, respectively.

For the tests using common factors, we consider three possibilities for the selection of the number of factors.
First, we calculate them assuming $m=2$, meaning that for DGP2 the number of common factors is correctly specified.
Second, we set $m=1$ for the case of under-specification.
Third, we select the number of common factors by minimizing the $IC_{p1}$ information criterion\footnote{The authors consider several information criteria.
In their simulations $IC_{p1}$ appears to be the best performing criterion under WCD.} of \cite{bai02} given by
\begin{equation}\label{eq:icformula}
IC_{p1} = V(m) + m\left( \frac{n+T}{nT} \right) \ln \left( \frac{nT}{n+T} \right)
\end{equation}
where $V(m)$ is defined by \eqref{eq128}.
The case of over-specification of the number of common factors is not separately considered because $IC_{p1}$ almost always over-estimates the number of common factors in small samples (see Table \ref{tab:icprop} below).

Before the discussion of the size and power of the robust tests, as a benchmark we refer to the results on the non-robust tests $S^{(1)}_{nT}$ and $C^{(1)}_{nT}$.
The size and power of these tests are reported in Table \ref{tab:nonrobusttest}.
As is expected, all tests are incorrectly sized.

\subsection{Size properties}

The results on the size of CD-robust overall EPA tests with DGP1 are given in Table \ref{tab:sizedgp1}.
The size of the kernel robust test $S^{(2)}_{nT}$ of the overall EPA hypothesis improves with either $T$ or $n$.
First, we focus on the results when the distance metric is correctly specified.
In the smallest samples with $T = 10$ and $n = 10$, this particular setting provides an empirical size of 9.9\%.
For $T = 100$ with $n = 10$ corresponding value equals 8.65\% whereas for $T = 10$ with $n = 100$ it is 8.4\%.
In the largest sample its size is 6.3\% which is close to the nominal value of 5\%.
When the distance between the panel units is incorrectly specified, the size of the test still improves with either dimension.
However, as expected the size distortions are slightly larger in this case.
In the largest smallest and largest sample sizes its size equals 12.5\% and 8.3\%, respectively.
The test which uses the partial sample estimator of the variance has similar size values.
However, its size improves only with $T$.
When $T = n = 10$ its size is slightly larger than that of the kernel robust test with a misspecified distance.
When $(T,n) = (10,100)$ the size distortion increases (13.4\%).
In the case of large $T$, however, it performs better than the kernel robust test with incorrect distance.
For instance, when $(T,n) = (100,100)$ its size equals 7.7\%.

The test $S^{(3)}_{nT}$ performs very well especially when $T$ is large and $n$ is small.
In most of the combinations of $T$ and $n$ it shows better properties than $S^{(2)}_{nT}$.
When $T$ is greater than 50, it is correctly sized {and} even when $T=30$ and $n=100$ its size is 6.9\% which makes it the preferred test over any version of $S^{(2)}_{nT}$.

The test $\underline{S}_{nT}^{(3)}$ shows good properties even though it wrongly assumes that the loss series contain common factors.
The test with $m=2$ performs similarly to $S^{(3)}_{nT}$ when $n$ is small but its performance is less good as $n$ gets large.
In general, the size distortion of the test is bigger when $m=1$.

The case where the number of common factors is chosen by IC requires some special attention.
The test performs well for small $n$ and $T$ but its performance drops as $n$ or $T$ gets large.
To understand this behavior of the test, we check the small sample properties of $IC_{p1}$.
The average number of common factors chosen by this criterion over simulations is reported in Table \ref{tab:icprop}.
It is seen that, when either of the dimension of the panel grows, the performance of the IC increases.
However, when one of the dimensions is small, this improvement is very slow.
In DGP1, only when one of the dimensions is greater than 50, the performance is in acceptable levels.
Once either $T$ or $n$ is greater than 50, the average number of factors selected over replications is either zero or very close to zero.
This means, in fact in these cases the test converges to the non-robust test.
Hence, the empirical size and power of the test equals the size and power of the non-robust test given in Table \ref{tab:nonrobusttest}.
Of course, in practice the number of common factors is rarely known to the researcher and the most realistic application of this test is this case which is based on IC.
However, it should not be understood that our testing procedure is best described by the performance in this case.
In practice, if the no CD hypothesis is rejected by a suitable test, and the researcher decides with the help of an IC that the loss differentials do not contain common factors, an EPA test which is robust to WCD has to be used; for instance $S^{(2)}_{nT}$ or $S^{(3)}_{nT}$.
As in the case of zero common factors $\underline{S}_{nT}^{(3)}$ is identical to $S^{(1)}_{nT}$, these tests have to be used only if the no CD hypothesis cannot be rejected before the application of EPA tests.

The results for the clustered EPA tests are reported in the right block of Table \ref{tab:sizedgp1}.
The kernel robust test $C^{(2)}_{nT}$ performs slightly worse than the overall test $S^{(2)}_{nT}$ and its performance improves rapidly with increases in the number of observations in either dimensions of the panel.
In the case of large samples the empirical sizes of the two tests are comparable.
Similar to the findings on the overall test, when the distance is misspecified, the size distortion in the test is only slightly higher and its performance gets better with increases of number of observations.
The performance of the test based on the partial sample estimator, namely $\underline{C}^{(2)}_{nT}$ is unsatisfactory in small samples, especially for small $n$.
For $T=10$, even for the largest $n$ we have, the empirical size of the test equals 19.4\%.
However its performance improves with $T$, and it reaches 5.5\% in the largest sample considered.

The test $C^{(3)}_{nT}$ is a very viable alternative to $C^{(2)}_{nT}$.
For small $T$, its size distortion is superior to that of $C^{(2)}_{nT}$ but even for $n=20$ it has good size properties.
However, contrary to $C^{(2)}_{nT}$, its performance improves only with $T$.
For instance, when $n=10$ and $T=20$, its size equals 9.9\% which is better than that of the kernel robust test (10.7\%).
When $n=100$ and $T=20$ the performance of the latter improves dramatically and reaches 6.3\% whereas that of $C^{(3)}_{nT}$ is 8.6\%.

Finally we focus on the size of the test using estimated common factors, $\underline{C}_{nT}^{(3)}$.
Irrespective of the choice of the number of common factors, when $T$ is small, the test suffers from more size distortions compared to $C^{(2)}_{nT}$ and $C^{(3)}_{nT}$.
However for large $T$ and small $n$ the test with $m=2$ has a similar performance to the others.
In fact for the smallest $n$ and largest $T$ that we consider it has an empirical size of 6\% which is better than that of $C^{(2)}_{nT}$ (9.7\%) and $C^{(3)}_{nT}$ (6.9\%).
When $m=2$ the improvement of the size of the test is much slower with the increase in $T$ and when the number of common factors is chosen by IC, it approaches the non-robust test, as expected.

The size results for DGP2 are reported in Table \ref{tab:sizedgp2}.
As expected, for this DGP the overall tests $S^{(2)}_{nT}$ and $\underline{S}^{(2)}_{nT}$ are grossly over-sized and their performance does not improve with increases in the sample size in any dimension.
The test $S^{(3)}_{nT}$ shows very good properties except when $T$ is very small, in particular when $T>30$.
Conclusions are similar for the factor-robust tests $\underline{S}_{nT}^{(3)}$.
In fact, this test performs better than $S^{(3)}_{nT}$ for all samples sizes considered.
A very important finding is that, even when the number of common factors is underspecified, the test performs very well.
We also see that the three versions of the test are equivalent in large samples.
For $T>10$ and $n>30$ these three tests have equal empirical size.

The findings concerning the clustered EPA tests robust to SCD are similar to those in the case of DGP1 with a few points worth mentioning.
The test $C^{(3)}_{nT}$ behaves in line with theoretical expectations such that it has lower size distortions for large $T$ and small $n$ and it performs slightly worse than the overall test $S^{(3)}_{nT}$.
The tests based on estimated common factors are found to be oversized in small samples when the number of common factors are chosen by the researcher.
The IC based version shows less size distortions.
For small $T$ it outperforms $C^{(3)}_{nT}$ overall.
Finally, in largest samples the IC based test and the test with $m=2$ have identical size.
This is expected because the information criterion $IC_{p1}$ consistently chooses the number of common factors when $n$ and $T$ are large, as seen in Table \ref{tab:icprop}.

{We conclude this section reporting the following two simulation exercises.
In the first, we evaluate the performance of the test statistics $S^{(3)}_{nT}$ and $\tilde{S}^{(3)}_{nT}$.
In the second exercise, we study the properties of the $J_n^D$ test statistic of QTZ.
We note that the modified test $\tilde{S}^{(3)}_{nT}$ is valid only for one step ahead forecasts.
Furthermore, it requires normality under SCD.
Table \ref{tab:nonnormal} reports the finite sample properties of $S^{(3)}_{nT}$ and $\tilde{S}^{(3)}_{nT}$ under the failure of normality assumption.
As can be seen, the test statistic $\tilde{S}^{(3)}_{nT}$ is perfectly sized irrespective of the sample size even when normality fails, under both DGPs.
Turning to the performance of $J_n^D$, we note that, the test is valid only under WCD (DGP1).
This is confirmed by the results reported in Table \ref{tab:tzsize}, where the test's size is very close to the nominal size irrespective of the sample size for DGP1, while it is grossly over-sized under DGP2.
}

To summarize, in the case of both DGPs the overall EPA hypothesis can be tested with a size close to the nominal value for almost all sample sizes.
In particular, it is found that the test $S^{(3)}_{nT}$ has very good properties in both DGPs.
{In addition, $\tilde{S}^{(3)}_{nT}$ can be used for one step ahead forecasts with size practically equal to the nominal size, irrespective of the type of CD in the loss differentials.}
For DGP1, for small $T$ and large $n$, the kernel robust test is preferred over the test based on the partial sample estimator given that the distance metric is correctly specified.
Finally, the test $C^{(3)}_{nT}$ is preferred over others, however, $\underline{C}_{nT}^{(3)}$ has also good properties when the number of common factors is overspecified.

\subsection{Power properties}

The power results of the tests for DGP1 under the homogeneous alternative hypothesis are given in Table \ref{tab:powerdgp1}.
In the previous subsection, we have seen that the size of the overall EPA tests $S^{(2)}_{nT}$, $\underline{S}^{(2)}_{nT}$ and $S^{(3)}_{nT}$ approach to the nominal level for this DGP.
Here, it is seen that the power of the tests $\underline{S}^{(2)}_{nT}$ and $S^{(2)}_{nT}$ converge to 100\%, for the latter the distance metric being unimportant.
Hence the test is consistent in all cases.
For moderate to large $T$, the test $S^{(3)}_{nT}$ is correctly sized.
We observe that its power is slightly lower compared to that of $S^{(2)}_{nT}$ in these sample sizes.
{For instance, when $T = 100$ and $n = 10$ the power of $S^{(2)}_{nT}$ equals 70.8\% whereas those of $S^{(3)}_{nT}$ and $\underline{S}^{(3)}_{nT}$ are 63.5\% and 57.2\%, respectively.}
{Hence,} even though they wrongly assume that there are common factors in the DGP, the power of the factor-robust tests $\underline{S}_{nT}^{(3)}$ are close to that of $S^{(3)}_{nT}$.
{To conclude, we see that, the factor robust tests have lower power under WCD, as expected.}

The previous results of the clustered EPA tests showed that in general they are correctly sized only for large $T$.
Here we see that their power is only slightly lower than the overall tests.
For instance, the power of the asymptotically correctly sized tests $S^{(2)}_{nT}$ and $C^{(2)}_{nT}$ are 19.1\% and 17.3\%, respectively and their power reaches 100\% in the largest sample.

Table \ref{tab:powerdgp2} reports the power results of the tests for DGP2 under the homogeneous alternative hypothesis.
For this DGP, we have seen that the tests $S^{(2)}_{nT}$ are over-sized even asymptotically.
Hence, we focus on the tests $S^{(3)}_{nT}$ and $\underline{S}_{nT}^{(3)}$.
It is seen that the power of both tests are very similar for all sample sizes.
For instance, for most of the sample sizes that we consider, the power of $C^{(3)}_{nT}$ equals the power of $\underline{C}_{nT}^{(3)}$ when the number of common factors is chosen by the IC.

{The power of the tests $S^{(3)}_{nT}$ and $\tilde{S}^{(3)}_{nT}$ under non-normality are reported in the second panel of Table \ref{tab:nonnormal}.
As is stated in the previous subsection, $\tilde{S}^{(3)}_{nT}$ improves greatly over $S^{(3)}_{nT}$ in terms of size.
We now see $\tilde{S}^{(3)}_{nT}$ has less power for any sample size compared to $S^{(3)}_{nT}$.
It has to be noted however that $\tilde{S}^{(3)}_{nT}$ performs still very well and its power equals 100\% as sample size increases.
}

To save space, we do not report the power results under the heterogeneous alternative hypothesis.
The main findings are summarized in Figure \ref{fig:sizeadjpow} where the power of $S^{(3)}_{nT}$, $C^{(3)}_{nT}$ and {$J_n^D$ are shown under DGP1 so that the assumptions of QTZ are satisfied.}
It can be seen that under the homogeneous alternative, the power of our tests $S^{(3)}_{nT}$ and $C^{(3)}_{nT}$ rapidly approach 100\%.
{Whereas the power of $J_n^D$ is very low compared to our proposed tests, although it increases with sample size.}
Under the heterogeneous alternative {on the other hand}, the power of the overall test $S^{(3)}_{nT}$ approaches the nominal size.
This is because the overall EPA hypothesis holds under this alternative hypothesis.
On the other hand, the expected value of the loss differentials is different from zero for clusters of panel units, hence the clustered EPA test $C^{(3)}_{nT}$ has power under this scenario.
{Under this alternative hypothesis, we see that the power of $J_n^D$ approaches zero.
Hence, it is inconsistent contrary to our proposed test $C^{(3)}_{nT}$.}
{
\section{Guidelines for Empirical Applications}\label{sec:imp}
}
The application of the panel EPA tests require some preliminary information on the cross-sectional and temporal dependence properties of the loss differentials.
First, the researcher has to determine whether the loss series contain CD, to choose between the non-robust tests $S^{(1)}_{nT}$ and $C^{(1)}_{nT}$, and the other tests which deal with CD.
If the data displays CD, one has to have some information on the type of the CD in loss differentials as WCD and SCD may require the use of different tests.
Exceptions to this are $S^{(3)}_{nT}$ and $C^{(3)}_{nT}$ which are shown to be performing very well under any type of CD in our simulations.
Second, to determine the time series kernel bandwidth parameter, one has to determine whether the loss differentials are autocorrelated or not.
This panel EPA testing approach is based on the following three steps.

\noindent \textbf{Step A}-- \textit{Analysis of CD}: Test the no CD hypothesis using a test such as that of \citet[][BP hereafter]{breusch80} or the modified and standardized version of it \citet[][Modified BP hereafter]{pesaran08}. If the no CD hypothesis is not rejected, proceed to Step B.1, otherwise calculate $IC_{p1}$ and proceed to Step B.2.

\noindent \textbf{Step B}-- \textit{Testing for no autocorrelation}: Consider the following empirical model for loss differentials:
\begin{equation}\label{eq:autocorrmod}
\Delta L_{it} = \pi_1 \Delta L_{i,t-1} + \pi_2 \Delta L_{i,t-2} + \dots + \pi_p \Delta L_{i,t-p} + a_i + \zeta_{it},
\end{equation}
and the hypothesis $H^{ac}_0: \pi_1 = \pi_2 = \dots = \pi_p$.
\begin{enumerate}
\item Run a fixed effects regression on \eqref{eq:autocorrmod} and test $H^{ac}_0$ using a Wald test with a variance robust to panel level heteroskedasticity.
\item Run a fixed effects regression on \eqref{eq:autocorrmod} and test $H^{ac}_0$ using a Wald test with a variance calculated by clustering on time index.
\end{enumerate}

If the no autocorrelation hypothesis is not rejected, set $d_T=1$, otherwise set $d_T>1$. If the no CD hypothesis is not rejected in Step A, proceed to Step C.1, otherwise proceed to Step C.2.

\noindent \textbf{Step C}-- \textit{Testing for EPA}:
\begin{enumerate}
\item To test the overall EPA hypothesis $H^a_{0}$ use $S^{(1)}_{nT}$, and for the clustered EPA hypothesis $H^{ca}_{0}$ use $C^{(1)}_{nT}$.
\item If $IC_{p1}$ indicates $m=0$, use $S^{(2)}_{nT}$, $\underline{S}^{(2)}_{nT}$ or $S^{(3)}_{nT}$ to test $H^a_{0}$, and $C^{(2)}_{nT}$, $\underline{C}^{(2)}_{nT}$ or $C^{(3)}_{nT}$ to test $H^{ca}_{0}$. If $IC_{p1}$ indicates $m>0$, use $S^{(3)}_{nT}$ or $\underline{S}_{nT}^{(3)}$ to test $H^a_{0}$, and $C^{(3)}_{nT}$ or $\underline{C}_{nT}^{(3)}$ to test $H^{ca}_{0}$.
\end{enumerate}

In Step A, we suggest two tests of CD.
The null hypothesis of BP test is the joint absence of CD between all pairs in the panel.
The statistic is distributed as $\chi^2_q$ with $q = n(n-1)/2$ under the null.
Hence, the test is more suitable for the cases of fixed and small $n$.
The Modified BP statistic is a bias corrected and standardized version of the LM test.
It is asymptotically normal under the null of no CD as $n \to \infty$ and is more suitable for large panels.
These CD tests can be used to test the hypothesis of no CD in a data set but they do not help to identify the type of the CD.
To see if the loss differentials contain common factors, we suggest the information criterion $IC_{p1}$ in \eqref{eq:icformula}.
As seen in our simulations (Table \ref{tab:icprop}), this IC performs quite well to choose between WCD and SCD.

In Step B, we analyze the serial correlation in loss differentials.
The test of the no autocorrelation hypothesis follows the analysis of the CD properties of the loss differentials because the variance computed for the Wald test of the no autocorrelation hypothesis depends on whether the loss series contain CD or not.
We suggest using a variance estimator calculated by clustering on the time index if the no CD hypothesis is not rejected in Step A.
This estimation corresponds to the \cite{driscoll98} variance estimator with a time series kernel bandwidth chosen such that the error terms in \eqref{eq:autocorrmod} are assumed to be serially uncorrelated.
For simplicity, here we do not distinguish between WCD and SCD, although this is possible by considering other variance estimators such as that of \cite{kim13}.
However, as noted by \cite{driscoll98}, their variance estimator is valid under WCD.
One important point in Step B is the determination of the lag length $p$.
\cite{han17} found that the BIC is inconsistent in panel autoregressions even in the absence of fixed effects.
We suggest to use the general-to-specific methodology that they propose.
The method starts with $p_{max}$ chosen by the researcher, and it continues by eliminating the biggest insignificant lag until a significant lag is found.
If the significance level of these tests is fixed by the researcher, the probability of overestimation of the lag length is nonzero.
To avoid this, we determine the significance level as $\alpha_{nT}= \exp\{\ln(0.25) \sqrt{nT}/10\}$, as suggested by \cite{han17}.

Finally, in Step C we test the EPA hypotheses $H^a_{0}$ or $H^{ca}_{0}$ on the basis of the outcome of the two previous steps.
{Of course, pre-testing as documented in this methodology may affect the properties of our proposed tests.
This is an interesting issue worth exploring but which goes beyond the scope of this paper.}
In the next section, we follow this methodology.

\section{Empirical Applications}\label{sec:app}

\subsection{Comparing OECD and IMF economic growth forecasts}\label{sec:app1}

We compare the OECD and IMF GDP growth forecasts using the EPA methodology described above.
{Our application can be seen as complementary to the application of QTZ comparing the IMF forecasts to Consensus Economics forecasts.}
The data for the IMF forecasts come from their Historical WEO Forecasts Database.
The database includes historical $h$-years ahead forecast values, $h=1,2,\dots,5$, for the GDP growth rate and covers up to 192 countries and starts from early 1990's.
We collected similar data from the past vintages of the Economic Outlook of the OECD.
The Economic Outlook contains only 1-year ahead forecasts.
Both organizations publish their forecasts twice a year.
In our application we focus on their summer forecasts made for the following year.
These forecasts are published in June every year by the OECD whereas IMF forecasts are published in April.
Publishing dates are close, hence the forecast errors are comparable.
To compute the forecast errors, we use {the GDP growth outturns by the IMF.
We also tried using the GDP growth published by the OECD and the results are unchanged (the Pearson correlation coefficient between the two outturn series equal 0.995 hence the difference is negligible).}
Eventually we constructed a balanced panel data set of {annual} GDP growth forecast errors of 29 OECD countries from the two organizations between 1998 and 2016.

We use the quadratic loss which is defined as 
\begin{equation*}
\Delta L^{(q)}_{it} = e_{1,it}^2 - e_{2,it}^2,
\end{equation*}
where, as is throughout this application, first organization is the OECD.
This loss function is arguably the most frequently used one {and is robust to measurement errors contrary to other loss functions such as absolute loss \citep{hoga22}.}

We begin the analysis by the DM tests applied to each country.
In the computations of the DM test statistics, we use a bandwidth parameter $d_T=1$ because we have 1-step ahead forecasts.
Note that below an autocorrelation test is used to confirm that the loss differentials are serially uncorrelated.\footnote{{It is worth noticing that we use the summer forecasts of the two organizations and they are expected to be autocorrelated by construction. Nevertheless, as the autocorrelation test we use indicates the absence of autocorrelation, its effect must be negligible. Below, for the panel tests we also use a different bandwidth for robustness and this expectation is confirmed.}}
The results of the DM tests are given in Table \ref{tab:dmbc} {where we report average loss differentials, DM test statistics and their $p$-values for each country over the period 1998-2016}.

{First, in terms of the sign of the average loss differentials, a considerable amount of heterogeneity can be observed in the sample.
We see that roughly half of the statistics are negative.
Second, most of these statistics are statistically insignificant with exceptions being Belgium, Spain, Hungary and Luxembourg.
For Belgium which is a country where the predictive ability of the IMF is superior, the EPA hypothesis can be rejected at 10\%.
For Spain OECD predictions, for Hungary IMF predictions outperform the other.
For Luxembourg we can reject the EPA hypothesis at the 10\% level and the predictive ability of the IMF is found to be superior for this country as well.}

{Finally, at the bottom right of Table \ref{tab:dmbc}, we report average loss differentials over clusters of countries and the period 1998-2016.
The average quadratic loss differential over all 29 OECD countries in the sample is 0.009.
This shows that, when we average over all countries, the difference between the {predictive ability} of the two institutions is positive but very small.

An interesting question is whether there are clusters of countries for which the forecast performance of the two institutions differ dramatically.
\cite{dreher08} test the hypothesis that the forecast performance of the IMF differs with respect to the direct influence of a country on the institution.
They use GDP of a country as a proxy of the political influence and find evidence that the forecast bias of the institution declines with GDP.
To see if we can find similar evidence in terms of the differences between the bias and efficiency of the forecasts made by the two institutions, we divide our country sample into the G7 and non-G7 countries.
The GDP of the G7 countries account for almost 70\% of the total GDP of all 29 countries in our sample in 2016, the last year of our data set.
Hence, if the OECD's forecast performance does not vary with a country's GDP but the performance of the IMF does, as found by \cite{dreher08}, we expect to have heterogeneity in average losses of the G7 and non-G7 countries.

The table shows that the average quadratic loss for G7 countries is 0.079.
For non-G7 countries this average is -0.014.
This shows that for G7 countries the IMF has a superior performance whereas for non-G7 countries the OECD does better in terms of their growth forecasts.
Hence, the forecast ability of the two institutions indeed varies with country clusters.
Below, we test if these averages are statistically different from zero using our tests.}

\textbf{Cross-sectional and temporal dependence in loss differentials.} As found in our Monte Carlo simulations, the increase in the number of cross-sections increases the power of EPA tests.
However, the gain from the usage of panels depends on the degree and the nature of CD.
Furthermore, the application of the correct EPA tests require some information on the cross-sectional and temporal dependence of the loss differentials, as we summarized in the previous subsection.
Hence, before proceeding to panel tests of EPA, we analyze the CD and autocorrelation properties of our dataset.

The results of BP and Modified BP tests {show that} the null hypothesis of no CD is rejected using any test in conventional significance levels.
{The results are omitted to save space but available from the authors.}
This means that the tests which allow for CD are more reliable for our data set.
Next, $IC_{p1}$ indicates existence of 6 common factors in both loss differential series.
Hence, we conclude that both series display SCD.
To determine the time series kernel bandwidth parameter, we estimate \eqref{eq:autocorrmod} with $p=1$ which was chosen by the general-to-specific methodology and check the significance of the common autoregressive parameter by clustering in the time index.
This autocorrelation test confirms that the loss differentials are serially uncorrelated, hence, {first} we set $d_T=1$.
{To see the impact of the change in the time series bandwidth, we also try $d_T=T^{1/3}$.}

To see the time series profile of the common factors in the loss differential series, we report the plot of the first 6 PCs {of the} loss differential series in panels (a) and (c) of Figure \ref{fig:lossdiff}.
The PCs are numbered in decreasing order with respect to their eigenvalues.
For better interpretation of the common factor estimates, we report a focus on the first three PCs of the quadratic loss differential.
Associated factor loadings estimates are reported in Table \ref{tab:lossdiff}.
To save space, estimated loadings for other PCs are dropped but they are available from the authors upon request.
The estimates of the common factors in loss differentials show the effect of the financial crisis.
The first {three} common factors in quadratic loss fall in 2009.
As can be seen in Table \ref{tab:lossdiff}, for 20 countries in the sample, the factor loading estimates are negative for the first common factor.
Hence, the OECD, had a superior predictive ability compared to the IMF in this year.
The second PC shows a similar pattern whereas the third PC has a movement in the opposite direction in the recovery period.
This PC therefore compensates the effect of the first two common factors for some countries.

\textbf{Panel tests for the EPA hypotheses.}
Here we apply the panel EPA tests to compare the performance of the two institutions.
Following the insights of the Monte Carlo results and the CD analysis, we apply the factor-robust tests $S^{(3)}_{nT}$, $C^{(3)}_{nT}$, $\underline{S}_{nT}^{(3)}$ and $\underline{C}_{nT}^{(3)}$.
As a benchmark, we also report the results from the tests assuming no CD, namely $S^{(1)}_{nT}$ and $C^{(1)}_{nT}$.

The results are given in Table \ref{tab:paneltests}.
The first part of the table reports the tests of the overall EPA hypothesis.
Hence, they are the significance tests of the overall difference reported in Table \ref{tab:dmbc} which is 0.009.
The test statistics for the non-robust test $S^{(1)}_{nT}$ are equal to 0.06 for both bandwidth parameters considered.
The conclusion does not change using the robust tests for which the test statistics are slightly weaker.
{Our results are in line with those obtained by QTZ on the comparison of IMF forecasts with Consensus Economic forecasts, in the sense that, they do not find a significant difference between the predictive ability of the IMF and Consensus Economics using EPA tests based on the summer forecasts of the two institutions.}

In the second part of the table the results for the clustered EPA tests are reported.
The clusters we consider are G7 countries and the non-G7 countries for which the average loss differentials are reported in Table \ref{tab:dmbc}.
It is seen that for these clusters, the clustered EPA hypothesis cannot be rejected using any test and any bandwidth parameter.
Hence, the predictive ability differences of the two institutions are insignificant.
%
%

\subsection{An Evaluation of the IMF Consumer Price Inflation Forecasts}\label{sec:app2}

In a second application, we challenge the quality of the IMF CPI forecasts by comparing them with random walk (RW) forecasts.
{In a similar exercise, QTZ compare IMF CPI inflation forecasts with those obtained from the AR(1) model.}
In this subsection, we summarize the application and its results.
The details can be found in Appendix \ref{sec:app3}.
The data on the IMF CPI forecasts used in this application come from the Fund's Historical WEO Forecasts Database.
The sample contains 127 countries over the period 1991-2019.
We restrict our attention to the period considered also by QTZ but we use the data from all countries available, contrary to their sample which covers 86 countries. 
In this application, in addition to the quadratic loss function, we use the absolute loss function defined as
\begin{equation*}
\Delta L^{(a)}_{it} = |e_{1,it}| - |e_{2,it}|,
\end{equation*}
where $e_{1,it}$ is the forecast error made by the IMF for country $i$ and year $t$, and $e_{2,it}$ is the RW forecast errors calculated using the CPI data published by the IMF.
We consider 2 different country clustering schemes.
The first one divides the sample of countries into OECD and non-OECD countries, and the second one divides them into G7 countries, non-G7 OECD countries and non-OECD countries.
In addition to the full sample analysis, we split the sample into pre- and post-global financial crises periods.

The results show that there is a small but statistically significant difference between the average absolute losses of the IMF and the RW forecasts in favor of the latter in the full sample.
When we divide the sample into pre- and post-crises periods, we see that these differences are driven by the differences in the pre-crisis period.
For the absolute loss, the clustered EPA hypothesis can be rejected in both pre- and post-crises periods.

{Using quadratic loss, QTZ find that the IMF CPI forecasts are significantly more accurate than the AR(1) forecasts.
Our results show that, overall, IMF CPI forecasts are less accurate than the RW forecasts.
However, for this loss function, neither overall EPA hypothesis nor clustered EPA hypothesis can be rejected in conventional levels using CD-robust tests.}

\section{Conclusions}\label{sec:conc}

In this paper, we proposed novel predictive ability test for panels, corresponding to two different equal predictive ability hypotheses.
Several overall EPA tests were developed to evaluate the hypothesis that the predictive ability of two forecasters is equal on average over all periods and units.
Consistent clustered EPA tests were also built which are able to test the hypothesis of whether two forecasters have equal predictive power for all clusters of units.
Our proposed tests are robust to different forms of cross-sectional dependence in the loss differentials, arising from weak and strong cross-sectional dependence.
The proposed tests are found to have appropriate size and power in a set of Monte Carlo simulations.
In particular, the overall EPA tests robust to arbitrary cross-sectional dependence are correctly sized.
Finally, we provided some useful three-step guideline on how to run the tests, which is then implemented in two applications.
We compare the prediction performance of two major organizations, the OECD and the IMF, on their historical economic growth forecasts.
In a second application, we evaluate the CPI forecasts of the IMF by comparing them with random walk forecasts.
We found evidence of strong cross-sectional dependence in loss differentials of forecast errors in both applications.
The results showed that there are only minor differences between the predictive ability of the OECD and the IMF in terms of their economic growth forecasts and they are statistically insignificant.
We found some evidence that the RW forecasts of the consumer price inflation are significantly more accurate than those made by the IMF.

\section*{Acknowledgements}

We thank participants in
the 18\textsuperscript{th} International Workshop on Spatial Econometrics and Statistics (AgroParisTech, Paris, France, 23-24 May 2019), in particular the discussant Paul Elhorst and Davide Fiaschi,
the 39\textsuperscript{th} International Symposium on Forecasting (Thessaloniki, Greece, 16-19 June 2019),
the 25\textsuperscript{th} International Panel Data Conference (Vilnius University, Vilnius, Lithuania, 4-5 July 2019),
the 22\textsuperscript{nd} Dynamic Econometrics Conference (Nuffield College, University of Oxford, Oxford, United Kingdom, 9-10 September 2019),
{the 20\textsuperscript{th} International Workshop on Spatial Econometrics and Statistics (Lille, France, 19-20 May 2022),
the 27\textsuperscript{th} International Panel Data Conference (Bertinoro, Italy, 16-19 June 2022),
and
the 16\textsuperscript{th} World Conference of Spatial Econometrics Association (Warsaw, Poland, 23-24 June 2022), in particular the discussant Ioanna Tziolas.}
The usual disclaimer applies.
This paper is a part of Oguzhan Akgun's PhD research project and it was developed while Giovanni Urga and Zhenlin Yang were visiting the University Paris Panth\'{e}on-Assas - CRED (France) in May and June 2019 and 2022, and during Oguzhan Akgun and Alain Pirotte's several visits at the Centre for Econometric Analysis (CEA) of Bayes Business School (formerly Cass).
We thank both institutions for financial support.

\bibliography{refs}

\begin{thebibliography}{37}
\expandafter\ifx\csname natexlab\endcsname\relax\def\natexlab#1{#1}\fi
\providecommand{\bibinfo}[2]{#2}
\ifx\xfnm\relax \def\xfnm[#1]{\unskip,\space#1}\fi
\bibitem[{Akgun et~al.(2020)Akgun, Pirotte, Urga \& Yang}]{akgun20}
\bibinfo{author}{Akgun, O.}, \bibinfo{author}{Pirotte, A.},
  \bibinfo{author}{Urga, G.}, \& \bibinfo{author}{Yang, Z.}
  (\bibinfo{year}{2020}).
\newblock \bibinfo{title}{Equal predictive ability tests for panel data with an
  application to {OECD} and {IMF} forecasts}.
\newblock {\it \bibinfo{journal}{arXiv preprint arXiv:2003.02803}\/}, .
\bibitem[{Andrews(1991)}]{andrews91}
\bibinfo{author}{Andrews, D.~W.} (\bibinfo{year}{1991}).
\newblock \bibinfo{title}{Heteroskedasticity and autocorrelation consistent
  covariance matrix estimation}.
\newblock {\it \bibinfo{journal}{Econometrica}\/},  {\it
  \bibinfo{volume}{59}\/}, \bibinfo{pages}{817--858}.
\bibitem[{Bai(2003)}]{bai03}
\bibinfo{author}{Bai, J.} (\bibinfo{year}{2003}).
\newblock \bibinfo{title}{Inferential theory for factor models of large
  dimensions}.
\newblock {\it \bibinfo{journal}{Econometrica}\/},  {\it
  \bibinfo{volume}{71}\/}, \bibinfo{pages}{135--171}.
\bibitem[{Bai \& Ng(2002)}]{bai02}
\bibinfo{author}{Bai, J.}, \& \bibinfo{author}{Ng, S.} (\bibinfo{year}{2002}).
\newblock \bibinfo{title}{Determining the number of factors in approximate
  factor models}.
\newblock {\it \bibinfo{journal}{Econometrica}\/},  {\it
  \bibinfo{volume}{70}\/}, \bibinfo{pages}{191--221}.
\bibitem[{Bai \& Ng(2006)}]{bai06}
\bibinfo{author}{Bai, J.}, \& \bibinfo{author}{Ng, S.} (\bibinfo{year}{2006}).
\newblock \bibinfo{title}{Confidence intervals for diffusion index forecasts
  and inference for factor-augmented regressions}.
\newblock {\it \bibinfo{journal}{Econometrica}\/},  {\it
  \bibinfo{volume}{74}\/}, \bibinfo{pages}{1133--1150}.
\bibitem[{Breusch \& Pagan(1980)}]{breusch80}
\bibinfo{author}{Breusch, T.~S.}, \& \bibinfo{author}{Pagan, A.~R.}
  (\bibinfo{year}{1980}).
\newblock \bibinfo{title}{The {L}agrange multiplier test and its applications
  to model specification in econometrics}.
\newblock {\it \bibinfo{journal}{The Review of Economic Studies}\/},  {\it
  \bibinfo{volume}{47}\/}, \bibinfo{pages}{239--253}.
\bibitem[{Chudik et~al.(2011)Chudik, Pesaran \& Tosetti}]{chudik11}
\bibinfo{author}{Chudik, A.}, \bibinfo{author}{Pesaran, M.~H.}, \&
  \bibinfo{author}{Tosetti, E.} (\bibinfo{year}{2011}).
\newblock \bibinfo{title}{Weak and strong cross-section dependence and
  estimation of large panels}.
\newblock {\it \bibinfo{journal}{Econometrics Journal}\/},  {\it
  \bibinfo{volume}{14}\/}, \bibinfo{pages}{45--90}.
\bibitem[{Clark \& McCracken(2001)}]{clark01}
\bibinfo{author}{Clark, T.~E.}, \& \bibinfo{author}{McCracken, M.~W.}
  (\bibinfo{year}{2001}).
\newblock \bibinfo{title}{Tests of equal forecast accuracy and encompassing for
  nested models}.
\newblock {\it \bibinfo{journal}{Journal of Econometrics}\/},  {\it
  \bibinfo{volume}{105}\/}, \bibinfo{pages}{85--110}.
\bibitem[{Clark \& McCracken(2015)}]{clark15}
\bibinfo{author}{Clark, T.~E.}, \& \bibinfo{author}{McCracken, M.~W.}
  (\bibinfo{year}{2015}).
\newblock \bibinfo{title}{Nested forecast model comparisons: {A} new approach
  to testing equal accuracy}.
\newblock {\it \bibinfo{journal}{Journal of Econometrics}\/},  {\it
  \bibinfo{volume}{186}\/}, \bibinfo{pages}{160--177}.
\bibitem[{Clark \& West(2007)}]{clark07}
\bibinfo{author}{Clark, T.~E.}, \& \bibinfo{author}{West, K.~D.}
  (\bibinfo{year}{2007}).
\newblock \bibinfo{title}{Approximately normal tests for equal predictive
  accuracy in nested models}.
\newblock {\it \bibinfo{journal}{Journal of Econometrics}\/},  {\it
  \bibinfo{volume}{138}\/}, \bibinfo{pages}{291--311}.
\bibitem[{Davies \& Lahiri(1995)}]{davies95}
\bibinfo{author}{Davies, A.}, \& \bibinfo{author}{Lahiri, K.}
  (\bibinfo{year}{1995}).
\newblock \bibinfo{title}{A new framework for analyzing survey forecasts using
  three-dimensional panel data}.
\newblock {\it \bibinfo{journal}{Journal of Econometrics}\/},  {\it
  \bibinfo{volume}{68}\/}, \bibinfo{pages}{205--228}.
\bibitem[{Diebold \& Mariano(1995)}]{diebold95}
\bibinfo{author}{Diebold, F.}, \& \bibinfo{author}{Mariano, R.}
  (\bibinfo{year}{1995}).
\newblock \bibinfo{title}{Comparing predictive accuracy}.
\newblock {\it \bibinfo{journal}{Journal of Business \& Economic
  Statistics}\/},  {\it \bibinfo{volume}{13}\/}, \bibinfo{pages}{253--63}.
\bibitem[{Dreher et~al.(2008)Dreher, Marchesi \& Vreeland}]{dreher08}
\bibinfo{author}{Dreher, A.}, \bibinfo{author}{Marchesi, S.}, \&
  \bibinfo{author}{Vreeland, J.~R.} (\bibinfo{year}{2008}).
\newblock \bibinfo{title}{The political economy of imf forecasts}.
\newblock {\it \bibinfo{journal}{Public Choice}\/},  {\it
  \bibinfo{volume}{137}\/}, \bibinfo{pages}{145--171}.
\bibitem[{Driscoll \& Kraay(1998)}]{driscoll98}
\bibinfo{author}{Driscoll, J.~C.}, \& \bibinfo{author}{Kraay, A.~C.}
  (\bibinfo{year}{1998}).
\newblock \bibinfo{title}{Consistent covariance matrix estimation with
  spatially dependent panel data}.
\newblock {\it \bibinfo{journal}{Review of Economics and Statistics}\/},  {\it
  \bibinfo{volume}{80}\/}, \bibinfo{pages}{549--560}.
\bibitem[{Giacomini \& White(2006)}]{giacomini06}
\bibinfo{author}{Giacomini, R.}, \& \bibinfo{author}{White, H.}
  (\bibinfo{year}{2006}).
\newblock \bibinfo{title}{Tests of conditional predictive ability}.
\newblock {\it \bibinfo{journal}{Econometrica}\/},  {\it
  \bibinfo{volume}{74}\/}, \bibinfo{pages}{1545--1578}.
\bibitem[{Gneiting(2011)}]{gneiting11}
\bibinfo{author}{Gneiting, T.} (\bibinfo{year}{2011}).
\newblock \bibinfo{title}{Making and evaluating point forecasts}.
\newblock {\it \bibinfo{journal}{Journal of the American Statistical
  Association}\/},  {\it \bibinfo{volume}{106}\/}, \bibinfo{pages}{746--762}.
\bibitem[{Han et~al.(2017)Han, Phillips \& Sul}]{han17}
\bibinfo{author}{Han, C.}, \bibinfo{author}{Phillips, P.~C.}, \&
  \bibinfo{author}{Sul, D.} (\bibinfo{year}{2017}).
\newblock \bibinfo{title}{Lag length selection in panel autoregression}.
\newblock {\it \bibinfo{journal}{Econometric Reviews}\/},  {\it
  \bibinfo{volume}{36}\/}, \bibinfo{pages}{225--240}.
\bibitem[{Hannan(1970)}]{hannan70}
\bibinfo{author}{Hannan, E.~J.} (\bibinfo{year}{1970}).
\newblock {\it \bibinfo{title}{Multiple time series}\/}.
\newblock \bibinfo{publisher}{John Wiley \& Sons}.
\bibitem[{Hoga \& Dimitriadis(2022)}]{hoga22}
\bibinfo{author}{Hoga, Y.}, \& \bibinfo{author}{Dimitriadis, T.}
  (\bibinfo{year}{2022}).
\newblock \bibinfo{title}{On testing equal conditional predictive ability under
  measurement error}.
\newblock {\it \bibinfo{journal}{Journal of Business \& Economic
  Statistics}\/},  {\it \bibinfo{volume}{0}\/}, \bibinfo{pages}{1--13}.
\bibitem[{Jansson(2002)}]{jansson02}
\bibinfo{author}{Jansson, M.} (\bibinfo{year}{2002}).
\newblock \bibinfo{title}{Consistent covariance matrix estimation for linear
  processes}.
\newblock {\it \bibinfo{journal}{Econometric Theory}\/},  {\it
  \bibinfo{volume}{18}\/}, \bibinfo{pages}{1449--1459}.
\bibitem[{Keane \& Runkle(1990)}]{keane90}
\bibinfo{author}{Keane, M.~P.}, \& \bibinfo{author}{Runkle, D.~E.}
  (\bibinfo{year}{1990}).
\newblock \bibinfo{title}{Testing the rationality of price forecasts: New
  evidence from panel data}.
\newblock {\it \bibinfo{journal}{The American Economic Review}\/},  {\it
  \bibinfo{volume}{80}\/}, \bibinfo{pages}{714--735}.
\bibitem[{Kelejian \& Prucha(1998)}]{kelejian98}
\bibinfo{author}{Kelejian, H.~H.}, \& \bibinfo{author}{Prucha, I.~R.}
  (\bibinfo{year}{1998}).
\newblock \bibinfo{title}{A generalized spatial two-stage least squares
  procedure for estimating a spatial autoregressive model with autoregressive
  disturbances}.
\newblock {\it \bibinfo{journal}{The Journal of Real Estate Finance and
  Economics}\/},  {\it \bibinfo{volume}{17}\/}, \bibinfo{pages}{99--121}.
\bibitem[{Kelejian \& Prucha(2007)}]{kelejian07}
\bibinfo{author}{Kelejian, H.~H.}, \& \bibinfo{author}{Prucha, I.~R.}
  (\bibinfo{year}{2007}).
\newblock \bibinfo{title}{{HAC} estimation in a spatial framework}.
\newblock {\it \bibinfo{journal}{Journal of Econometrics}\/},  {\it
  \bibinfo{volume}{140}\/}, \bibinfo{pages}{131--154}.
\bibitem[{Kim \& Sun(2013)}]{kim13}
\bibinfo{author}{Kim, M.~S.}, \& \bibinfo{author}{Sun, Y.}
  (\bibinfo{year}{2013}).
\newblock \bibinfo{title}{Heteroskedasticity and spatiotemporal dependence
  robust inference for linear panel models with fixed effects}.
\newblock {\it \bibinfo{journal}{Journal of Econometrics}\/},  {\it
  \bibinfo{volume}{177}\/}, \bibinfo{pages}{85--108}.
\bibitem[{Mariano \& Preve(2012)}]{mariano12}
\bibinfo{author}{Mariano, R.~S.}, \& \bibinfo{author}{Preve, D.}
  (\bibinfo{year}{2012}).
\newblock \bibinfo{title}{Statistical tests for multiple forecast comparison}.
\newblock {\it \bibinfo{journal}{Journal of econometrics}\/},  {\it
  \bibinfo{volume}{169}\/}, \bibinfo{pages}{123--130}.
\bibitem[{Moscone \& Tosetti(2012)}]{moscone12}
\bibinfo{author}{Moscone, F.}, \& \bibinfo{author}{Tosetti, E.}
  (\bibinfo{year}{2012}).
\newblock \bibinfo{title}{{HAC} estimation in spatial panels}.
\newblock {\it \bibinfo{journal}{Economics Letters}\/},  {\it
  \bibinfo{volume}{117}\/}, \bibinfo{pages}{60--65}.
\bibitem[{Moscone \& Tosetti(2015)}]{moscone15}
\bibinfo{author}{Moscone, F.}, \& \bibinfo{author}{Tosetti, E.}
  (\bibinfo{year}{2015}).
\newblock \bibinfo{title}{Robust estimation under error cross section
  dependence}.
\newblock {\it \bibinfo{journal}{Economics Letters}\/},  {\it
  \bibinfo{volume}{133}\/}, \bibinfo{pages}{100--104}.
\bibitem[{Pain et~al.(2014)Pain, Lewis, Dang, Jin \& Richardson}]{pain14}
\bibinfo{author}{Pain, N.}, \bibinfo{author}{Lewis, C.}, \bibinfo{author}{Dang,
  T.-T.}, \bibinfo{author}{Jin, Y.}, \& \bibinfo{author}{Richardson, P.}
  (\bibinfo{year}{2014}).
\newblock {\it \bibinfo{title}{{OECD} Forecasts During and After the Financial
  Crisis}\/}.
\newblock \bibinfo{type}{Working Paper} \bibinfo{number}{1107} OECD Economics
  Department.
\bibitem[{Pesaran(2015)}]{pesaran15}
\bibinfo{author}{Pesaran, M.~H.} (\bibinfo{year}{2015}).
\newblock \bibinfo{title}{Testing weak cross-sectional dependence in large
  panels}.
\newblock {\it \bibinfo{journal}{Econometric Reviews}\/},  {\it
  \bibinfo{volume}{34}\/}, \bibinfo{pages}{1089--1117}.
\bibitem[{Pesaran et~al.(2008)Pesaran, Ullah \& Yamagata}]{pesaran08}
\bibinfo{author}{Pesaran, M.~H.}, \bibinfo{author}{Ullah, A.}, \&
  \bibinfo{author}{Yamagata, T.} (\bibinfo{year}{2008}).
\newblock \bibinfo{title}{A bias-adjusted {LM} test of error cross-section
  independence}.
\newblock {\it \bibinfo{journal}{The Econometrics Journal}\/},  {\it
  \bibinfo{volume}{11}\/}, \bibinfo{pages}{105--127}.
\bibitem[{Phillips \& Magdalinos(2009)}]{phillips09}
\bibinfo{author}{Phillips, P.~C.}, \& \bibinfo{author}{Magdalinos, T.}
  (\bibinfo{year}{2009}).
\newblock \bibinfo{title}{Unit root and cointegrating limit theory when
  initialization is in the infinite past}.
\newblock {\it \bibinfo{journal}{Econometric Theory}\/},  {\it
  \bibinfo{volume}{25}\/}, \bibinfo{pages}{1682--1715}.
\bibitem[{Phillips \& Solo(1992)}]{phillips92}
\bibinfo{author}{Phillips, P.~C.}, \& \bibinfo{author}{Solo, V.}
  (\bibinfo{year}{1992}).
\newblock \bibinfo{title}{Asymptotics for linear processes}.
\newblock {\it \bibinfo{journal}{The Annals of Statistics}\/},  {\it
  \bibinfo{volume}{20}\/}, \bibinfo{pages}{971--1001}.
\bibitem[{Qu et~al.(2022)Qu, Timmermann \& Zhu}]{qu22}
\bibinfo{author}{Qu, R.}, \bibinfo{author}{Timmermann, A.}, \&
  \bibinfo{author}{Zhu, Y.} (\bibinfo{year}{2022}).
\newblock \bibinfo{title}{Comparing forecasting performance with panel data}.
\newblock \bibinfo{note}{Unpublished manuscript.}
\bibitem[{Stock \& Watson(2002)}]{stock02}
\bibinfo{author}{Stock, J.~H.}, \& \bibinfo{author}{Watson, M.~W.}
  (\bibinfo{year}{2002}).
\newblock \bibinfo{title}{Forecasting using principal components from a large
  number of predictors}.
\newblock {\it \bibinfo{journal}{Journal of the American Statistical
  Association}\/},  {\it \bibinfo{volume}{97}\/}, \bibinfo{pages}{1167--1179}.
\bibitem[{Vuong(1989)}]{vuong89}
\bibinfo{author}{Vuong, Q.~H.} (\bibinfo{year}{1989}).
\newblock \bibinfo{title}{Likelihood ratio tests for model selection and
  non-nested hypotheses}.
\newblock {\it \bibinfo{journal}{Econometrica}\/},  {\it
  \bibinfo{volume}{57}\/}, \bibinfo{pages}{307--333}.
\bibitem[{West(1996)}]{west96}
\bibinfo{author}{West, K.~D.} (\bibinfo{year}{1996}).
\newblock \bibinfo{title}{Asymptotic inference about predictive ability}.
\newblock {\it \bibinfo{journal}{Econometrica}\/},  {\it
  \bibinfo{volume}{64}\/}, \bibinfo{pages}{1067--1084}.
\bibitem[{White(2001)}]{white01}
\bibinfo{author}{White, H.} (\bibinfo{year}{2001}).
\newblock {\it \bibinfo{title}{Asymptotic theory for econometricians}\/}.
\newblock \bibinfo{publisher}{Academic Press, Revised Edition}.

\end{thebibliography}


\newgeometry{right=1.8cm, left=1.8cm, top=2.7cm, bottom=2.7cm}

\begin{spacing}{0.40}
\setlength\extrarowheight{0pt}
{\normalsize
\begin{landscape}
\begin{table}
  \centering
  \caption{Small Sample Properties of the Non-Robust Tests $S^{(1)}_{n,T}$ and $C^{(1)}_{n,T}$}
  \begin{threeparttable}
    \begin{tabular}{ccccccccccccc}
    \toprule
          & \multicolumn{5}{c}{Overall EPA Tests: $S^{(1)}_{n,T}$} &       &       & \multicolumn{5}{c}{Clustered EPA Tests: $C^{(1)}_{n,T}$} \\
\cmidrule{1-6}\cmidrule{8-13}    \textit{n\textbackslash{}T} & 10    & 20    & 30    & 50    & 100   &       & \textit{n\textbackslash{}T} & 10    & 20    & 30    & 50    & 100 \\
    \midrule
    \multicolumn{13}{c}{DGP1: Size} \\
    \midrule
    10    & 13.6  & 13.1  & 12.1  & 12.7  & 13.4  &       & 10    & 15.4  & 16.0  & 14.5  & 14.8  & 14.8 \\
    20    & 13.1  & 12.5  & 12.2  & 10.5  & 11.5  &       & 20    & 14.8  & 15.0  & 14.1  & 12.5  & 13.2 \\
    30    & 11.4  & 11.7  & 11.9  & 11.4  & 10.5  &       & 30    & 14.1  & 12.5  & 12.6  & 12.0  & 11.1 \\
    50    & 14.3  & 11.6  & 10.4  & 11.1  & 10.1  &       & 50    & 16.0  & 12.9  & 11.4  & 12.1  & 12.2 \\
    100   & 13.5  & 11.1  & 12.8  & 12.4  & 11.4  &       & 100   & 12.9  & 10.6  & 11.6  & 10.9  & 10.2 \\
    \midrule
    \multicolumn{13}{c}{DGP2: Size} \\
    \midrule
    10    & 48.0  & 47.3  & 44.9  & 42.9  & 48.5  &       & 10    & 41.9  & 40.6  & 38.4  & 34.8  & 42.1 \\
    20    & 62.2  & 62.2  & 60.4  & 58.4  & 56.8  &       & 20    & 56.3  & 55.7  & 55.4  & 52.7  & 51.3 \\
    30    & 67.7  & 64.3  & 65.1  & 65.6  & 65.4  &       & 30    & 63.2  & 59.9  & 60.7  & 61.2  & 60.0 \\
    50    & 72.2  & 72.4  & 72.0  & 74.2  & 73.0  &       & 50    & 69.3  & 68.9  & 67.6  & 70.0  & 68.6 \\
    100   & 79.9  & 80.8  & 81.9  & 79.2  & 80.0  &       & 100   & 76.4  & 75.9  & 77.4  & 75.6  & 76.5 \\
    \midrule
    \multicolumn{13}{c}{DGP1: Power} \\
    \midrule
    10    & 24.2  & 33.1  & 38.2  & 54.8  & 76.3  &       & 10    & 23.6  & 29.8  & 34.2  & 49.3  & 71.7 \\
    20    & 32.0  & 49.2  & 61.4  & 78.3  & 95.9  &       & 20    & 28.7  & 44.8  & 57.3  & 73.2  & 93.5 \\
    30    & 41.0  & 61.1  & 74.7  & 89.5  & 99.4  &       & 30    & 38.2  & 54.0  & 67.7  & 85.8  & 98.7 \\
    50    & 55.5  & 78.3  & 91.1  & 98.5  & 100.0 &       & 50    & 51.7  & 73.1  & 87.3  & 97.6  & 100.0 \\
    100   & 79.9  & 96.4  & 99.3  & 100.0 & 100.0 &       & 100   & 73.0  & 94.3  & 98.7  & 100.0 & 100.0 \\
    \midrule
    \multicolumn{13}{c}{DGP2: Power} \\
    \midrule
    10    & 95.4  & 99.6  & 100.0 & 100.0 & 100.0 &       & 10    & 95.6  & 99.6  & 100.0 & 100.0 & 100.0 \\
    20    & 98.2  & 99.8  & 100.0 & 100.0 & 100.0 &       & 20    & 98.5  & 99.9  & 100.0 & 100.0 & 100.0 \\
    30    & 98.8  & 100.0 & 100.0 & 100.0 & 100.0 &       & 30    & 99.1  & 100.0 & 100.0 & 100.0 & 100.0 \\
    50    & 99.1  & 100.0 & 100.0 & 100.0 & 100.0 &       & 50    & 99.6  & 100.0 & 100.0 & 100.0 & 100.0 \\
    100   & 99.5  & 100.0 & 100.0 & 100.0 & 100.0 &       & 100   & 100.0 & 100.0 & 100.0 & 100.0 & 100.0 \\
    \bottomrule
    \end{tabular}%
    \begin{tablenotes}
	\item Note: Overall EPA Tests are introduced in Section \ref{sec:overalltests} and Clustered EPA Tests are in Section \ref{sec:jointtests}. The nominal size is 5\%. The power is calculated under homogeneous alternative hypothesis.
	\end{tablenotes}
    \end{threeparttable}
  \label{tab:nonrobusttest}%
\end{table}%
\end{landscape}
}
\end{spacing}
\clearpage
\begin{spacing}{0.5}
\setlength\extrarowheight{0pt}
{\normalsize
\begin{landscape}
\begin{threeparttable}

    \begin{longtable}{ccccccccccccccc}
\caption{Size - DGP1: No Common Factors, Spatial Dependence} \label{tab:sizedgp1} \\
    \toprule
    
    \multicolumn{7}{c}{Overall EPA Tests}                 &       & \multicolumn{7}{c}{Clustered EPA Tests} \\
\cmidrule{1-7}\cmidrule{9-15}          & \textit{n\textbackslash{}T} & 10    & 20    & 30    & 50    & 100   &       &       & \textit{n\textbackslash{}T} & 10    & 20    & 30    & 50    & 100 \\
\midrule
	\endfirsthead
	\multicolumn{15}{l}{\textit{Table \ref{tab:sizedgp1} cont'd.}} \\
\midrule

    \multicolumn{7}{c}{Overall EPA Tests}                 &       & \multicolumn{7}{c}{Clustered EPA Tests} \\
\cmidrule{1-7}\cmidrule{9-15}          & \textit{n\textbackslash{}T} & 10    & 20    & 30    & 50    & 100   &       &       & \textit{n\textbackslash{}T} & 10    & 20    & 30    & 50    & 100 \\
\midrule

	\endhead

    \midrule
    
    \endfoot
    
    \bottomrule
    
    \endlastfoot
    $S^{(2)}_{n,T}$ & 10    & 9.9   & 9.1   & 8.2   & 9.0   & 8.7   &       & $C^{(2)}_{n,T}$ & 10    & 11.0  & 10.7  & 9.5   & 9.5   & 9.7 \\
          & 20    & 8.7   & 8.0   & 7.9   & 6.9   & 7.2   &       &       & 20    & 10.0  & 9.0   & 8.8   & 7.7   & 7.8 \\
          & 30    & 7.8   & 6.9   & 8.4   & 6.4   & 6.9   &       &       & 30    & 9.3   & 7.5   & 7.9   & 6.7   & 6.3 \\
          & 50    & 9.7   & 6.9   & 6.4   & 7.3   & 6.5   &       &       & 50    & 11.5  & 7.6   & 6.7   & 7.1   & 7.3 \\
          & 100   & 8.4   & 6.9   & 7.4   & 6.8   & 6.3   &       &       & 100   & 8.6   & 6.3   & 7.7   & 7.1   & 6.1 \\
    $S^{(2)}_{n,T} [ms]$ & 10    & 12.5  & 11.1  & 10.3  & 10.8  & 11.0  &       & $C^{(2)}_{n,T} [ms]$ & 10    & 13.3  & 13.0  & 11.7  & 11.7  & 11.6 \\
          & 20    & 11.1  & 10.3  & 10.2  & 8.8   & 9.3   &       &       & 20    & 12.3  & 11.1  & 11.2  & 10.2  & 10.2 \\
          & 30    & 9.7   & 9.2   & 10.5  & 8.6   & 8.4   &       &       & 30    & 11.1  & 9.3   & 10.1  & 9.2   & 8.3 \\
          & 50    & 11.8  & 9.2   & 7.9   & 9.6   & 8.4   &       &       & 50    & 13.0  & 9.8   & 8.8   & 8.8   & 9.3 \\
          & 100   & 9.8   & 8.4   & 9.7   & 9.0   & 8.3   &       &       & 100   & 8.3   & 6.1   & 8.0   & 7.2   & 6.1 \\
        $\underline{S}_{n,T}^{(2)}$  & 10    & 12.9  & 9.1   & 7.7   & 7.0   & 6.3   &         &  $\underline{C}_{n,T}^{(2)}$  & 10    & 23.5  & 17.8  & 15.2  & 13.9  & 12.5 \\
               & 20    & 12.2  & 10.0  & 9.1   & 6.8   & 6.6   &         &         & 20    & 18.2  & 12.1  & 9.8   & 8.1   & 7.3 \\
               & 30    & 11.4  & 8.7   & 9.2   & 6.6   & 6.8   &         &         & 30    & 23.2  & 15.5  & 14.3  & 11.6  & 9.2 \\
               & 50    & 13.1  & 9.6   & 7.7   & 7.5   & 6.8   &         &         & 50    & 24.9  & 15.9  & 12.1  & 10.7  & 10.5 \\
               & 100   & 13.4  & 9.9   & 10.0  & 9.0   & 7.7   &         &         & 100   & 19.4  & 9.8   & 9.3   & 7.3   & 5.5 \\
    $S^{(3)}_{n,T}$    & 10    & 8.3   & 7.2   & 5.9   & 5.8   & 5.9   &       & $C^{(3)}_{n,T}$     & 10    & 13.8  & 9.9   & 7.4   & 6.3   & 6.9 \\
          & 20    & 9.7   & 7.4   & 6.9   & 6.0   & 5.7   &       &       & 20    & 14.2  & 9.2   & 7.9   & 6.6   & 5.8 \\
          & 30    & 8.7   & 6.9   & 7.5   & 5.3   & 5.2   &       &       & 30    & 13.9  & 8.4   & 7.7   & 6.3   & 5.0 \\
          & 50    & 10.9  & 7.4   & 5.3   & 5.9   & 5.7   &       &       & 50    & 16.9  & 9.3   & 6.7   & 6.9   & 5.8 \\
          & 100   & 10.0  & 7.2   & 7.0   & 6.5   & 5.4   &       &       & 100   & 15.7  & 8.6   & 8.6   & 7.9   & 5.7 \\
    $\underline{S}_{n,T}^{(3)} [m=2]$ & 10    & 8.9   & 8.2   & 6.2   & 5.4   & 4.5   &       & $\underline{C}_{n,T}^{(3)} [m=2]$    & 10    & 14.9  & 12.4  & 8.6   & 6.8   & 6.0 \\
          & 20    & 10.7  & 9.0   & 8.2   & 6.9   & 6.8   &       &       & 20    & 16.5  & 12.1  & 10.5  & 9.0   & 7.6 \\
          & 30    & 9.4   & 8.3   & 8.6   & 7.4   & 6.9   &       &       & 30    & 14.1  & 11.1  & 10.7  & 9.0   & 7.7 \\
          & 50    & 13.0  & 8.7   & 7.7   & 8.1   & 7.7   &       &       & 50    & 16.9  & 12.0  & 9.8   & 9.8   & 9.7 \\
          & 100   & 11.7  & 9.3   & 10.6  & 10.5  & 9.0   &       &       & 100   & 13.9  & 9.6   & 11.0  & 9.8   & 8.8 \\
    $\underline{S}_{n,T}^{(3)} [m=1]$ & 10    & 10.3  & 9.4   & 8.1   & 7.2   & 7.2   &       & $\underline{C}_{n,T}^{(3)} [m=1]$ & 10    & 16.6  & 14.3  & 11.0  & 10.2  & 10.6 \\
          & 20    & 11.9  & 10.1  & 9.6   & 8.6   & 8.1   &       &       & 20    & 16.9  & 13.5  & 12.1  & 10.6  & 9.7 \\
          & 30    & 10.3  & 9.5   & 9.5   & 8.9   & 8.4   &       &       & 30    & 13.9  & 12.3  & 11.6  & 9.9   & 8.8 \\
          & 50    & 13.1  & 9.7   & 9.1   & 9.6   & 9.0   &       &       & 50    & 17.1  & 12.7  & 10.6  & 10.8  & 10.6 \\
          & 100   & 11.9  & 10.0  & 11.2  & 11.6  & 10.1  &       &       & 100   & 13.6  & 10.1  & 10.9  & 10.4  & 9.5 \\
    $\underline{S}_{n,T}^{(3)} [ic]$ & 10    & 8.2   & 6.8   & 5.8   & 5.6   & 7.9   &       & $\underline{C}_{n,T}^{(3)} [ic]$ & 10    & 14.4  & 10.1  & 7.4   & 6.4   & 8.4 \\
          & 20    & 9.6   & 8.2   & 10.0  & 9.8   & 11.4  &       &       & 20    & 15.0  & 11.3  & 12.2  & 11.9  & 13.2 \\
          & 30    & 8.8   & 9.6   & 11.5  & 11.2  & 10.5  &       &       & 30    & 14.8  & 11.3  & 12.0  & 11.8  & 11.1 \\
          & 50    & 11.2  & 11.3  & 10.3  & 11.1  & 10.1  &       &       & 50    & 17.3  & 12.9  & 11.5  & 12.1  & 12.2 \\
          & 100   & 11.2  & 11.1  & 12.8  & 12.4  & 11.4  &       &       & 100   & 15.1  & 10.6  & 11.6  & 10.9  & 10.2 \\
          \bottomrule
\end{longtable}%
\begin{tablenotes} \normalsize
\item Note: See the note of Table \ref{tab:nonrobusttest}. [\textit{ms}] indicates that the test uses a misspecified distance metric. For $\underline{S}_{n,T}^{(3)}$, the number of common factors is shown in brackets. [\textit{ic}] means that the number of common factors is chosen by IC.
\end{tablenotes}
\end{threeparttable}
\end{landscape}
}
\end{spacing}

\begin{spacing}{0.5}
\setlength\extrarowheight{-0pt}
{\normalsize
\begin{landscape}
\begin{threeparttable}

    \begin{longtable}{ccccccccccccccc}
\caption{Size - DGP2: Common Factors, Spatial Dependence} \label{tab:sizedgp2} \\
    \toprule
    
    \multicolumn{7}{c}{Overall EPA Tests}                 &       & \multicolumn{7}{c}{Clustered EPA Tests} \\
\cmidrule{1-7}\cmidrule{9-15}          & \textit{n\textbackslash{}T} & 10    & 20    & 30    & 50    & 100   &       &       & \textit{n\textbackslash{}T} & 10    & 20    & 30    & 50    & 100 \\
\midrule
	\endfirsthead
\multicolumn{15}{l}{\textit{Table \ref{tab:sizedgp2} cont'd.}} \\
\midrule

    \multicolumn{7}{c}{Overall EPA Tests}                 &       & \multicolumn{7}{c}{Clustered EPA Tests} \\
\cmidrule{1-7}\cmidrule{9-15}          & \textit{n\textbackslash{}T} & 10    & 20    & 30    & 50    & 100   &       &       & \textit{n\textbackslash{}T} & 10    & 20    & 30    & 50    & 100 \\
\midrule

	\endhead

    \midrule
    
    \endfoot
    
    \bottomrule
        
    \endlastfoot
    $S^{(2)}_{n,T}$ & 10    & 30.7  & 30.5  & 26.6  & 24.7  & 29.6  &       & $C^{(2)}_{n,T}$ & 10    & 24.1  & 22.2  & 19.6  & 17.0  & 20.7 \\
          & 20    & 30.0  & 29.4  & 28.1  & 25.9  & 25.0  &       &       & 20    & 21.5  & 20.6  & 19.5  & 17.3  & 17.1 \\
          & 30    & 37.3  & 34.7  & 35.6  & 34.8  & 32.6  &       &       & 30    & 29.3  & 25.7  & 27.4  & 25.6  & 24.1 \\
          & 50    & 47.4  & 43.6  & 42.9  & 44.4  & 42.7  &       &       & 50    & 39.0  & 35.1  & 33.9  & 35.2  & 32.7 \\
          & 100   & 59.3  & 59.1  & 58.0  & 56.3  & 58.0  &       &       & 100   & 52.7  & 51.9  & 52.2  & 49.6  & 49.9 \\
    $S^{(2)}_{n,T} [ms]$ & 10    & 37.7  & 36.1  & 33.5  & 31.3  & 38.1  &       & $C^{(2)}_{n,T} [ms]$ & 10    & 30.5  & 29.9  & 25.9  & 24.7  & 29.1 \\
          & 20    & 45.5  & 44.9  & 43.5  & 41.4  & 40.3  &       &       & 20    & 36.9  & 35.5  & 35.2  & 32.4  & 31.6 \\
          & 30    & 52.8  & 49.5  & 51.2  & 50.8  & 50.0  &       &       & 30    & 46.0  & 42.2  & 43.8  & 42.0  & 40.3 \\
          & 50    & 60.9  & 59.3  & 58.9  & 60.2  & 59.8  &       &       & 50    & 55.7  & 51.9  & 51.9  & 53.3  & 52.0 \\
          & 100   & 68.2  & 67.3  & 67.6  & 66.4  & 67.4  &       &       & 100   & 61.2  & 61.0  & 60.0  & 58.5  & 59.8 \\
        $\underline{S}_{n,T}^{(2)}$  & 10    & 24.3  & 23.0  & 20.3  & 18.9  & 22.4  &         &  $\underline{C}_{n,T}^{(2)}$  & 10    & 35.3  & 29.2  & 26.4  & 23.1  & 25.3 \\
               & 20    & 34.8  & 33.7  & 32.9  & 30.5  & 29.0  &         &         & 20    & 33.7  & 28.8  & 26.5  & 24.2  & 23.0 \\
               & 30    & 40.4  & 37.6  & 38.4  & 37.4  & 35.9  &         &         & 30    & 46.9  & 39.9  & 40.9  & 39.0  & 35.5 \\
               & 50    & 45.7  & 41.8  & 41.3  & 42.0  & 40.5  &         &         & 50    & 50.8  & 44.3  & 42.0  & 43.8  & 39.3 \\
               & 100   & 54.0  & 54.3  & 54.2  & 51.6  & 52.3  &         &         & 100   & 56.8  & 53.1  & 53.6  & 50.2  & 49.3 \\
    $S^{(3)}_{n,T}$    & 10    & 10.2  & 8.5   & 6.0   & 6.3   & 5.7   &       & $C^{(3)}_{n,T}$    & 10    & 16.9  & 9.8   & 7.5   & 6.7   & 7.1 \\
          & 20    & 9.4   & 7.4   & 6.7   & 5.2   & 5.9   &       &       & 20    & 15.4  & 10.1  & 8.0   & 6.5   & 6.2 \\
          & 30    & 10.3  & 6.5   & 7.7   & 5.8   & 5.4   &       &       & 30    & 15.7  & 8.8   & 8.3   & 6.7   & 6.0 \\
          & 50    & 8.7   & 6.9   & 6.5   & 5.5   & 5.0   &       &       & 50    & 15.2  & 9.3   & 7.3   & 6.8   & 4.8 \\
          & 100   & 8.5   & 7.1   & 5.3   & 5.7   & 4.7   &       &       & 100   & 14.9  & 9.0   & 7.5   & 7.2   & 5.2 \\
    $\underline{S}_{n,T}^{(3)} [m=2]$ & 10    & 9.9   & 8.0   & 5.6   & 6.0   & 5.3   &       & $\underline{C}_{n,T}^{(3)} [m=2]$    & 10    & 20.0  & 12.9  & 11.0  & 9.1   & 9.8 \\
          & 20    & 9.3   & 7.3   & 6.3   & 5.0   & 5.4   &       &       & 20    & 21.1  & 15.8  & 15.4  & 13.8  & 12.8 \\
          & 30    & 10.1  & 6.5   & 7.7   & 5.5   & 5.4   &       &       & 30    & 22.0  & 16.8  & 16.6  & 16.0  & 14.6 \\
          & 50    & 8.6   & 6.8   & 6.4   & 5.4   & 4.9   &       &       & 50    & 23.8  & 18.6  & 15.8  & 15.9  & 14.3 \\
          & 100   & 8.5   & 7.1   & 5.2   & 5.7   & 4.6   &       &       & 100   & 13.3  & 8.7   & 7.3   & 7.4   & 5.4 \\
    $\underline{S}_{n,T}^{(3)} [m=1]$ & 10    & 9.8   & 7.7   & 5.6   & 5.8   & 5.2   &       & $\underline{C}_{n,T}^{(3)} [m=1]$ & 10    & 25.3  & 20.3  & 18.7  & 16.8  & 15.6 \\
          & 20    & 9.3   & 7.2   & 6.2   & 4.9   & 5.3   &       &       & 20    & 24.8  & 19.2  & 18.5  & 16.2  & 15.8 \\
          & 30    & 10.1  & 6.4   & 7.6   & 5.5   & 5.3   &       &       & 30    & 24.2  & 18.6  & 18.2  & 17.2  & 15.1 \\
          & 50    & 8.6   & 6.7   & 6.4   & 5.4   & 4.9   &       &       & 50    & 25.5  & 19.4  & 16.6  & 16.3  & 14.5 \\
          & 100   & 8.6   & 7.1   & 5.2   & 5.7   & 4.6   &       &       & 100   & 12.4  & 8.8   & 7.2   & 7.7   & 5.3 \\
    $\underline{S}_{n,T}^{(3)} [ic]$ & 10    & 10.1  & 8.3   & 6.0   & 6.2   & 5.5   &       & $\underline{C}_{n,T}^{(3)} [ic]$ & 10    & 16.7  & 9.5   & 7.3   & 6.1   & 6.3 \\
          & 20    & 9.3   & 7.3   & 6.4   & 5.1   & 5.5   &       &       & 20    & 15.9  & 10.3  & 9.1   & 8.0   & 7.7 \\
          & 30    & 10.2  & 6.5   & 7.7   & 5.5   & 5.4   &       &       & 30    & 16.4  & 11.4  & 13.8  & 13.7  & 13.5 \\
          & 50    & 8.6   & 6.8   & 6.4   & 5.4   & 4.9   &       &       & 50    & 16.7  & 17.5  & 15.2  & 15.9  & 14.3 \\
          & 100   & 8.5   & 7.1   & 5.2   & 5.7   & 4.6   &       &       & 100   & 14.4  & 8.8   & 7.3   & 7.4   & 5.4 \\
          \bottomrule
\end{longtable}%
\begin{tablenotes} \normalsize
\item Note: See the note of Table \ref{tab:sizedgp1}.
\end{tablenotes}
\end{threeparttable}
\end{landscape}
}
\end{spacing}

\begin{spacing}{0.5}
\setlength\extrarowheight{0pt}
{\normalsize
\begin{landscape}
\begin{table}
  \centering
  \caption{Small Sample Properties of $S^{(3)}_{n,T}$ and $\tilde{S}^{(3)}_{n,T}$ Under Deviations from Normality}
  \begin{threeparttable}
    \begin{tabular}{ccccccccccccccc}
    \toprule
          & \multicolumn{6}{c}{DGP1}                      &       & \multicolumn{7}{c}{DGP2} \\
\cmidrule{1-7}\cmidrule{9-15}          & \textit{n\textbackslash{}T} & 10    & 20    & 30    & 50    & 100   &       &       & \textit{n\textbackslash{}T} & 10    & 20    & 30    & 50    & 100 \\
    \midrule
    \multicolumn{15}{c}{\textit{Size}} \\
    \midrule
    $S^{(3)}_{n,T}$    & 10    & 8.6   & 6.3   & 6.4   & 5.1   & 5.1   &       & $S^{(3)}_{n,T}$    & 10    & 9.3   & 7.6   & 6.8   & 5.8   & 5.7 \\
          & 20    & 8.4   & 7.4   & 5.0   & 6.0   & 5.3   &       &       & 20    & 10.2  & 7.3   & 6.9   & 5.9   & 5.3 \\
          & 30    & 10.5  & 6.6   & 6.3   & 6.2   & 5.7   &       &       & 30    & 9.6   & 7.7   & 6.7   & 5.2   & 5.6 \\
          & 50    & 8.3   & 7.6   & 6.1   & 6.0   & 5.5   &       &       & 50    & 10.1  & 7.2   & 7.4   & 5.9   & 6.6 \\
          & 100   & 8.9   & 7.0   & 5.8   & 5.3   & 5.0   &       &       & 100   & 8.1   & 6.2   & 5.6   & 5.1   & 5.6 \\
    $\tilde{S}^{(3)}_{n,T}$ & 10    & 4.1   & 3.9   & 5.0   & 4.4   & 4.8   &       & $\tilde{S}^{(3)}_{n,T}$ & 10    & 4.6   & 5.2   & 5.3   & 5.3   & 5.5 \\
          & 20    & 3.6   & 4.6   & 3.7   & 5.2   & 4.8   &       &       & 20    & 4.9   & 5.4   & 5.1   & 4.8   & 4.9 \\
          & 30    & 4.8   & 4.7   & 5.1   & 5.4   & 5.1   &       &       & 30    & 5.2   & 5.5   & 5.6   & 4.4   & 5.3 \\
          & 50    & 4.4   & 5.2   & 4.6   & 5.2   & 5.4   &       &       & 50    & 5.3   & 5.0   & 6.1   & 5.4   & 5.9 \\
          & 100   & 4.4   & 4.5   & 4.5   & 4.6   & 4.7   &       &       & 100   & 3.9   & 4.0   & 4.6   & 4.3   & 5.1 \\
    \midrule
    \multicolumn{15}{c}{\textit{Power}} \\
    \midrule
    $S^{(3)}_{n,T}$    & 10    & 13.4  & 14.9  & 20.8  & 27.8  & 45.3  &       & $S^{(3)}_{n,T}$    & 10    & 72.8  & 93.5  & 98.2  & 100.0 & 100.0 \\
          & 20    & 21.3  & 25.9  & 33.6  & 49.3  & 73.5  &       &       & 20    & 74.4  & 95.6  & 99.2  & 100.0 & 100.0 \\
          & 30    & 25.2  & 34.6  & 46.7  & 63.8  & 87.2  &       &       & 30    & 77.6  & 95.3  & 98.9  & 100.0 & 100.0 \\
          & 50    & 32.0  & 49.4  & 63.1  & 83.6  & 98.0  &       &       & 50    & 77.9  & 95.4  & 99.1  & 100.0 & 100.0 \\
          & 100   & 52.5  & 75.4  & 88.0  & 98.0  & 100.0 &       &       & 100   & 78.9  & 96.8  & 99.3  & 100.0 & 100.0 \\
    $\tilde{S}^{(3)}_{n,T}$ & 10    & 7.0   & 11.2  & 17.0  & 25.4  & 43.9  &       & $\tilde{S}^{(3)}_{n,T}$ & 10    & 60.5  & 90.6  & 97.4  & 100.0 & 100.0 \\
          & 20    & 12.1  & 20.8  & 29.5  & 46.2  & 72.2  &       &       & 20    & 62.2  & 93.3  & 98.7  & 100.0 & 100.0 \\
          & 30    & 14.9  & 28.2  & 42.2  & 62.0  & 86.6  &       &       & 30    & 65.8  & 93.2  & 98.8  & 100.0 & 100.0 \\
          & 50    & 20.7  & 41.8  & 59.5  & 82.1  & 97.6  &       &       & 50    & 65.9  & 93.8  & 98.9  & 100.0 & 100.0 \\
          & 100   & 38.3  & 68.9  & 85.4  & 97.6  & 100.0 &       &       & 100   & 66.4  & 95.3  & 99.2  & 100.0 & 100.0 \\
    \bottomrule
    \end{tabular}%
    \begin{tablenotes}
	\item Note: The tests are introduced in Section \ref{sec:overalltests}. The nominal size is 5\%. The power is calculated under homogeneous alternative hypothesis.
	\end{tablenotes}
    \end{threeparttable}
  \label{tab:nonnormal}%
\end{table}%
\end{landscape}
}
\end{spacing}

\begin{table}[htbp]
  \centering
  \caption{Size Properties of the Clustered Test $J_n^D$ of QTZ}
    \begin{tabular}{cccccccccccc}
    \toprule
          & \multicolumn{5}{c}{DGP1}              &       & \multicolumn{5}{c}{DGP2} \\
\cmidrule{1-6}\cmidrule{8-12}    \textit{n\textbackslash{}T} & 10    & 20    & 30    & 50    & 100   &       & 10    & 20    & 30    & 50    & 100 \\
    \midrule
    10    & 5.5   & 5.0   & 5.4   & 5.1   & 5.6   &       & 15.4  & 15.8  & 15.7  & 15.1  & 15.7 \\
    20    & 5.6   & 5.1   & 5.8   & 5.0   & 5.2   &       & 19.5  & 20.3  & 20.7  & 20.3  & 19.9 \\
    30    & 5.5   & 5.6   & 5.3   & 5.5   & 5.5   &       & 23.0  & 22.7  & 23.3  & 23.9  & 22.9 \\
    50    & 5.9   & 5.3   & 6.0   & 5.6   & 4.4   &       & 28.6  & 28.6  & 29.3  & 30.5  & 27.8 \\
    100   & 5.5   & 5.2   & 6.7   & 6.1   & 6.3   &       & 47.0  & 48.5  & 46.8  & 46.2  & 47.9 \\
    \bottomrule
    \end{tabular}%
  \label{tab:tzsize}%
\end{table}%

\begin{spacing}{0.5}
\setlength\extrarowheight{-0pt}
{\normalsize
\begin{landscape}
\begin{threeparttable}

    \begin{longtable}{ccccccccccccccc}
\caption{Power Under Homogeneous Alternative – DGP 1: No Common Factors, Spatial Dependence} \label{tab:powerdgp1} \\
    \toprule
    
    \multicolumn{7}{c}{Overall EPA Tests}                 &       & \multicolumn{7}{c}{Clustered EPA Tests} \\
\cmidrule{1-7}\cmidrule{9-15}          & \textit{n\textbackslash{}T} & 10    & 20    & 30    & 50    & 100   &       &       & \textit{n\textbackslash{}T} & 10    & 20    & 30    & 50    & 100 \\
\midrule
	\endfirsthead
	\multicolumn{15}{l}{\textit{Table \ref{tab:powerdgp1} cont'd.}} \\
\midrule

    \multicolumn{7}{c}{Overall EPA Tests}                 &       & \multicolumn{7}{c}{Clustered EPA Tests} \\
\cmidrule{1-7}\cmidrule{9-15}          & \textit{n\textbackslash{}T} & 10    & 20    & 30    & 50    & 100   &       &       & \textit{n\textbackslash{}T} & 10    & 20    & 30    & 50    & 100 \\
\midrule

	\endhead

    \midrule
    
    \endfoot
    
    \bottomrule
        
    \endlastfoot
    $S^{(2)}_{n,T}$ & 10    & 19.1  & 26.6  & 30.8  & 47.6  & 70.8  &       & $C^{(2)}_{n,T}$ & 10    & 17.3  & 22.3  & 26.8  & 40.3  & 64.1 \\
          & 20    & 24.6  & 40.1  & 53.1  & 71.2  & 93.2  &       &       & 20    & 21.7  & 34.3  & 45.7  & 63.7  & 88.4 \\
          & 30    & 33.6  & 51.9  & 67.2  & 85.2  & 98.9  &       &       & 30    & 29.5  & 44.6  & 57.8  & 78.4  & 97.3 \\
          & 50    & 48.6  & 71.6  & 87.8  & 97.5  & 100.0 &       &       & 50    & 42.8  & 63.5  & 81.1  & 94.8  & 100.0 \\
          & 100   & 71.2  & 94.1  & 98.4  & 100.0 & 100.0 &       &       & 100   & 63.4  & 89.8  & 96.8  & 100.0 & 100.0 \\
    $S^{(2)}_{n,T} [ms]$ & 10    & 21.8  & 29.9  & 34.9  & 51.6  & 73.8  &       & $C^{(2)}_{n,T} [ms]$ & 10    & 20.0  & 26.1  & 30.6  & 45.1  & 68.3 \\
          & 20    & 28.7  & 45.3  & 57.3  & 75.1  & 95.0  &       &       & 20    & 24.9  & 39.6  & 51.9  & 68.5  & 91.1 \\
          & 30    & 37.6  & 56.3  & 71.2  & 87.9  & 99.3  &       &       & 30    & 33.9  & 49.6  & 62.8  & 82.4  & 98.3 \\
          & 50    & 52.4  & 75.2  & 89.8  & 98.1  & 100.0 &       &       & 50    & 47.0  & 68.0  & 84.6  & 96.2  & 100.0 \\
          & 100   & 74.3  & 95.2  & 98.7  & 100.0 & 100.0 &       &       & 100   & 67.4  & 91.8  & 97.8  & 100.0 & 100.0 \\
        $\underline{S}_{n,T}^{(2)}$  & 10    & 20.3  & 23.3  & 27.0  & 41.0  & 64.2  &         &  $\underline{C}_{n,T}^{(2)}$  & 10    & 31.5  & 30.9  & 33.0  & 45.5  & 67.7 \\
               & 20    & 27.0  & 38.7  & 52.0  & 68.8  & 92.0  &         &         & 20    & 30.9  & 36.8  & 45.5  & 61.2  & 87.0 \\
               & 30    & 36.2  & 52.9  & 65.6  & 83.4  & 98.6  &         &         & 30    & 43.7  & 52.3  & 65.1  & 82.6  & 97.8 \\
               & 50    & 50.6  & 72.5  & 85.8  & 97.1  & 100.0 &         &         & 50    & 56.0  & 70.9  & 83.9  & 95.4  & 100.0 \\
               & 100   & 73.3  & 94.5  & 98.5  & 100.0 & 100.0 &         &         & 100   & 66.6  & 86.8  & 94.6  & 99.6  & 100.0 \\
    $S^{(3)}_{n,T}$    & 10    & 16.5  & 21.7  & 25.4  & 39.1  & 63.5  &       & $C^{(3)}_{n,T}$    & 10    & 18.5  & 19.9  & 22.3  & 32.0  & 53.5 \\
          & 20    & 23.8  & 37.3  & 48.9  & 66.2  & 90.4  &       &       & 20    & 26.0  & 32.5  & 42.3  & 57.8  & 84.9 \\
          & 30    & 31.8  & 48.8  & 63.0  & 82.2  & 98.3  &       &       & 30    & 33.1  & 42.9  & 54.3  & 73.2  & 96.3 \\
          & 50    & 47.0  & 67.7  & 84.3  & 96.3  & 100.0 &       &       & 50    & 45.5  & 60.7  & 78.4  & 92.8  & 100.0 \\
          & 100   & 68.5  & 92.2  & 97.8  & 99.9  & 100.0 &       &       & 100   & 66.2  & 88.4  & 95.8  & 99.8  & 100.0 \\
    $\underline{S}_{n,T}^{(3)} [m=2]$ & 10    & 17.4  & 20.8  & 25.1  & 36.0  & 57.2  &       & $\underline{C}_{n,T}^{(3)} [m=2]$    & 10    & 21.0  & 21.8  & 23.5  & 31.0  & 47.4 \\
          & 20    & 25.5  & 38.6  & 50.4  & 67.1  & 90.1  &       &       & 20    & 28.0  & 37.8  & 46.3  & 61.4  & 84.9 \\
          & 30    & 34.3  & 52.3  & 66.9  & 84.1  & 98.1  &       &       & 30    & 34.4  & 47.8  & 58.8  & 79.0  & 96.9 \\
          & 50    & 50.3  & 72.1  & 87.3  & 97.3  & 100.0 &       &       & 50    & 48.7  & 67.6  & 81.9  & 95.5  & 100.0 \\
          & 100   & 73.0  & 94.4  & 98.7  & 100.0 & 100.0 &       &       & 100   & 68.6  & 92.0  & 97.5  & 99.9  & 100.0 \\
    $\underline{S}_{n,T}^{(3)} [m=1]$ & 10    & 19.4  & 24.0  & 28.3  & 41.5  & 61.9  &       & $\underline{C}_{n,T}^{(3)} [m=1]$ & 10    & 23.4  & 25.6  & 27.3  & 38.7  & 57.2 \\
          & 20    & 28.3  & 42.8  & 53.5  & 71.8  & 92.1  &       &       & 20    & 28.3  & 40.2  & 50.8  & 66.8  & 88.3 \\
          & 30    & 36.3  & 54.7  & 70.1  & 86.3  & 98.8  &       &       & 30    & 36.1  & 50.6  & 62.7  & 81.3  & 97.5 \\
          & 50    & 52.4  & 74.4  & 89.1  & 97.7  & 100.0 &       &       & 50    & 50.4  & 69.9  & 84.0  & 96.4  & 100.0 \\
          & 100   & 75.5  & 95.3  & 98.9  & 100.0 & 100.0 &       &       & 100   & 70.4  & 92.8  & 97.9  & 100.0 & 100.0 \\
    $\underline{S}_{n,T}^{(3)} [ic]$ & 10    & 16.3  & 21.0  & 23.8  & 36.8  & 64.9  &       & $\underline{C}_{n,T}^{(3)} [ic]$ & 10    & 19.1  & 19.7  & 20.9  & 30.6  & 56.4 \\
          & 20    & 23.8  & 38.7  & 56.0  & 76.3  & 95.8  &       &       & 20    & 26.6  & 35.3  & 51.6  & 71.1  & 93.3 \\
          & 30    & 31.6  & 56.0  & 73.1  & 89.0  & 99.4  &       &       & 30    & 33.5  & 49.9  & 65.7  & 85.4  & 98.7 \\
          & 50    & 48.2  & 77.0  & 90.9  & 98.5  & 100.0 &       &       & 50    & 47.0  & 71.8  & 87.1  & 97.6  & 100.0 \\
          & 100   & 71.8  & 96.4  & 99.3  & 100.0 & 100.0 &       &       & 100   & 68.0  & 94.2  & 98.6  & 100.0 & 100.0 \\
          \bottomrule
\end{longtable}%
\begin{tablenotes} \normalsize
\item Note: See the note of Table \ref{tab:sizedgp1}.
\end{tablenotes}
\end{threeparttable}
\end{landscape}
}
\end{spacing}

\begin{spacing}{0.5}
\setlength\extrarowheight{-0pt}
{\normalsize
\begin{landscape}
\begin{threeparttable}

    \begin{longtable}{ccccccccccccccc}
\caption{Power Under Homogeneous Alternative – DGP 2: Common Factors, Spatial Dependence} \label{tab:powerdgp2} \\
    \toprule
    
    \multicolumn{7}{c}{Overall EPA Tests}                 &       & \multicolumn{7}{c}{Clustered EPA Tests} \\
\cmidrule{1-7}\cmidrule{9-15}          & \textit{n\textbackslash{}T} & 10    & 20    & 30    & 50    & 100   &       &       & \textit{n\textbackslash{}T} & 10    & 20    & 30    & 50    & 100 \\
\midrule
	\endfirsthead
	\multicolumn{15}{l}{\textit{Table \ref{tab:powerdgp2} cont'd.}} \\
	\midrule
    \multicolumn{7}{c}{Overall EPA Tests}                 &       & \multicolumn{7}{c}{Clustered EPA Tests} \\
\cmidrule{1-7}\cmidrule{9-15}          & \textit{n\textbackslash{}T} & 10    & 20    & 30    & 50    & 100   &       &       & \textit{n\textbackslash{}T} & 10    & 20    & 30    & 50    & 100 \\
\midrule

	\endhead

    \midrule
    
    \endfoot
    
    \bottomrule
        
    \endlastfoot
    $S^{(2)}_{n,T}$ & 10    & 91.8  & 98.9  & 99.9  & 100.0 & 100.0 &       & $C^{(2)}_{n,T}$ & 10    & 90.0  & 98.6  & 99.9  & 100.0 & 100.0 \\
          & 20    & 92.8  & 99.2  & 100.0 & 100.0 & 100.0 &       &       & 20    & 91.1  & 99.2  & 100.0 & 100.0 & 100.0 \\
          & 30    & 95.6  & 99.5  & 100.0 & 100.0 & 100.0 &       &       & 30    & 94.9  & 99.6  & 100.0 & 100.0 & 100.0 \\
          & 50    & 97.1  & 99.9  & 100.0 & 100.0 & 100.0 &       &       & 50    & 97.1  & 100.0 & 100.0 & 100.0 & 100.0 \\
          & 100   & 98.5  & 100.0 & 100.0 & 100.0 & 100.0 &       &       & 100   & 99.9  & 100.0 & 100.0 & 100.0 & 100.0 \\
    $S^{(2)}_{n,T} [ms]$ & 10    & 93.8  & 99.3  & 100.0 & 100.0 & 100.0 &       & $C^{(2)}_{n,T} [ms]$ & 10    & 93.0  & 99.3  & 99.9  & 100.0 & 100.0 \\
          & 20    & 96.3  & 99.6  & 100.0 & 100.0 & 100.0 &       &       & 20    & 95.9  & 99.7  & 100.0 & 100.0 & 100.0 \\
          & 30    & 97.6  & 99.8  & 100.0 & 100.0 & 100.0 &       &       & 30    & 97.4  & 99.8  & 100.0 & 100.0 & 100.0 \\
          & 50    & 98.3  & 100.0 & 100.0 & 100.0 & 100.0 &       &       & 50    & 98.9  & 100.0 & 100.0 & 100.0 & 100.0 \\
          & 100   & 99.2  & 100.0 & 100.0 & 100.0 & 100.0 &       &       & 100   & 99.7  & 100.0 & 100.0 & 100.0 & 100.0 \\
        $\underline{S}_{n,T}^{(2)}$  & 10    & 88.8  & 98.3  & 99.9  & 100.0 & 100.0 &         &  $\underline{C}_{n,T}^{(2)}$  & 10    & 93.1  & 98.8  & 100.0 & 100.0 & 100.0 \\
               & 20    & 93.6  & 99.3  & 100.0 & 100.0 & 100.0 &         &         & 20    & 95.6  & 99.6  & 100.0 & 100.0 & 100.0 \\
               & 30    & 95.5  & 99.6  & 100.0 & 100.0 & 100.0 &         &         & 30    & 98.1  & 100.0 & 100.0 & 100.0 & 100.0 \\
               & 50    & 96.3  & 99.9  & 100.0 & 100.0 & 100.0 &         &         & 50    & 99.3  & 100.0 & 100.0 & 100.0 & 100.0 \\
               & 100   & 98.4  & 99.9  & 100.0 & 100.0 & 100.0 &         &         & 100   & 100.0 & 100.0 & 100.0 & 100.0 & 100.0 \\
    $S^{(3)}_{n,T}$    & 10    & 74.5  & 93.1  & 99.0  & 100.0 & 100.0 &       & $C^{(3)}_{n,T}$    & 10    & 81.3  & 94.9  & 99.2  & 100.0 & 100.0 \\
          & 20    & 74.9  & 94.8  & 99.4  & 100.0 & 100.0 &       &       & 20    & 85.7  & 98.2  & 99.6  & 100.0 & 100.0 \\
          & 30    & 77.7  & 95.6  & 99.3  & 100.0 & 100.0 &       &       & 30    & 89.4  & 99.2  & 99.8  & 100.0 & 100.0 \\
          & 50    & 78.1  & 95.4  & 99.5  & 100.0 & 100.0 &       &       & 50    & 94.3  & 99.8  & 100.0 & 100.0 & 100.0 \\
          & 100   & 79.0  & 96.6  & 99.7  & 100.0 & 100.0 &       &       & 100   & 99.7  & 100.0 & 100.0 & 100.0 & 100.0 \\
    $\underline{S}_{n,T}^{(3)} [m=2]$ & 10    & 73.8  & 92.8  & 98.9  & 100.0 & 100.0 &       & $\underline{C}_{n,T}^{(3)} [m=2]$    & 10    & 82.4  & 95.1  & 99.3  & 100.0 & 100.0 \\
          & 20    & 74.6  & 94.6  & 99.3  & 100.0 & 100.0 &       &       & 20    & 88.5  & 98.4  & 99.7  & 100.0 & 100.0 \\
          & 30    & 77.4  & 95.6  & 99.2  & 100.0 & 100.0 &       &       & 30    & 91.8  & 99.4  & 99.8  & 100.0 & 100.0 \\
          & 50    & 77.9  & 95.4  & 99.5  & 100.0 & 100.0 &       &       & 50    & 96.3  & 99.9  & 100.0 & 100.0 & 100.0 \\
          & 100   & 78.9  & 96.6  & 99.7  & 100.0 & 100.0 &       &       & 100   & 99.7  & 100.0 & 100.0 & 100.0 & 100.0 \\
    $\underline{S}_{n,T}^{(3)} [m=1]$ & 10    & 73.6  & 92.6  & 98.9  & 100.0 & 100.0 &       & $\underline{C}_{n,T}^{(3)} [m=1]$ & 10    & 86.1  & 96.2  & 99.3  & 100.0 & 100.0 \\
          & 20    & 74.2  & 94.5  & 99.3  & 100.0 & 100.0 &       &       & 20    & 90.1  & 98.9  & 99.7  & 100.0 & 100.0 \\
          & 30    & 77.4  & 95.6  & 99.2  & 100.0 & 100.0 &       &       & 30    & 92.8  & 99.5  & 99.9  & 100.0 & 100.0 \\
          & 50    & 77.8  & 95.4  & 99.5  & 100.0 & 100.0 &       &       & 50    & 97.0  & 99.9  & 100.0 & 100.0 & 100.0 \\
          & 100   & 78.9  & 96.6  & 99.7  & 100.0 & 100.0 &       &       & 100   & 99.8  & 100.0 & 100.0 & 100.0 & 100.0 \\
    $\underline{S}_{n,T}^{(3)} [ic]$ & 10    & 74.3  & 93.0  & 99.0  & 100.0 & 100.0 &       & $\underline{C}_{n,T}^{(3)} [ic]$ & 10    & 81.3  & 94.7  & 99.2  & 100.0 & 100.0 \\
          & 20    & 74.9  & 94.7  & 99.4  & 100.0 & 100.0 &       &       & 20    & 86.0  & 98.0  & 99.7  & 100.0 & 100.0 \\
          & 30    & 77.6  & 95.6  & 99.2  & 100.0 & 100.0 &       &       & 30    & 89.7  & 99.3  & 99.8  & 100.0 & 100.0 \\
          & 50    & 78.1  & 95.4  & 99.5  & 100.0 & 100.0 &       &       & 50    & 94.5  & 99.9  & 100.0 & 100.0 & 100.0 \\
          & 100   & 79.0  & 96.6  & 99.7  & 100.0 & 100.0 &       &       & 100   & 99.7  & 100.0 & 100.0 & 100.0 & 100.0 \\
          \bottomrule
\end{longtable}%
\begin{tablenotes} \normalsize
\item Note: See the note of Table \ref{tab:sizedgp1}.
\end{tablenotes}
\end{threeparttable}
\end{landscape}
}
\end{spacing}

\begin{table}[htbp]
  \centering
  \caption{Small Sample Properties of the Information Criterion $IC_{p1}$}
    \begin{tabular}{cccccccccccc}
    \toprule
          & \multicolumn{5}{c}{DGP1}              &       & \multicolumn{5}{c}{DGP2} \\
\cmidrule{1-6}\cmidrule{8-12}    \textit{n\textbackslash{}T} & 10    & 20    & 30    & 50    & 100   &       & 10    & 20    & 30    & 50    & 100 \\
    \midrule
    10    & 5.00  & 5.00  & 4.95  & 4.57  & 2.57  &       & 5.00  & 5.00  & 5.00  & 5.00  & 5.00 \\
    20    & 5.00  & 3.88  & 1.09  & 0.20  & 0.01  &       & 5.00  & 5.52  & 4.48  & 3.73  & 3.28 \\
    30    & 4.99  & 1.12  & 0.25  & 0.06  & 0.00  &       & 5.00  & 4.07  & 2.80  & 2.45  & 2.28 \\
    50    & 4.94  & 0.15  & 0.05  & 0.01  & 0.00  &       & 5.00  & 2.25  & 2.08  & 2.04  & 2.01 \\
    100   & 3.62  & 0.01  & 0.00  & 0.00  & 0.00  &       & 4.98  & 1.94  & 1.99  & 2.00  & 2.00 \\
    \bottomrule
    \end{tabular}%
  \label{tab:icprop}%
\end{table}%

\begin{table}[htbp!]
\caption{Average Loss Differentials for the Economic Growth Forecasts, DM Test Statistics and Their $p$-values, 1998-2016 (OECD \textit{vs.} IMF)}
\label{tab:dmbc}\centering
\resizebox*{!}{22cm}{        
\begin{threeparttable}
    \begin{tabular}{ccccc}
    \toprule
    {Country} & {Statistics} &       & {Country} & {Statistics} \\
    \midrule
    Australia   & -0.052 &       & Iceland   & -0.484 \\
          & (-0.451) &       &       & (-0.459) \\
          & [0.652] &       &       & [0.646] \\
    Austria   & -0.035 &       & Italy$\dagger$   & 0.492 \\
          & (-0.293) &       &       & (1.047) \\
          & [0.770] &       &       & [0.295] \\
    Belgium   & \textbf{0.555} &       & Japan$\dagger$   & 0.191 \\
          & \textbf{(1.701)} &       &       & (1.040) \\
          & \textbf{[0.089]} &       &       & [0.298] \\
    Canada$\dagger$   & 0.273 &       & South Korea   & 0.571 \\
          & (1.504) &       &       & (0.564) \\
          & [0.133] &       &       & [0.573] \\
    Switzerland   & 0.272 &       & Luxembourg   & \textbf{1.951} \\
          & (1.131) &       &       & \textbf{(1.757)} \\
          & [0.258] &       &       & \textbf{[0.079]} \\
    Czech Republic   & -1.295 &       & Mexico   & 0.009 \\
          & (-1.014) &       &       & (0.007) \\
          & [0.311] &       &       & [0.994] \\
    Germany$\dagger$   & -0.627 &       & Netherlands   & 0.405 \\
          & (-1.125) &       &       & (1.549) \\
          & [0.260] &       &       & [0.121] \\
    Denmark   & -0.150 &       & Norway   & -0.165 \\
          & (-0.771) &       &       & (-0.614) \\
          & [0.440] &       &       & [0.539] \\
    Spain   & \textbf{-0.657} &       & New Zealand   & -0.463 \\
          & \textbf{(-1.726)} &       &       & (-1.443) \\
          & \textbf{[0.084]} &       &       & [0.149] \\
    Finland   & 0.062 &       & Poland   & -0.477 \\
          & (0.079) &       &       & (-0.903) \\
          & [0.937] &       &       & [0.367] \\
    France$\dagger$   & 0.216 &       & Portugal   & 0.062 \\
          & (1.505) &       &       & (0.128) \\
          & [0.132] &       &       & [0.898] \\
    United Kingdom$\dagger$   & -0.143 &       & Sweden   & 0.269 \\
          & (-0.740) &       &       & (0.501) \\
          & [0.459] &       &       & [0.616] \\
    Greece   & -2.111 &       & Türkiye   & -0.746 \\
          & (-1.465) &       &       & (-0.306) \\
          & [0.143] &       &       & [0.760] \\
    Hungary   & \textbf{1.303} &       & United States$\dagger$   & 0.150 \\
          & \textbf{(1.873)} &       &       & (0.460) \\
          & \textbf{[0.061]} &       &       & [0.645] \\
\cmidrule{4-5}    Ireland   & 0.875 &       & \textit{Average} & 0.009 \\
          & (0.761) &       & \textit{Average (G7)} & 0.079 \\
    \textcolor[rgb]{ 1,  1,  1}{Average (Non-G7)} & [0.447] &       & \textit{Average (Non-G7)} & -0.014 \\
    \bottomrule
    \end{tabular}%
    \begin{tablenotes}
    	\item Note: $^\dagger$ G7 countries. DM statistics in parentheses are calculated as $S_{i,T}^{(0)}=\sqrt{T} (\Delta \bar{L}_{i,T}/\hat{\sigma}_{i,T}) \overset{d}{\rightarrow} N(0,1)$ where $\hat{\sigma}_{i,T}^{2}=\frac{1}{T}\sum_{t=1}^{T} \Delta \tilde{L}^2_{it}$. Differences significant at 10\% are shown in bold. $p$-values are in brackets.
    \end{tablenotes}
  \end{threeparttable} 
  }
\end{table}

\begin{table}[htbp]
  \centering
  \caption{PC Estimates of the Factor Loadings in the Loss Differentials of the Economic Growth Forecasts (OECD \textit{vs.} IMF)}
    \begin{tabular}{cccc}
    \toprule
    Countries & PC1   & PC2   & PC3 \\
    \midrule
    Australia   & -0.05 & -0.20 & 0.26 \\
    Austria   & -0.14 & 0.27  & -0.21 \\
    Belgium   & 0.37  & 0.26  & -1.10 \\
    Canada   & 0.26  & 0.14  & -0.20 \\
    Switzerland   & -0.08 & -0.48 & -0.42 \\
    Czech Republic   & 0.07  & -2.97 & -1.71 \\
    Germany   & -1.01 & -1.66 & 0.65 \\
    Denmark   & 0.24  & 0.08  & 0.09 \\
    Spain   & 0.17  & 1.01  & 0.95 \\
    Finland   & -1.06 & -2.69 & 0.61 \\
    France   & -0.08 & -0.09 & -0.28 \\
    United Kingdom   & -0.08 & 0.09  & 0.58 \\
    Greece   & -1.12 & 3.76  & -3.58 \\
    Hungary   & -0.03 & -0.17 & -2.38 \\
    Ireland   & -1.55 & -3.53 & 0.18 \\
    Iceland   & 0.79  & -0.86 & 1.42 \\
    Italy   & -0.65 & -1.29 & -0.85 \\
    Japan   & 0.04  & -0.34 & 0.06 \\
    South Korea   & -2.56 & -2.31 & 1.10 \\
    Luxembourg   & -1.04 & -1.67 & -3.53 \\
    Mexico   & -2.13 & -3.24 & -1.47 \\
    Netherlands   & -0.08 & -0.70 & -0.50 \\
    Norway   & 0.43  & 0.28  & 0.80 \\
    New Zealand   & 0.18  & 0.08  & 0.66 \\
    Poland   & -0.78 & 0.66  & -1.04 \\
    Portugal   & -0.36 & -0.10 & -0.42 \\
    Sweden   & -0.31 & -0.56 & -1.33 \\
    Türkiye   & -10.32 & 2.03  & 0.93 \\
    United States   & -0.50 & -0.92 & 0.09 \\
    \bottomrule
    \end{tabular}%
  \label{tab:lossdiff}%
\end{table}%

\begin{table}[htbp!]
  \centering
  \caption{Panel Tests of EPA for the Economic Growth Forecasts (OECD \textit{vs.} IMF)}  
\begin{threeparttable}
    \begin{tabular}{ccccccc}
    \toprule
          & \multicolumn{2}{c}{Overall EPA Tests} &       &       & \multicolumn{2}{c}{Clustered EPA Tests} \\
    \midrule
    Test  & $d_T = 0$ & $d_T =  T^{1/4}$ &       & Test  & $d_T = 0$ & $d_T =  T^{1/4}$ \\
\cmidrule{1-3}\cmidrule{5-7}    $S^{(1)}_{n,T}$    & 0.06  & 0.06  &       & $C^{(1)}_{n,T}$    & 0.41  & 0.50 \\
          & (0.95) & (0.95) &       &       & (0.82) & (0.78) \\
    $S^{(3)}_{n,T}$    & 0.04  & 0.04  &       & $C^{(3)}_{n,T}$    & 0.42  & 0.59 \\
          & (0.97) & (0.97) &       &       & (0.81) & (0.74) \\
    $\underline{S}^{(3)}_{n,T}$  & 0.04  & 0.04  &       & $\underline{C}^{(3)}_{n,T}$  & 0.42  & 0.41 \\
          & (0.97) & (0.97) &       &       & (0.81) & (0.81) \\
    \bottomrule
    \end{tabular}%
    \begin{tablenotes}
    	\item Note: The values shown in parentheses are $p$-values.
    \end{tablenotes}
  \end{threeparttable}
  \label{tab:paneltests}%
\end{table}

\begin{figure}[htbp]
    \centering
    \begin{subfigure}[b]{0.45\textwidth} 
        \centering \includegraphics[width=\textwidth]{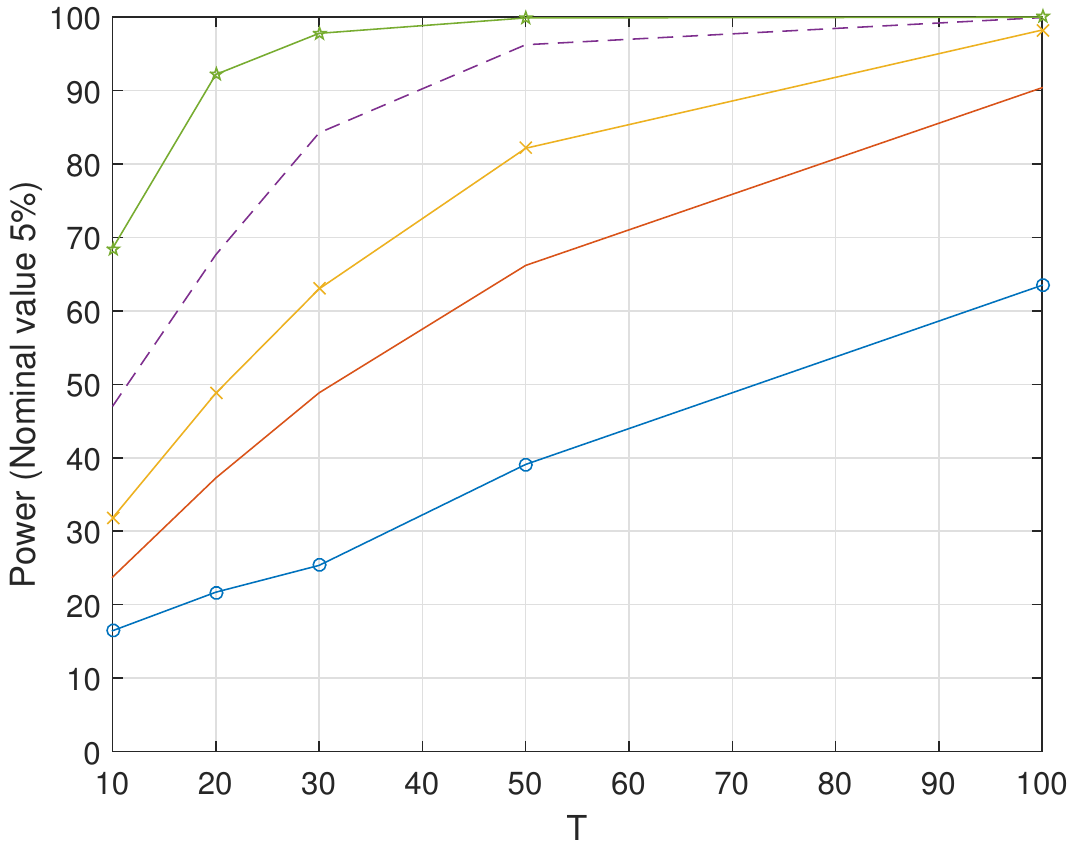}
        \caption{$S^{(3)}_{n,T}$, Homogeneous Alternative}\label{fig:sizeadjpow11}
    \end{subfigure}
    ~ 
    \begin{subfigure}[b]{0.45\textwidth}
        \centering \includegraphics[width=\textwidth]{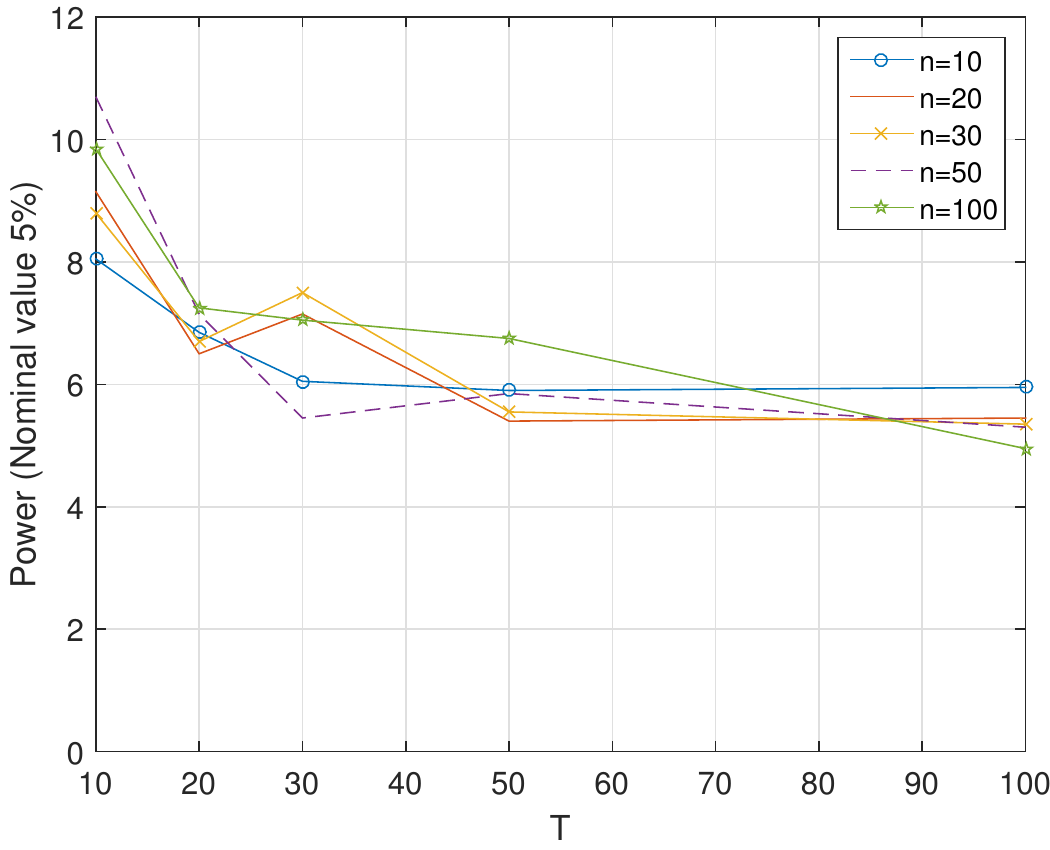}
        \caption{$S^{(3)}_{n,T}$, Heterogeneous Alternative}\label{fig:sizeadjpow12}
    \end{subfigure}
    
    	~ 
	
    \begin{subfigure}[b]{0.45\textwidth}
        \centering \includegraphics[width=\textwidth]{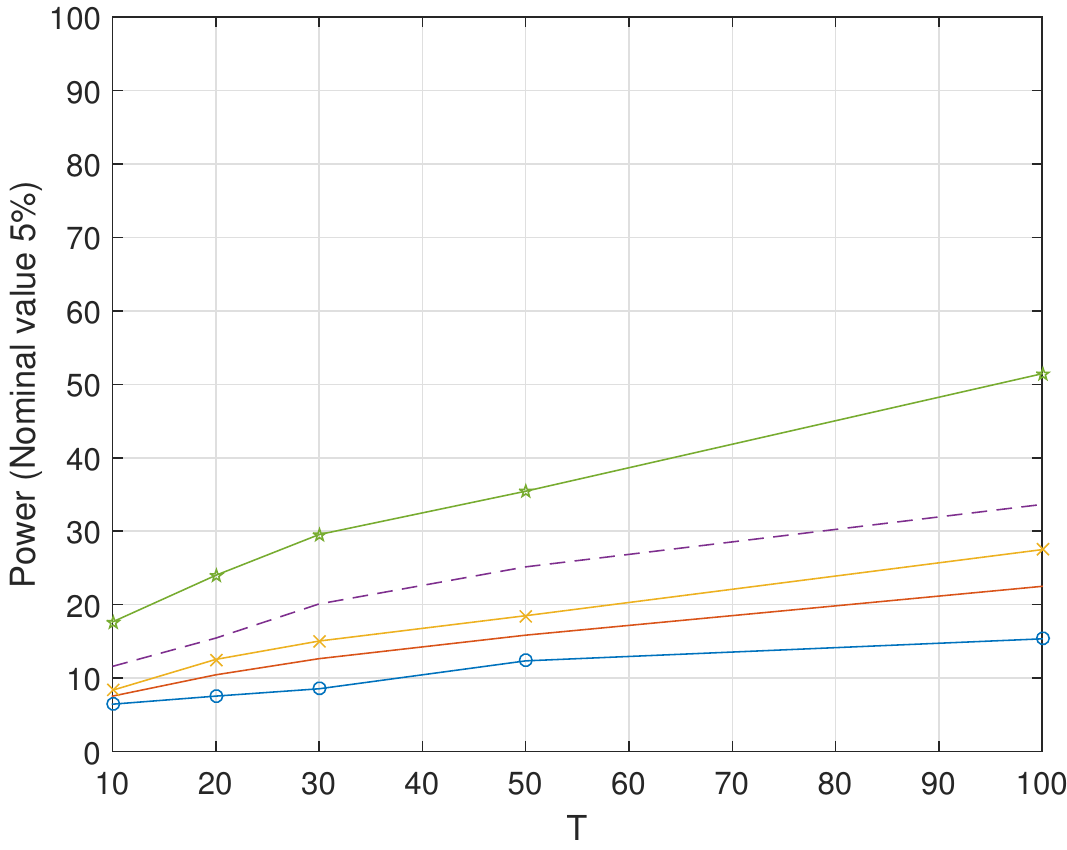}
        \caption{$J_n^D$, Homogeneous Alternative}\label{fig:sizeadjpow13}
    \end{subfigure}
    ~ 
    \begin{subfigure}[b]{0.45\textwidth}
        \centering \includegraphics[width=\textwidth]{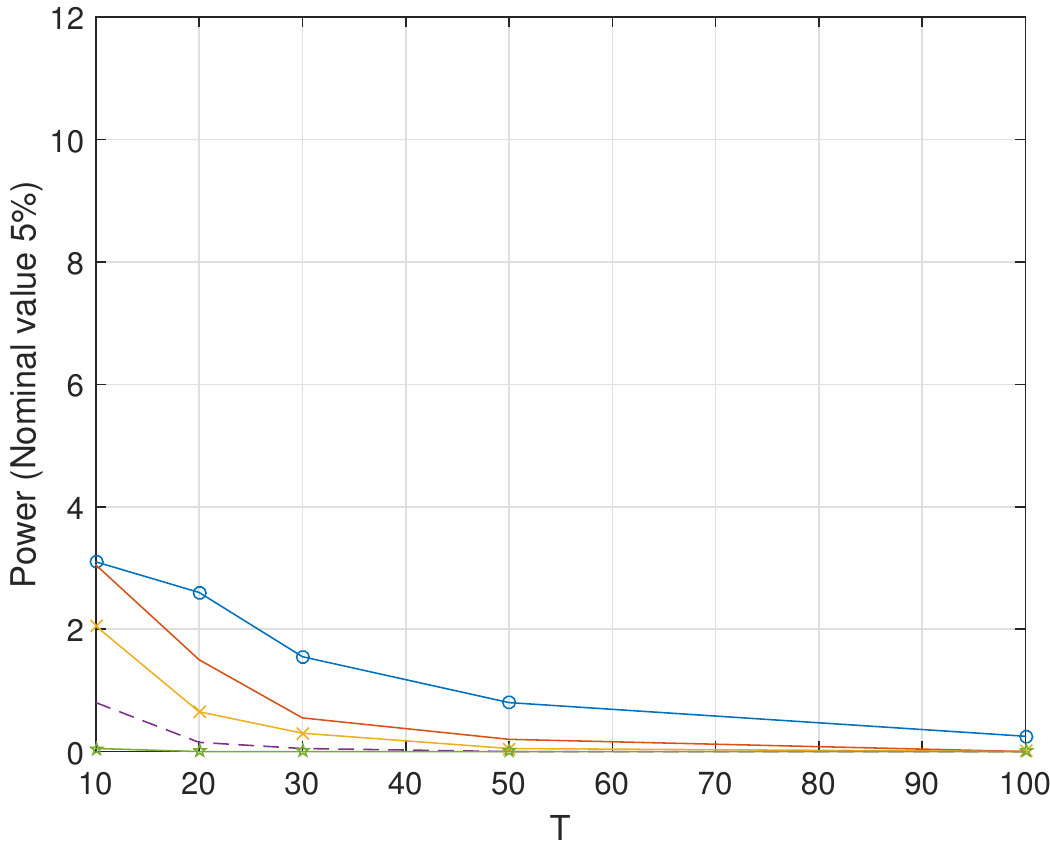}
        \caption{$J_n^D$, Heterogeneous Alternative}\label{fig:sizeadjpow14}
    \end{subfigure}
    
	~ 
	
    \begin{subfigure}[b]{0.45\textwidth}
        \centering \includegraphics[width=\textwidth]{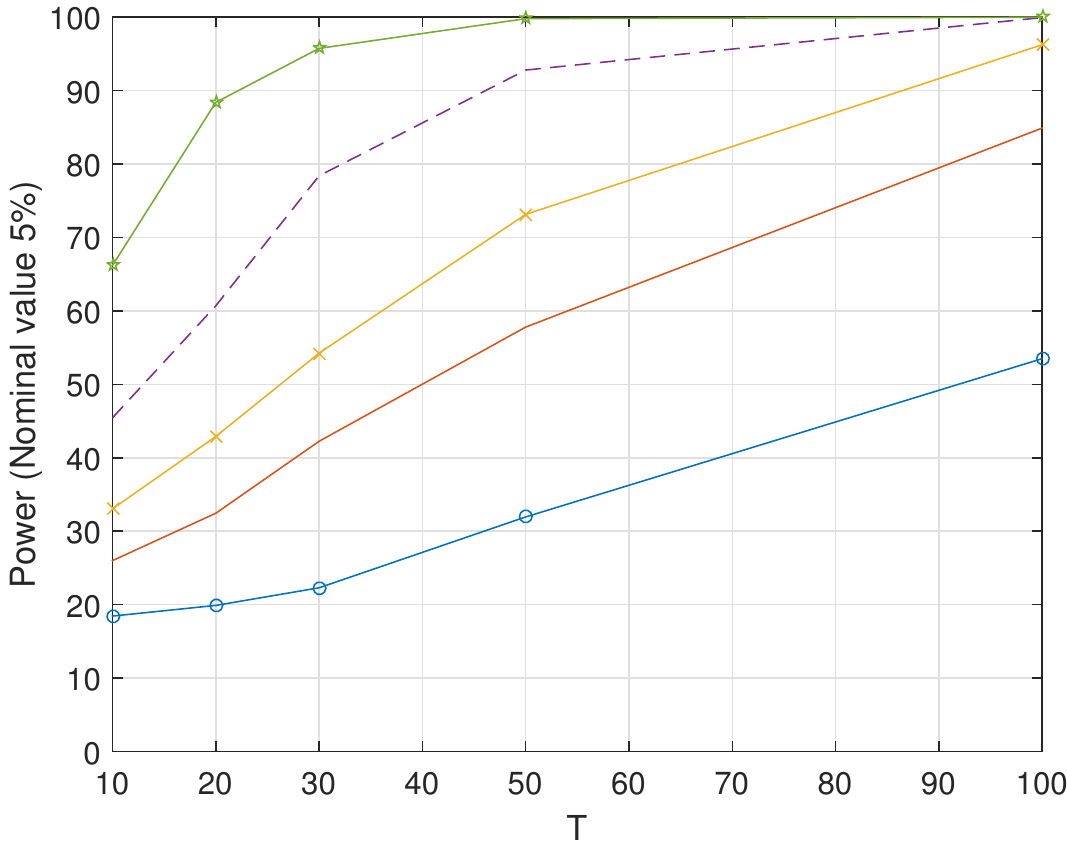}
        \caption{$C^{(3)}_{n,T}$, Homogeneous Alternative}\label{fig:sizeadjpow15}
    \end{subfigure}
    ~ 
    \begin{subfigure}[b]{0.45\textwidth}
        \centering \includegraphics[width=\textwidth]{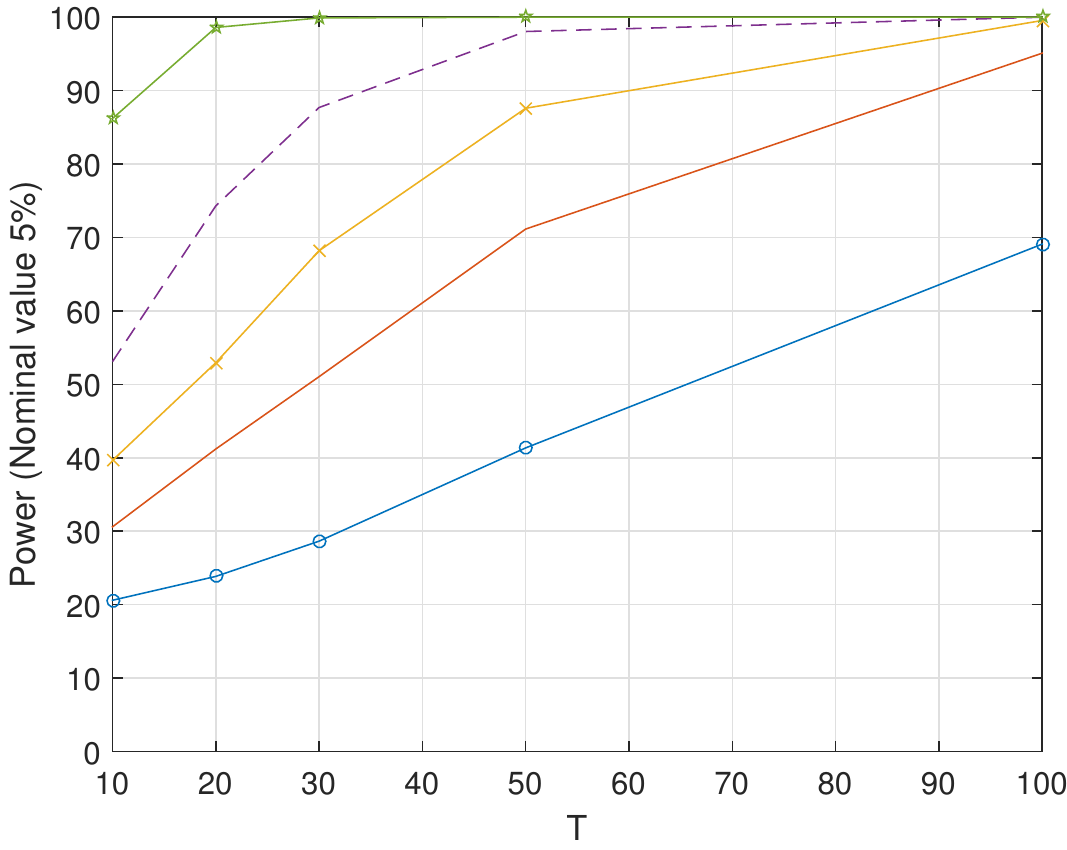}
        \caption{$C^{(3)}_{n,T}$, Heterogeneous Alternative}\label{fig:sizeadjpow16}
    \end{subfigure}
        \caption{Power of Selected Tests Under Different Alternative Hypotheses for {DGP1} (5\% Nominal Size)}\label{fig:sizeadjpow}
\end{figure}

\begin{figure}[htbp]
    \centering
    \begin{subfigure}[b]{0.45\textwidth}
        \centering \includegraphics[width=\textwidth]{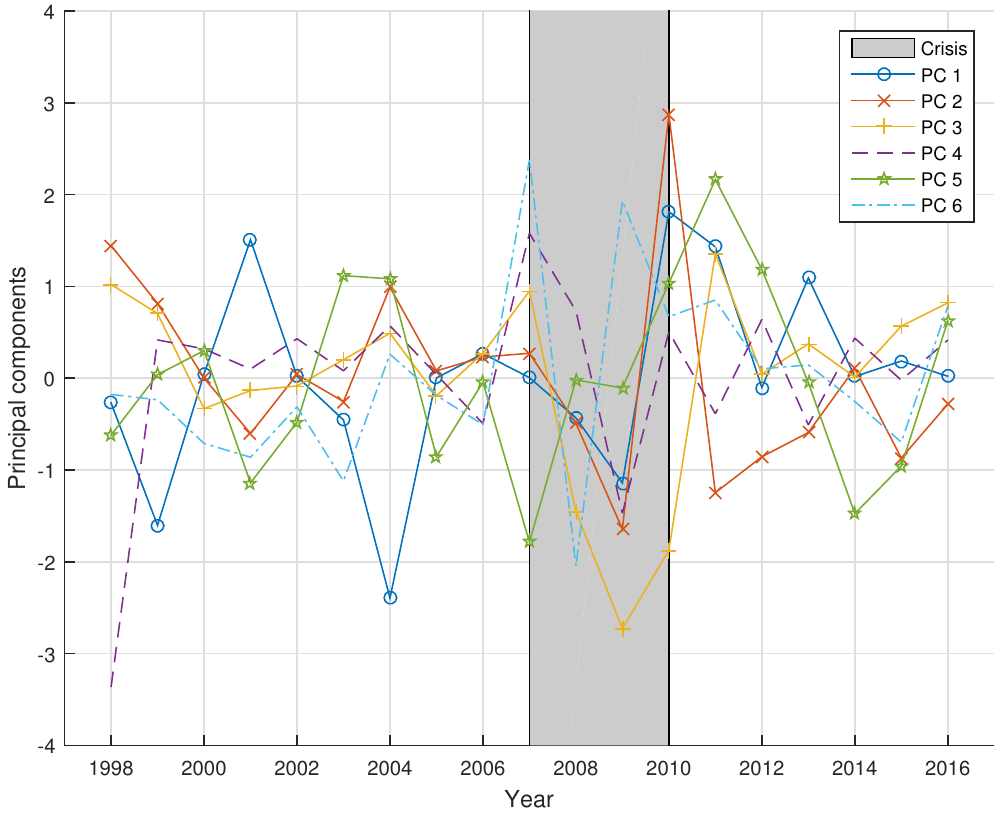}
        \caption{First 6 PCs in Quadratic Loss Differentials}\label{fig:Fig_23}
    \end{subfigure}
    ~ 
    \begin{subfigure}[b]{0.45\textwidth}
        \centering \includegraphics[width=\textwidth]{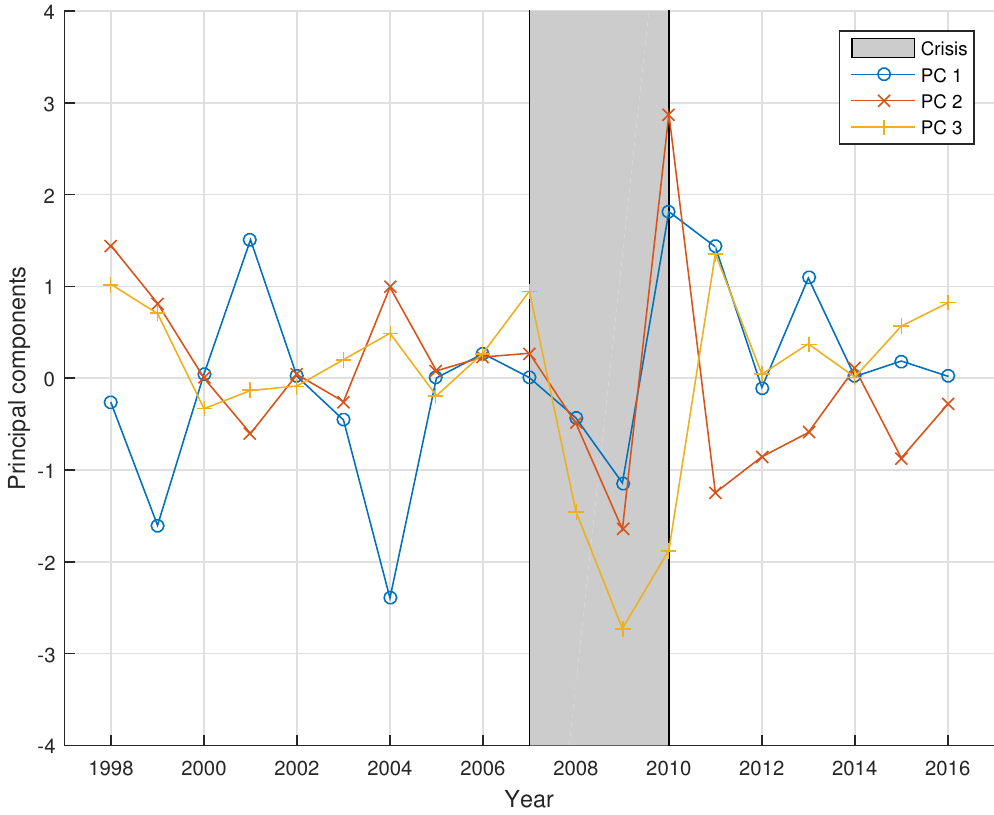}
        \caption{Focus on First 3 PCs}\label{fig:Fig_24}
    \end{subfigure}
    \caption{PC Estimates of the Common Factors in the Loss Differentials of the Economic Growth Forecasts Errors (OECD \textit{vs.} IMF)}\label{fig:lossdiff}
\end{figure}

\clearpage

\begin{appendices}

\noindent {\bf {\Large Appendices}}

\section{Loss Differentials Specification: A Justification}\label{appendixloss}

In this appendix, we present the derivation of the model for the loss differentials starting from a pure common factor model for the forecast errors assuming two most commonly used loss functions, namely absolute loss and quadratic loss.
The forecast error $e_{l,it}$ of the forecaster $l=1,2$, is assumed to be given by
\begin{equation*}
e_{l,it} = c_{li} + \boldsymbol{\theta}'_{li} \mathbf{g}_{lt} + u_{l,it}, \quad i=1,2,\dots,n, \quad t=1,2,\dots,T,
\end{equation*}
where $\mathbf{g}_{lt} = (g_{l1,t},\dots,g_{lm_l,t})'$ is an $m_l \times 1$ vector of common factors, $\boldsymbol{\theta}_{li} = (\theta_{li,1},\dots,\theta_{li,m_l})'$ is their respective factor loadings vector, and $u_{l,it}$ is an error term which can in general be serially and cross-sectionally weakly correlated.
Since our objective in this section is to demonstrate the validity of the common factor model for the loss differentials, we will not focus on the weak-dependence in the errors.
We assume that $\mathrm{E}(u_{l,it} | \mathbf{g}_t) = 0$ and $\mathrm{E}(u^2_{l,it} | \mathbf{g}_t) = \sigma_l^2$ for $l=1,2$ where $\mathbf{g}_t = (\mathbf{g}'_{1t},\mathbf{g}'_{2t})'$.
{A similar factor model for forecast errors has been used by QTZ but they impose $c_{li} = 0$ for all $i=1,\dots,n$ and $l=1,2$.}

The absolute loss differential is given by $d_{1,it} = |e_{1,it}| - |e_{2,it}|$. Then, we have $\mathrm{E}(d_{1,it}|\mathbf{g}_t) = \mu_i + \boldsymbol{\lambda}'_{i} \mathbf{f}_{t}$ where $\boldsymbol{\lambda}_{i} = (\boldsymbol{\theta}'_{1i},\boldsymbol{\theta}'_{2i})'$, $\mu_i = \mathrm{E}[\mathrm{sign}(e_{1,it})]c_{1i} - \mathrm{E}[\mathrm{sign}(e_{2,it})]c_{2i}$ and
\begin{equation*}
    \mathbf{f}'_{t} = \left\{\begin{array}{lr}
        (\mathbf{g}'_{1t},-\mathbf{g}'_{2t})', & \text{if } e_{1,it} \geq 0 \text{ and } e_{2,it} \geq 0, \\
        (\mathbf{g}'_{1t},\mathbf{g}'_{2t})', & \text{if } e_{1,it} \geq 0 \text{ and } e_{2,it} < 0, \\
		(-\mathbf{g}'_{1t},-\mathbf{g}'_{2t})', & \text{if } e_{1,it} < 0 \text{ and } e_{2,it} \geq 0, \\
        (-\mathbf{g}'_{1t},\mathbf{g}'_{2t})', & \text{if } e_{1,it} < 0 \text{ and } e_{2,it} < 0.
        \end{array} \right.\
\end{equation*}
Hence, the model for the loss differentials is obtained with $m = m_1 + m_2$ in the case of absolute loss function. The quadratic loss differential is defined as $d_{2,it} = e^2_{1,it} - e^2_{2,it}$. To simplify the notation, let us assume that $c_{li} = 0$ for each $l=1,2$. The squared forecast error then satisfies
\begin{equation*}
\begin{split}
e^2_{l,it} =& \boldsymbol{\theta}'_{li} \mathbf{g}_{lt} \mathbf{g}'_{lt}\boldsymbol{\theta}_{li} + 2 \boldsymbol{\theta}'_{li} \mathbf{g}_{lt} u_{l,it} + u^2_{l,it} \\
=& \sum_{k_1 = 1}^{m_l} \sum_{k_2 = 1}^{m_l} \theta_{li,k_1} \theta_{li,k_2} g_{lk_1,t}g_{lk_2,t} + 2 \boldsymbol{\theta}'_{li} \mathbf{g}_{lt} u_{l,it} + u^2_{l,it} \\
=& \sum_{k = 1}^{m^2_l} \gamma_{li,k} h_{lk,t} + 2 \boldsymbol{\theta}'_{li} \mathbf{g}_{lt} u_{l,it} + u^2_{l,it} \\
=& \boldsymbol{\gamma}'_{li} \boldsymbol{\iota}_{lt} + 2 \boldsymbol{\theta}'_{li} \mathbf{g}_{lt} u_{l,it} + u^2_{l,it},
\end{split}
\end{equation*}
with straightforward definitions of $\gamma_{li,k}$ and $h_{lk,t}$ which are the $k$th elements of $m^2_l \times 1$ vectors $\boldsymbol{\gamma}_{li}$ and $\boldsymbol{\iota}_{lt}$, respectively. Then, we obtain the model for the loss differentials as $\mathrm{E}(d_{2,it}|\mathbf{g}_t) = \mu_i + \boldsymbol{\lambda}'_{i} \mathbf{f}_{t}$ with $\mu_i = \sigma_1^2 - \sigma_2^2$,  $\boldsymbol{\lambda}_{i} = (\boldsymbol{\gamma}'_{1i},-\boldsymbol{\gamma}'_{2i})'$, $\mathbf{f}_{t} = (\boldsymbol{\iota}'_{1t},\boldsymbol{\iota}'_{2t})'$ and $m = m^2 _1 + m^2_2$.

\section{Proofs}

In this appendix, we present the proofs of Theorems \ref{res5} and \ref{res6}. Propositions \ref{res1}-\ref{res4} are special cases of these theorems.
Let $\mathbf{H}_{n}$ be an $n \times G$ matrix which has $\boldsymbol{\iota}_{g_i}'$, the $g_i$th column of $\mathbf{I}_{G}$, in its $i$th row with $g_i \in \{1,2,\dots,G\}$ being a variable which states the cluster which $i$th unit belongs to.
Define also $\boldsymbol{\epsilon}_{n,t} = (\epsilon_{1t},\epsilon_{2t},\dots,\epsilon_{nt})'$, $\mathbf{c}_{n} = \mathrm{diag}(c_{1},c_{2},\dots,c_{n})'$, $c_{i} = \sum_{h=0}^{\infty} c_{ih}$, $\boldsymbol{\Lambda}_n = (\boldsymbol{\lambda}_1,\boldsymbol{\lambda}_2,\dots,\boldsymbol{\lambda}_n)'$, $\mathbf{C} = \sum_{h=0}^{\infty} \mathbf{C}_{h}$, $\tilde{\mathbf{f}}_{t} = \mathbf{f}_{t} - \frac{1}{T} \sum_{t=1}^{T} \mathbf{f}_{t}$, $\tilde{\boldsymbol{\epsilon}}_{n,t} = \boldsymbol{\epsilon}_{n,t} - \frac{1}{T} \sum_{t=1}^{T} \boldsymbol{\epsilon}_{n,t}$, $\mathbf{D}_n = \mathrm{diag}(n_1,n_2,\dots,n_G)$ and  ${\gamma}_{ij,d_{ts}} = \mathrm{E}(\epsilon_{it} \epsilon_{js})$.

Define
\[
\begin{split}
\mathbf{V}_{1,nT} &= \frac{1}{n} \sum_{i=1}^{n} \boldsymbol{\iota}_{g_i} \boldsymbol{\iota}_{g_i}^{\prime} \bar{\gamma}_{i,T}, \\
\mathbf{V}_{2,nT} &= \frac{1}{n} \sum_{i,j=1}^{n} \boldsymbol{\iota}_{g_i} \boldsymbol{\iota}_{g_j}^{\prime} \mathbf{r}_{i.}' \bar{{\boldsymbol{\gamma}}}_{nT} \mathbf{r}_{j.}, \\
\mathbf{V}_{3,nT} &= \frac{1}{n^2} \sum_{i,j=1}^{n} \boldsymbol{\iota}_{g_i} \boldsymbol{\iota}_{g_j}^{\prime} \boldsymbol{\lambda}'_i \bar{\boldsymbol{\Gamma}}_{T} \boldsymbol{\lambda}_j + \frac{1}{n^2} \sum_{i,j=1}^{n} \boldsymbol{\iota}_{g_i} \boldsymbol{\iota}_{g_j}^{\prime} \mathbf{r}_{i.}' \bar{{\boldsymbol{\gamma}}}_{nT} \mathbf{r}_{j.}.
\end{split}
\]
To prove Theorems 1 and 2, we need the following lemmas.

\begin{lemma}\label{lemma_1}
Suppose Assumptions \ref{ass_epsilon}, \ref{ass_r1} and \ref{ass_comcomp} hold. Then, as $T,n \rightarrow \infty$,
\begin{enumerate}[label=(\roman*)]
\item\label{lemma_10} $\frac{1}{\sqrt{nT}} \sum_{t=1}^T \mathbf{H}_{n}' \boldsymbol{\epsilon}_{n,t} \overset{d}{\rightarrow} N(\mathbf{0},\mathbf{V}_1)$, where $\mathbf{V}_1 = \lim_{(n,T)\to\infty} \mathbf{V}_{1,nT}$,
\item\label{lemma_11} $\frac{1}{\sqrt{nT}} \sum_{t=1}^T \mathbf{H}_{n}' \mathbf{R}_n \boldsymbol{\epsilon}_{n,t} \overset{d}{\rightarrow} N(\mathbf{0},\mathbf{V}_2)$, where $\mathbf{V}_2 = \lim_{(n,T)\to\infty} \mathbf{V}_{2,nT}$,
\item\label{lemma_12} $\frac{1}{n\sqrt{T}} \sum_{t=1}^T \mathbf{H}_{n}' \boldsymbol{\Lambda}_n \mathbf{f}_{t} \overset{d}{\rightarrow} N(\mathbf{0},\mathbf{V}_3)$, where $\mathbf{V}_3 = \lim_{(n,T)\to\infty} \mathbf{V}_{3,nT}$.
\end{enumerate}
\end{lemma}

\begin{lemma}\label{lemma_2}
Suppose Assumptions \ref{ass_epsilon}, \ref{ass_tkernel} and \ref{ass_comcomp} hold. Then, for any fixed $n$, as $T \rightarrow \infty$
\begin{enumerate}[label=(\roman*)]
\item\label{lemma_22} $T^{-1} \sum_{t,s=1}^{T} k_{T}\left( {d_{ts}/d_{T}}\right)  {\mathbf{f}}_{t} {\mathbf{f}}'_{s} - \bar{\boldsymbol{\Gamma}}_{T} = o_p(1)$,
\item\label{lemma_21} $T^{-1} \sum_{t,s=1}^{T} k_{T}\left( {d_{ts}/d_{T}}\right)  \tilde{\mathbf{f}}_{t} \tilde{\mathbf{f}}'_{s} - \bar{\boldsymbol{\Gamma}}_{T} = o_p(1)$,
\item\label{lemma_24} $T^{-1} \sum_{t,s=1}^{T} k_{T}\left( {d_{ts}/d_{T}}\right)  \tilde{\mathbf{f}}_{t} \tilde{\boldsymbol{\epsilon}}'_{n,s} = o_p(1)$,
\item\label{lemma_28} $T^{-1} \sum_{t,s=1}^{T} k_{T}\left( {d_{ts}/d_{T}}\right)  \tilde{\boldsymbol{\epsilon}}_{n,t} \tilde{\boldsymbol{\epsilon}}'_{n,t} - \bar{\boldsymbol{\gamma}}_{nT} = o_p(1)$.
\end{enumerate}
\end{lemma}


\begin{lemma}\label{lemma_4}
Suppose Assumptions \ref{ass_epsilon}, \ref{ass_r1}, \ref{ass_comcomp} and \ref{ass_commoncomprate} hold. Then,
\begin{enumerate}[label=(\roman*)]
\item\label{lemma_43}$\frac{1}{n^2 T} \sum_{i =1}^n \sum_{t,s=1}^{T} k_{T}\left( {d_{ts}/d_{T}}\right) \widehat{\varepsilon}_{it} \widehat{\varepsilon}_{is} = o_p \left( 1 \right)$,
\item\label{lemma_44}$\frac{1}{n^2} \sum_{i,j =1}^n \left[ \frac{1}{T} \sum_{t,s=1}^{T} k_{T}\left( {d_{ts}/d_{T}}\right) \left( \widehat{\boldsymbol{\lambda}}'_i \widehat{\mathbf{f}}_{t} \widehat{\boldsymbol{\lambda}}'_j \widehat{\mathbf{f}}_{s} - \boldsymbol{\lambda}'_i  {\mathbf{f}}_{t} {\mathbf{f}}'_{s} \boldsymbol{\lambda}_j   \right) \right] = o_p \left( 1 \right)$.
\end{enumerate}
\end{lemma}

\noindent \textbf{Proof of Lemma \ref{lemma_1}.} (i) is a special case of (ii) with $\mathbf{R}_n = \mathbf{I}_n$, hence we focus on the two last results. We start by applying the Beveridge-Nelson (BN) decomposition \citep[see, for instance,][]{phillips92} to each component of $\boldsymbol{\epsilon}_{n,t}$. We have $\epsilon_{it} = c_{i}\psi_{it} + \tilde{\psi}_{i,t-1} + \tilde{\psi}_{it}
$, where $\tilde{\psi}_{it} = \sum_{h=0}^{\infty} \tilde{c}_{ih} \psi_{i,t-h}$, $\tilde{c}_{ih} = \sum_{j=h+1}^{\infty} {c}_{ij}$. Then, we can write
\begin{equation*}
\begin{split}
\frac{1}{\sqrt{nT}} \sum_{t=1}^T \mathbf{H}_{n}' \mathbf{R}_n \boldsymbol{\epsilon}_{n,t} =& \frac{1}{\sqrt{nT}} \sum_{t=1}^T \mathbf{H}_{n}' \mathbf{R}_n \mathbf{c}_{n} \boldsymbol{\psi}_{n,t} + \frac{1}{\sqrt{nT}} \mathbf{H}_{n}' \mathbf{R}_n \tilde{\boldsymbol{\psi}}_{n,1} - \frac{1}{\sqrt{nT}} \mathbf{H}_{n}' \mathbf{R}_n \tilde{\boldsymbol{\psi}}_{n,T} \\
=& A_1 + A_2 + A_3,
\end{split}
\end{equation*}
where $\boldsymbol{\psi}_{n,t} = (\psi_{1t},\psi_{2t},\dots,\psi_{nt})'$ and $\tilde{\boldsymbol{\psi}}_{n,t} = (\tilde{\psi}_{1t},\tilde{\psi}_{2t},\dots,\tilde{\psi}_{nt})'$. The variance of the first term is $\mathrm{Var} \left( A_1 \right) = n^{-1} \sum_{i,j=1}^{n} \boldsymbol{\iota}_{g_i} \mathbf{r}_{i.}' \mathbf{c}_{n} \mathbf{c}'_{n} \mathbf{r}_{j.} \boldsymbol{\iota}'_{g_j}$. A typical element of this matrix, say $lw$th, satisfies
\begin{equation*}
\begin{split}
\left|\frac{1}{n} \sum_{i \in G_l} \sum_{j \in G_w} \mathbf{r}_{i.}' \mathbf{c}_{i} \mathbf{c}'_{j} \mathbf{r}_{j.} \right| =& \left|\frac{1}{n} \sum_{i \in G_l} \sum_{j \in G_w} \sum_{k=1}^{n} r_{ik} r_{jk} \sum_{h=0}^{\infty}  \sum_{h'=0}^{\infty} c_{kh} c_{kh'} \right| \\
\leq& \frac{1}{n} \sum_{i \in G_l} \sum_{j \in G_w} \sum_{k=1}^{n} |r_{ik}| |r_{jk}| \sum_{h=0}^{\infty}  \sum_{h'=0}^{\infty} |c_{kh}| |c_{kh'}| \\
\leq& \frac{1}{n} \sum_{k=1}^{n} \sum_{i \in G_l} |r_{ik}| \sum_{j \in G_w}  |r_{jk}| \sum_{h=0}^{\infty} |c_{kh}| \sum_{h'=0}^{\infty}  |c_{kh'}| = O(1),
\end{split}
\end{equation*}
where we used the fact that absolute summability is implied by the summability condition we impose on coefficients $c_{ih}$ in Assumption \ref{ass_epsilon}. Hence we showed that $A_1$ is $O_p(1)$. The variances of the second and the third terms are $\mathrm{Var} \left( A_2 \right) = (nT)^{-1} \sum_{i,j=1}^{n} \boldsymbol{\iota}_{g_i} \mathbf{r}_{i.}' \mathrm{E} ( \tilde{\boldsymbol{\psi}}_{n,1} \tilde{\boldsymbol{\psi}}_{n,1}') \mathbf{r}_{j.} \boldsymbol{\iota}'_{g_j}$ and $\mathrm{Var} \left( A_3 \right) = (nT)^{-1} \sum_{i,j=1}^{n} \boldsymbol{\iota}_{g_i} \mathbf{r}_{i.}' \mathrm{E} (\tilde{\boldsymbol{\psi}}_{n,T} \tilde{\boldsymbol{\psi}}_{n,T}') \mathbf{r}_{j.} \boldsymbol{\iota}'_{g_j}$.
respectively. For \ref{lemma_11} to hold, we need to show that these are asymptotically negligible. This holds if elements of $\mathrm{E} ( \tilde{\boldsymbol{\psi}}_{n,1} \tilde{\boldsymbol{\psi}}_{n,1}')$ and $\mathrm{E} ( \tilde{\boldsymbol{\psi}}_{n,T} \tilde{\boldsymbol{\psi}}_{n,T}')$ are finite which holds under Assumption \ref{ass_epsilon} as shown by the BN Lemma in \cite{phillips92}. Then the $lw$th elements of $\mathrm{Var} \left( A_2 \right)$ and $\mathrm{Var} \left( A_3 \right)$ satisfy $(nT)^{-1} \left|\sum_{i \in G_l} \sum_{j \in G_w} \mathbf{r}_{i.}' \mathbf{r}_{j.} O(1)\right|$ which is $O(T^{-1})$ under Assumption \ref{ass_r1}. It follows that the second and the third terms are dominated by the first one. Since the variance of the first term is $O(1)$, it satisfies a central limit theorem for triangular arrays \citep[][]{kelejian98} and \ref{lemma_11} follows.

To prove \ref{lemma_12}, we apply the multivariate generalization of the BN decomposition \citep[see, for instance,][p. 985]{phillips92} to $\mathbf{f}_{t}$. We have
\begin{equation*}
\begin{split}
\frac{1}{n\sqrt{T}} \sum_{t=1}^T \mathbf{H}_{n}' \boldsymbol{\Lambda}_n \mathbf{f}_{t} =& \frac{1}{n\sqrt{T}} \sum_{t=1}^T \mathbf{H}_{n}' \boldsymbol{\Lambda}_n \mathbf{C} \boldsymbol{\Psi}_{t} + \frac{1}{n\sqrt{T}} \mathbf{H}_{n}' \boldsymbol{\Lambda}_n \tilde{\boldsymbol{\Psi}}_{1} - \frac{1}{n\sqrt{T}} \mathbf{H}_{n}' \boldsymbol{\Lambda}_n \tilde{\boldsymbol{\Psi}}_{T} \\
=& B_1 + B_2 + B_3,
\end{split}
\end{equation*}
where $\tilde{\boldsymbol{\Psi}}_{t} = \sum_{h=0}^{\infty} \tilde{\mathbf{C}}_{h} \boldsymbol{\Psi}_{t-h}$, $\tilde{\mathbf{C}}_{h} = \sum_{j=h+1}^{\infty} {\mathbf{C}}_{j}$. The variance of the first term is given by
$
\mathrm{Var} \left( B_1 \right) = \frac{1}{n^2} \sum_{i,j=1}^{n} \boldsymbol{\iota}_{g_i} \boldsymbol{\lambda}_{i}' \mathbf{C} \mathbf{C}' \boldsymbol{\lambda}_{j} \boldsymbol{\iota}'_{g_j}
$.
The $lw$th element of this matrix satisfies
\begin{equation*}
\left| \frac{1}{n^2} \sum_{i \in G_l} \sum_{j \in G_w} \boldsymbol{\lambda}_{i}' \mathbf{C} \mathbf{C}' \boldsymbol{\lambda}_{j} \right| \leq \frac{1}{n^2} \sum_{i \in G_l} \sum_{j \in G_w} \left| \boldsymbol{\lambda}_{i}' \mathbf{C} \mathbf{C}' \boldsymbol{\lambda}_{j} \right| \leq \frac{1}{n^2} \sum_{i \in G_l} \sum_{j \in G_w} || \mathbf{C} ||^2 || \boldsymbol{\lambda}_{i} || || \boldsymbol{\lambda}_{j} || = O(1),
\end{equation*}
where we used the fact that $|| \mathbf{C} ||^2 \leq \sum_{h=0}^{\infty} || \mathbf{C}_{h} ||^2 < \infty$. It follows that $B_1$ is $O_p(1)$. The variances of the second and the third terms are
$
\mathrm{Var} \left( B_2 \right) = \frac{1}{n^2T} \mathbf{H}_{n}' \boldsymbol{\Lambda}_n \mathrm{E}(\tilde{\boldsymbol{\Psi}}_{1} \tilde{\boldsymbol{\Psi}}'_{1}) \boldsymbol{\Lambda}'_n\mathbf{H}_{n}
$ and
$
\mathrm{Var} \left( B_3 \right) = \frac{1}{n^2T} \mathbf{H}_{n}' \boldsymbol{\Lambda}_n \mathrm{E}(\tilde{\boldsymbol{\Psi}}_{T} \tilde{\boldsymbol{\Psi}}'_{T}) \boldsymbol{\Lambda}'_n\mathbf{H}_{n}
$,
respectively. Similar to the reasoning above, we have $\mathrm{E}(\tilde{\boldsymbol{\Psi}}_{1} \tilde{\boldsymbol{\Psi}}'_{1}) < \infty$ and $\mathrm{E}(\tilde{\boldsymbol{\Psi}}_{T} \tilde{\boldsymbol{\Psi}}'_{T}) < \infty$, under the summability condition we impose on the coefficients of the process $\mathbf{f}_T$ \citep[see, for instance][]{phillips09}. Furthermore, the $lw$th element of the matrix $\frac{1}{n^2} \mathbf{H}_{n}' \boldsymbol{\Lambda}_n \boldsymbol{\Lambda}'_n\mathbf{H}_{n}$ satisfies $\left|n^{-2} \sum_{i \in G_l} \sum_{j \in G_w}  \sum_{k,k'=1}^{m} \lambda_{ik} \lambda_{jk'} \right| = O(1)$. Hence, $B_2$ and $B_3$ are $o_p(1)$. Since the the first term is $O_p(1)$, it dominates the second and the third terms and a central limit theorem for triangular arrays \citep[][]{kelejian98} applies to the first term. Now, since the second term in $\mathbf{V}_{3,nT}$ is $O(n^{-1})$ from \ref{lemma_11}, the asymptotic variance of $\mathbf{V}_{3,nT}$ equals the asymptotic variance of $B_1$ which leads to \ref{lemma_12}.
\medskip

\noindent \textbf{Proof of Lemma \ref{lemma_2}.} \ref{lemma_22} follows from the proof of Theorem 2 of \cite{jansson02} under Assumptions \ref{ass_epsilon} and \ref{ass_tkernel}. \ref{lemma_21} follows immediately from their proof noting that $T^{-1} \sum_{t=1}^{T} \mathbf{f}_{t} = o_p(1)$ under Assumption \ref{ass_comcomp}\ref{ass_comcomp_factors}. \ref{lemma_24} holds again by their proof, as by Assumption \ref{ass_comcomp}\ref{ass_comcomp_factors} ${\mathbf{f}}_{t}$ and ${\boldsymbol{\epsilon}}_{n,s}$ are independent for every $t,s$, and noting that $T^{-1} \sum_{t=1}^{T} {\boldsymbol{\epsilon}}_{n,t} = o_p(1)$. Similarly, \ref{lemma_28} is a result of their proof under Assumption \ref{ass_epsilon}.
\medskip

\noindent \textbf{Proof of Lemma \ref{lemma_4}.}
We have $\widehat{\varepsilon}_{it} = \Delta L_{it} - \Delta \bar{L}_{i,T} - \widehat{\boldsymbol{\lambda}}'_i \widehat{\mathbf{f}}_{t} = ({\mu}_i - \Delta \bar{L}_{i,T}) + (\boldsymbol{\lambda}'_i  {\mathbf{f}}_{t} - \widehat{\boldsymbol{\lambda}}'_i \widehat{\mathbf{f}}_{t}) + {\varepsilon}_{it}$.
By Assumptions \ref{ass_epsilon} and \ref{ass_comcomp}, ${\mu}_i - \Delta \bar{L}_{i,T} = O_p({T}^{-1/2})$, ${\varepsilon}_{it} = O_p (1)$ and by Assumption \ref{ass_commoncomprate} $\boldsymbol{\lambda}'_i  {\mathbf{f}}_{t} - \widehat{\boldsymbol{\lambda}}'_i \widehat{\mathbf{f}}_{t} = O_p(\delta_{nT}^{-1})$.
After trivial algebra, this gives $\widehat{\varepsilon}_{it} \widehat{\varepsilon}_{is} = {\varepsilon}_{it} {\varepsilon}_{is} + O_p(\delta_{nT}^{-1})$.
Then $({n^2 T})^{-1} \sum_{i =1}^n \sum_{t,s=1}^{T} k_{T}\left( {d_{ts}/d_{T}}\right) \widehat{\varepsilon}_{it} \widehat{\varepsilon}_{is} = ({n^2 T})^{-1} \sum_{i =1}^n \sum_{t,s=1}^{T} k_{T}\left( {d_{ts}/d_{T}}\right) O_p(\delta_{nT}^{-1}) + ({n^2 T})^{-1} \sum_{i =1}^n \sum_{t,s=1}^{T} k_{T}\left( {d_{ts}/d_{T}}\right) {\varepsilon}_{it} {\varepsilon}_{is} = A_1 + A_2$, say.
We have $A_1 = {n^{-2}} \sum_{i =1}^n O_p(\delta_{nT}^{-1}) \left[ T^{-1} \sum_{h=-T+1}^{T-1} k_{T} \left({h/d_T}\right) \right]$ where the term in brackets is $O(d_T/T)$ by Assumption \ref{ass_tkernel}, which in turn gives $A_1 = O(n^{-1})O_p(\delta_{nT}^{-1})O(d_T/T)=o_p(1)$.
For the second term, we find $\mathrm{E}(A_2) = \frac{1}{n^2 T} \sum_{i =1}^n \sum_{t,s=1}^{T} k_{T}\left( {d_{ts}/d_{T}}\right) \mathbf{r}_{i.}' \boldsymbol{\gamma}_{n,d_{ts}} \mathbf{r}_{i.} = n^{-2} \sum_{i,k=1}^n r_{ik}^2 \left[ T^{-1} \sum_{h=-T+1}^{T-1} k_{T} \left({h/d_T}\right) {\gamma}_{k,h} \right]$.
By Assumption \ref{ass_epsilon} the spectral density function of $\epsilon_{it}$, $f_i(\cdot)$, exists and is bounded.
By Theorem 2 of \cite{jansson02}, under Assumption \ref{ass_tkernel}, $\lim_{T \to \infty} T^{-1} \sum_{h=-T+1}^{T-1} k_{T} \left({h/d_T}\right) {\gamma}_{i,h} = T^{-1} 2 \pi f_i(0) = O(T^{-1})$ for each $i$.
Then we have $\mathrm{E}(A_2)=O(T^{-1})n^{-2} \sum_{i,k=1}^n r_{ik}^2 = O[(nT)^{-1}]$.
For the variance, we find
\begin{equation}\label{varA2}
\begin{aligned}[b]
\mathrm{Var}(A_2) &= \mathrm{E} \left[ \frac{1}{n^4 T^2} \sum_{i,j=1}^n \sum_{t_1,s_1,t_2,s_2=1}^{T}  k_{T} \left( \frac{d_{t_1 s_1}}{d_T}  \right) k_{T} \left( \frac{d_{t_2 s_2}}{d_T}  \right) {\varepsilon}_{it_1} {\varepsilon}_{is_1} {\varepsilon}_{jt_2} {\varepsilon}_{js_2} \right] \\
&= \frac{1}{n^4 T^2} \sum_{i,j=1}^n \sum_{t_1,s_1,t_2,s_2=1}^{T}  k_{T} \left( \frac{d_{t_1 s_1}}{d_T}  \right) k_{T} \left( \frac{d_{t_2 s_2}}{d_T}  \right) \mathrm{E} ( {\varepsilon}_{it_1} {\varepsilon}_{is_1} {\varepsilon}_{jt_2} {\varepsilon}_{js_2} ) \\
&= \frac{1}{n^4 T^2} \sum_{i,j=1}^n \sum_{t_1,s_1,t_2,s_2=1}^{T}  k_{T} \left( \frac{d_{t_1 s_1}}{d_T}  \right) k_{T} \left( \frac{d_{t_2 s_2}}{d_T}  \right) \mathrm{E}( \mathbf{r}_{i.}' {\boldsymbol{\epsilon}}_{n,t_1} {\boldsymbol{\epsilon}}'_{n,s_1} \mathbf{r}_{i.} \mathbf{r}_{j.}' {\boldsymbol{\epsilon}}_{n,t_2} {\boldsymbol{\epsilon}}'_{n,s_2} \mathbf{r}_{j.} ) \\
&= \frac{1}{n^4 T^2} \sum_{i,j=1}^n \sum_{t_1,s_1,t_2,s_2=1}^{T}  k_{T} \left( \frac{d_{t_1 s_1}}{d_T}  \right) k_{T} \left( \frac{d_{t_2 s_2}}{d_T}  \right) \mathrm{E} \left( \sum_{l_1,l_2=1}^n r_{il_1} r_{il_2} \epsilon_{l_1,t_1} \epsilon_{l_2,s_1} \sum_{l_3,l_4=1}^n r_{jl_3} r_{jl_4} \epsilon_{l_3,t_2} \epsilon_{l_4,s_2} \right) \\
&= \frac{1}{n^4} \sum_{i,j,l_1,l_2,l_3,l_4=1}^n r_{il_1} r_{il_2} r_{jl_3} r_{jl_4} \left[ \frac{1}{T^2} \sum_{t_1,s_1,t_2,s_2=1}^{T} k_{T} \left( \frac{d_{t_1 s_1}}{d_T}  \right) k_{T} \left( \frac{d_{t_2 s_2}}{d_T}  \right) \mathrm{E}(\epsilon_{l_1,t_1} \epsilon_{l_2,s_1}\epsilon_{l_3,t_2} \epsilon_{l_4,s_2}) \right].
\end{aligned}
\end{equation}
In general, the expectation in the last line can be written as
\begin{equation*}
\begin{split}
\mathrm{E}(\epsilon_{l_1,t_1} \epsilon_{l_2,s_1}\epsilon_{l_3,t_2} \epsilon_{l_4,s_2}) =& {\gamma}_{l_1l_2,d_{t_1s_1}}{\gamma}_{l_3l_4,d_{t_2s_2}} + {\gamma}_{l_1l_3,d_{t_1s_2}}{\gamma}_{l_2l_4,d_{t_1s_2}}  \\
&+ {\gamma}_{l_1l_4,d_{t_1s_2}}{\gamma}_{l_2l_3,d_{t_1s_2}} + \kappa_{l_1l_2l_3l_4}(t_1,s_1,t_2,s_2),
\end{split}
\end{equation*}
where $\kappa(\cdot)$ is the cumulant of the fourth order between $\epsilon_{l_1,t_1}$, $\epsilon_{l_2,s_1}$, $\epsilon_{l_3,t_2}$ and $\epsilon_{l_4,s_2}$ \cite[see][p. 23]{hannan70}. In our case, by Assumption \ref{ass_epsilon} $\epsilon_{it}$ are independent over $i$, hence, the cumulant is null. Furthermore, the covariances in the expression are null unless $l_i = l_j$, $i,j=1,2,3,4$. Using these in \eqref{varA2}, we find
\begin{equation*}
\begin{split}
\mathrm{Var}(A_2) =& \frac{1}{n^4} \sum_{i,j=1}^n \sum_{l_1=1}^n r^2_{il_1} \left[ \frac{1}{T} \sum_{h_1=-T+1}^{T-1} k_{T} \left( \frac{h_1}{d_T} \right) {\gamma}_{l_1,h_1} \right] \sum_{l_2=1}^n r^2_{jl_2}  \left[ \frac{1}{T} \sum_{h_2=-T+1}^{T-1} k_{T} \left( \frac{h_2}{d_T} \right) {\gamma}_{l_2,h_2} \right] \\
&+ \frac{1}{n^4} \sum_{i,j=1}^n \sum_{l_1=1}^n r^2_{il_1} \left[ \frac{1}{T} \sum_{h_1=-T+1}^{T-1} k_{T} \left( \frac{h_1}{d_T} \right) {\gamma}_{l_1,h_1} \right] \sum_{l_3=1}^n r^2_{jl_3}  \left[ \frac{1}{T} \sum_{h_2=-T+1}^{T-1} k_{T} \left( \frac{h_2}{d_T} \right) {\gamma}_{l_3,h_2} \right] \\
&+ \frac{1}{n^4} \sum_{i,j=1}^n \sum_{l_1=1}^n r^2_{il_1} \left[ \frac{1}{T} \sum_{h_1=-T+1}^{T-1} k_{T} \left( \frac{h_1}{d_T} \right) {\gamma}_{l_1,h_1} \right] \sum_{l_4=1}^n r^2_{jl_4}  \left[ \frac{1}{T} \sum_{h_2=-T+1}^{T-1} k_{T} \left( \frac{h_2}{d_T} \right) {\gamma}_{l_4,h_2} \right] \\
=& A_{21} + A_{22} + A_{23}.
\end{split}
\end{equation*}
All terms in the brackets are $O(T^{-1})$.
Hence, $A_{21} = O(T^{-2}) n^{-4} \sum_{i,j=1}^n \sum_{l_1=1}^n r^2_{il_1} \sum_{l_2=1}^n r^2_{jl_2} = O[(nT)^{-2}]$. Similarly, $A_{22} = A_{23} = O[(nT)^{-2}]$. As a result, $A_2 = O_p[(nT)^{-1}] = o_p(1)$.

For (ii), we write $\widehat{\boldsymbol{\lambda}}'_i \widehat{\mathbf{f}}_{t} \widehat{\mathbf{f}}'_{s} \widehat{\boldsymbol{\lambda}}_j - \boldsymbol{\lambda}'_i  {\mathbf{f}}_{t} {\mathbf{f}}'_{s} \boldsymbol{\lambda}_j   = (\widehat{\boldsymbol{\lambda}}'_i \widehat{\mathbf{f}}_{t} - \boldsymbol{\lambda}'_i  {\mathbf{f}}_{t}) \widehat{\mathbf{f}}'_{s} \widehat{\boldsymbol{\lambda}}_j  + \boldsymbol{\lambda}'_i  {\mathbf{f}}_{t}( \widehat{\mathbf{f}}'_{s} \widehat{\boldsymbol{\lambda}}_j - {\mathbf{f}}'_{s}  \boldsymbol{\lambda}_j )$.
By Assumption \ref{ass_comcomp} $\boldsymbol{\lambda}'_i  {\mathbf{f}}_{t}$, and by Assumption \ref{ass_commoncomprate} $\widehat{\boldsymbol{\lambda}}'_j \widehat{\mathbf{f}}_{s}$ are $O_p(1)$.
This implies that $\widehat{\boldsymbol{\lambda}}'_i \widehat{\mathbf{f}}_{t} \widehat{\mathbf{f}}'_{s} \widehat{\boldsymbol{\lambda}}_j - \boldsymbol{\lambda}'_i  {\mathbf{f}}_{t} {\mathbf{f}}'_{s} \boldsymbol{\lambda}_j   = O_p(\delta^{-1}_{nT})$ by Assumption \ref{ass_commoncomprate}.
Then for the expression in the statement we obtain $\frac{1}{n^2} \sum_{i,j =1}^n O_p(\delta^{-1}_{nT}) \left[ \frac{1}{T} \sum_{h=-T+1}^{T-1} k_{T} \left({h/d_T}\right) \right] = O_p(\delta^{-1}_{nT})O(d_T/T)=o_p(1)$.
\medskip

\noindent \textbf{Proof of Proposition \ref{res1}.} This is a special case of Theorem \ref{res5} with $\mathbf{R}_n = \mathbf{I}_n$ and $G=1$.
\medskip

\noindent \textbf{Proof of Proposition \ref{res2}.} This is a special case of Theorem \ref{res5} with $G=1$.
\medskip

\noindent \textbf{Proof of Proposition \ref{res3}.} This is a special case of Theorem \ref{res6} with $G=1$.
\medskip

\noindent \textbf{Proof of Proposition \ref{res4}.} This is a special case of Theorem \ref{res5} with $\mathbf{R}_n = \mathbf{I}_n$.
\medskip

\noindent \textbf{Proof of Theorem \ref{res5}.}
Define $\Delta\mathbf{L}_{n,t} = (\Delta L_{1,t},\Delta L_{2,t},\dots,\Delta L_{n,t})'$. We have
\begin{equation*}
\begin{split}
\Delta\bar{\mathbf{L}}_{nT} =& \mathbf{D}^{-1}_n \frac{1}{T} \sum_{t=1}^T \mathbf{H}_{n}' \Delta\mathbf{L}_{n,t} = \mathbf{D}^{-1}_n \frac{1}{T} \sum_{t=1}^T \mathbf{H}_{n}' (\boldsymbol{\mu} + \mathbf{R}_n \boldsymbol{\epsilon}_{n,t}) = \bar{\boldsymbol{\mu}}_{n} + \mathbf{D}^{-1}_n \frac{1}{T} \sum_{t=1}^T \mathbf{H}_{n}' \mathbf{R}_n \boldsymbol{\epsilon}_{n,t},
\end{split}
\end{equation*}
where $\boldsymbol{\mu} = (\mu_1,\mu_2,\dots,\mu_n)'$.
It follows that
\begin{equation*}
\sqrt{nT} (\Delta\bar{\mathbf{L}}_{nT} - \bar{\boldsymbol{\mu}}_{n}) = \left(\frac{\mathbf{D}_n}{n}\right)^{-1} \frac{1}{\sqrt{nT}} \sum_{t=1}^T \mathbf{H}_{n}' \mathbf{R}_n \boldsymbol{\epsilon}_{n,t}.
\end{equation*}
Then $\sqrt{nT}\boldsymbol{\Omega}_{2,nT}^{-1/2}(\Delta\bar{\mathbf{L}}_{nT} - \bar{\boldsymbol{\mu}}_{n})\overset{D}{\rightarrow} N(\mathbf{0},\mathbf{I}_{G})$, where $\boldsymbol{\Omega}_{2,nT} = \sum_{i,j=1}^{n} \frac{n}{n_{g_i} n_{g_j}}  \boldsymbol{\iota}_{g_i}\boldsymbol{\iota}_{g_j}^{\prime} \mathbf{r}_{i.}' \bar{\boldsymbol{\gamma}}_{nT} \mathbf{r}_{j.}$ by Lemma \ref{lemma_1}\ref{lemma_11} and noting that $n^{-1}\mathbf{D}_n$ converges to a finite and nonsingular matrix under Assumption \ref{ass_groupnumbers}.
The matrix $\boldsymbol{\Omega}_{2,nT}$ can be written as $\boldsymbol{\Omega}_{2,nT} = (\mathbf{D}_n/n)^{-1} \mathbf{V}_{2,nT} (\mathbf{D}_n/n)^{-1}$.
Since $\mathbf{D}_n$ is known, estimation of $\boldsymbol{\Omega}_{2,nT}$, requires only the estimation of $\mathbf{V}_{2,nT}$.
Similarly, we write $\widehat{\boldsymbol{\Omega}}_{2,nT} = (\mathbf{D}_n/n)^{-1} \widehat{\mathbf{V}}_{2,nT} (\mathbf{D}_n/n)^{-1}$
where
\begin{equation*}
\widehat{\mathbf{V}}_{2,nT} = \frac{1}{nT}\sum_{i,j=1}^{n} \sum_{t,s=1}^{T} k_{T}\left( \frac{d_{ts}}{d_{T}}\right) k_{S}\left( \frac{d_{ij}}{d_{n}}\right)\boldsymbol{\iota}_{g_i}\boldsymbol{\iota}_{g_j}^{\prime} \Delta \tilde{L}_{it}\Delta \tilde{L}_{js}.
\end{equation*}
The $lw$th element of the matrices $\mathbf{V}_{2,nT}$ and $\widehat{\mathbf{V}}_{2,nT}$ are
\begin{equation}\label{v2_lw}
v^{lw}_{2,nT} = \frac{1}{n} \sum_{i \in G_l} \sum_{j \in G_w} \mathbf{r}_{i.}' \bar{\boldsymbol{\gamma}}_{nT} \mathbf{r}_{j.},
\end{equation}
and
\begin{equation*}
\hat{v}^{lw}_{2,nT} = \frac{1}{n T} \sum_{i \in G_l} \sum_{j \in G_w} \sum_{t,s=1}^{T} k_{S}\left( \frac{d_{ij}}{d_{n}}\right) k_{T}\left( \frac{d_{ts}}{d_{T}}\right)  \Delta \tilde{L}_{it}\Delta \tilde{L}_{js},
\end{equation*}
respectively.
We will show that $\hat{v}^{lw}_{2,nT} - v^{lw}_{2,nT} = o_p (1)$ which leads to the first result in the proposition.
We have $\Delta \tilde{L}_{it} = \mathbf{r}_{i.}' \tilde{\boldsymbol{\epsilon}}_{n,t}$ which gives
$
\Delta \tilde{L}_{it} \Delta \tilde{L}_{js} = \mathbf{r}_{i.}' \tilde{\boldsymbol{\epsilon}}_{n,t} \tilde{\boldsymbol{\epsilon}}'_{n,s} \mathbf{r}_{j.}
$.
Then,
\begin{equation*}
\begin{split}
\hat{v}^{lw}_{2,nT} - v^{lw}_{2,nT} =& \frac{1}{n} \sum_{i \in G_l} \sum_{j \in G_w} k_{S}\left( \frac{d_{ij}}{d_{n}}\right) \mathbf{r}_{i.}'  \left[ \frac{1}{T} \sum_{t,s=1}^{T} k_{T}\left( \frac{d_{ts}}{d_{T}}\right)  \tilde{\boldsymbol{\epsilon}}_{n,t} \tilde{\boldsymbol{\epsilon}}'_{n,s} \right] \mathbf{r}_{j.} - \frac{1}{n} \sum_{i \in G_l} \sum_{j \in G_w} \mathbf{r}_{i.}' \bar{\boldsymbol{\gamma}}_{nT} \mathbf{r}_{j.} \\
=& \frac{1}{n} \sum_{i \in G_l} \sum_{j \in G_w} k_{S}\left( \frac{d_{ij}}{d_{n}}\right) \mathbf{r}_{i.}'  \left[ \frac{1}{T} \sum_{t,s=1}^{T} k_{T}\left( \frac{d_{ts}}{d_{T}}\right)  \tilde{\boldsymbol{\epsilon}}_{n,t} \tilde{\boldsymbol{\epsilon}}'_{n,s} - \bar{\boldsymbol{\gamma}}_{nT} \right] \mathbf{r}_{j.} \\
&- \frac{1}{n} \sum_{i \in G_l} \sum_{j \in G_w} \left[ 1 - k_{S}\left( \frac{d_{ij}}{d_{n}}\right) \right] \mathbf{r}_{i.}' \bar{\boldsymbol{\gamma}}_{nT} \mathbf{r}_{j.}.
\end{split}
\end{equation*}
Since $\bar{\boldsymbol{\gamma}}_{nT} = O(1)$, and $T^{-1} \sum_{t,s=1}^{T} k_{T}( d_{ts}/d_{T})  \tilde{\boldsymbol{\epsilon}}_{n,t} \tilde{\boldsymbol{\epsilon}}'_{n,s} - \bar{\boldsymbol{\gamma}}_{nT} = o_p(1)$ by Lemma \ref{lemma_2}\ref{lemma_28}, it suffices to show that $\frac{1}{n} \sum_{i \in G_l} \sum_{j \in G_w} k_{S}\left( {d_{ij}/d_{n}}\right) \mathbf{r}_{i.}' \mathbf{r}_{j.} = O(1)$ and $\frac{1}{n} \sum_{i \in G_l} \sum_{j \in G_w} [1 - k_{S}\left( {d_{ij}/d_{n}}\right)] \mathbf{r}_{i.}' \mathbf{r}_{j.} = o(1)$ in order to prove the consistency of $\hat{v}^{lw}_{2,nT}$.
Starting with the latter, we have
\begin{equation*}
\begin{split}
\left| \frac{1}{n} \sum_{i \in G_l} \sum_{j \in G_w} \left[ 1 - k_{S}\left( \frac{d_{ij}}{d_{n}}\right) \right] \mathbf{r}_{i.}' \mathbf{r}_{j.} \right| &\leq \frac{1}{n} \sum_{i \in G_l} \sum_{j \in G_w} \left|1 - k_{S}\left( \frac{d_{ij}}{d_{n}}\right) \right| |\mathbf{r}_{i.}' \mathbf{r}_{j.}| \\
&\leq \frac{1}{n d_n^{\rho_s}} \sum_{i \in G_l} \sum_{j \in G_w} |\mathbf{r}_{i.}' \mathbf{r}_{j.}| {d_{ij}^{\rho_s}} = O \left(\frac{1}{d_n^{\rho_s}} \right) = o(1),
\end{split}
\end{equation*}
where the last equality follows by the assumptions that $d_n \to \infty$ and $\rho_s \geq 1$.
For the first term, write
\begin{equation*}
\begin{split}
\frac{1}{n} \sum_{i \in G_l} \sum_{j \in G_w} k_{S}\left( \frac{d_{ij}}{d_{n}}\right) \mathbf{r}_{i.}' \mathbf{r}_{j.} = \frac{1}{n} \sum_{i \in G_l} \sum_{j \in G_w} \mathbf{r}_{i.}' \mathbf{r}_{j.} -  \frac{1}{n} \sum_{i \in G_l} \sum_{j \in G_w} \left[ 1 - k_{S}\left( \frac{d_{ij}}{d_{n}}\right) \right] \mathbf{r}_{i.}' \mathbf{r}_{j.} = O(1),
\end{split}
\end{equation*}
which follows from the fact that the first term is $O(1)$ and the second term is $o(1)$ by the previous equation.
Hence, $\hat{v}^{lw}_{2,nT} - v^{lw}_{2,nT} = o_p (1)$ and the result follows.
This completes the proof of (i).

To show the consistency of $\underline{\widehat{\boldsymbol{\Omega}}}_{2,nT}$, we  first write $\underline{\widehat{\boldsymbol{\Omega}}}_{2,nT} = (\mathbf{D}_n/n)^{-1} \underline{\widehat{\mathbf{V}}}_{2,nT} (\mathbf{D}_n/n)^{-1}$, where
\begin{equation*}
\underline{\widehat{\mathbf{V}}}_{2,nT} = \frac{1}{\underline{n} T}\sum_{i=1}^{\underline{n}_{{g_i}}} \sum_{j=1}^{\underline{n}_{{g_j}}} \sum_{t,s=1}^{T} k_{T}\left( \frac{d_{ts}}{d_{T}}\right) \boldsymbol{\iota}_{g_i}\boldsymbol{\iota}_{g_j}^{\prime} \Delta \tilde{L}_{it}\Delta \tilde{L}_{js}.
\end{equation*}
The $lw$th element of this matrix, corresponding to clusters $l$ and $w$ is
\begin{equation*}
\underline{\hat{v}}^{lw}_{2,nT} = \frac{1}{\underline{n} T} \sum_{i \in \underline{G}_l} \sum_{j \in \underline{G}_w} \mathbf{r}_{i.}' \left[ \frac{1}{T} \sum_{t,s=1}^{T} k_{T}\left( \frac{d_{ts}}{d_{T}}\right)  \tilde{\boldsymbol{\epsilon}}_{n,t} \tilde{\boldsymbol{\epsilon}}'_{n,t} \right] \mathbf{r}_{j.},
\end{equation*}
where $\underline{G}_g$, $g=1,\dots,G$, is the set of indices in cluster $k$ which are used in the calculation of the partial variance estimate. This set has a cardinality of $\underline{n}_{g}$.
Using this expression together with \eqref{v2_lw}, we can write
\begin{equation}\label{bkt_yine}
\underline{\hat{v}}^{lw}_{2,nT} - v^{lw}_{2,nT} = (\underline{\hat{v}}^{lw}_{2,nT} - \underline{v}^{lw}_{2,nT}) + (\underline{v}^{lw}_{2,nT} - v^{lw}_{2,nT}).
\end{equation}
where $\underline{v}^{lw}_{2,nT} = \frac{1}{\underline{n}} \sum_{i \in \underline{G}_l} \sum_{j \in \underline{G}_w} \mathbf{r}_{i.}' \bar{\boldsymbol{\gamma}}_{nT} \mathbf{r}_{j.}$. Our objective is to show that both terms in parentheses approach to zero.
We have
\begin{equation*}
\begin{split}
\underline{\hat{v}}^{lw}_{2,nT} - \underline{v}^{lw}_{2,nT} = \frac{1}{\underline{n}} \sum_{i \in \underline{G}_l} \sum_{j \in \underline{G}_w} \mathbf{r}_{i.}' \left[ \frac{1}{T} \sum_{t,s=1}^{T} k_{T}\left( \frac{d_{ts}}{d_{T}}\right)  \tilde{\boldsymbol{\epsilon}}_{n,t} \tilde{\boldsymbol{\epsilon}}'_{n,t} - \bar{\boldsymbol{\gamma}}_{nT} \right] \mathbf{r}_{j.}.
\end{split}
\end{equation*}
The term in brackets is $o_p(1)$ by Lemma \ref{lemma_2}\ref{lemma_28}. Then it is sufficient to show that $\frac{1}{\underline{n}} \sum_{i \in \underline{G}_l} \sum_{j \in \underline{G}_w} \mathbf{r}_{i.}' \mathbf{r}_{j.}$ is bounded.
We have
\begin{equation*}
\left| \frac{1}{\underline{n}} \sum_{i \in \underline{G}_l} \sum_{j \in \underline{G}_w} \mathbf{r}_{i.}' \mathbf{r}_{j.} \right| = \left| \frac{1}{\underline{n}} \sum_{i \in \underline{G}_l} \sum_{j \in \underline{G}_w} \sum_{k=1}^n r_{ik} r_{jk} \right| \leq \frac{1}{\underline{n}} \sum_{i \in \underline{G}_l} \sum_{k=1}^n |r_{ik}| \sum_{j \in \underline{G}_w} |r_{jk}| = \frac{\underline{n}_{l}}{\underline{n}} O(1),
\end{equation*}
which is $O(1)$ because $\underline{n}_{l}/\underline{n} \to \tau_l \in (0,1)$.
For the second term in \eqref{bkt_yine} we note that, by definition, $\lim_{(n,T)\to\infty} \underline{v}^{lw}_{2,nT} = \lim_{(n,T)\to\infty} v^{lw}_{2,nT} = v^{lw}_{2}$ where $v^{lw}_{2}$ is the $lw$th element of the matrix $\mathbf{V}_2$ defined in Lemma  \ref{lemma_1}\ref{lemma_11}.
\medskip

\noindent \textbf{Proof of Theorem \ref{res6}.} We have
\begin{equation*}
\Delta\bar{\mathbf{L}}_{nT} = \bar{\boldsymbol{\mu}}_{n} + \mathbf{D}^{-1}_n \frac{1}{T} \sum_{t=1}^T \mathbf{H}_{n}' \mathbf{R}_n \boldsymbol{\epsilon}_{n,t} + \mathbf{D}^{-1}_n \frac{1}{T} \sum_{t=1}^T \mathbf{H}^{-1}_n \boldsymbol{\Lambda}_n \mathbf{f}_{t},
\end{equation*}
from which, we obtain
\begin{equation*}
\sqrt{T} (\Delta\bar{\mathbf{L}}_{nT} - \bar{\boldsymbol{\mu}}_{n}) = \left(\frac{\mathbf{D}_n}{n}\right)^{-1} \frac{1}{n\sqrt{T}} \sum_{t=1}^T \mathbf{H}_{n}' \mathbf{R}_n \boldsymbol{\epsilon}_{n,t} + \left(\frac{\mathbf{D}_n}{n}\right)^{-1} \frac{1}{n\sqrt{T}} \sum_{t=1}^T \mathbf{H}_{n}' \boldsymbol{\Lambda}_n \mathbf{f}_{t}.
\end{equation*}
From Lemma \ref{lemma_1}\ref{lemma_11} it follows that $\frac{1}{n\sqrt{T}} \sum_{t=1}^T \mathbf{H}_{n}' \mathbf{R}_n \boldsymbol{\epsilon}_{n,t}= O_p(n^{-1/2})$.
Then, $\sqrt{T}\boldsymbol{\Omega}_{3,nT}^{-1/2}(\Delta\bar{\mathbf{L}}_{nT} - \bar{\boldsymbol{\mu}}_{n})\overset{D}{\rightarrow} N(\mathbf{0},\mathbf{I}_{G})$, where
$
\boldsymbol{\Omega}_{3,nT} = \sum_{i,j=1}^{n} \frac{1}{n_{g_i} n_{g_j}} \boldsymbol{\iota}_{g_i}\boldsymbol{\iota}_{g_j}^{\prime} \left( \boldsymbol{\lambda}'_i \bar{\boldsymbol{\Gamma}}_{T} \boldsymbol{\lambda}_j + \mathbf{r}_{i.}' \bar{\boldsymbol{\gamma}}_{nT} \mathbf{r}_{j.} \right)
$ by noting that $n^{-1}\mathbf{D}_n$ converges to a finite and nonsingular matrix under Assumption \ref{ass_groupnumbers} and Lemma \ref{lemma_1}\ref{lemma_12}. 
The matrix $\boldsymbol{\Omega}_{3,nT}$ can be written as $\boldsymbol{\Omega}_{3,nT} = (\mathbf{D}_n/n)^{-1} \mathbf{V}_{3,nT} (\mathbf{D}_n/n)^{-1}$.
As in the proof of Theorem \ref{res5}, estimation of $\boldsymbol{\Omega}_{2,nT}$, requires only the estimation of $\mathbf{V}_{2,nT}$ because $\mathbf{D}_n$ is known.
Write $\widehat{\boldsymbol{\Omega}}_{3,nT} = (\mathbf{D}_n/n)^{-1} \widehat{\mathbf{V}}_{3,nT} (\mathbf{D}_n/n)^{-1}$
where
\begin{equation*}
\widehat{\mathbf{V}}_{3,nT} = \frac{1}{n^2 T} \sum_{i,j=1}^{n} \sum_{t,s=1}^{T} k_{T}\left( \frac{d_{ts}}{d_{T}}\right) \boldsymbol{\iota}_{g_i} \boldsymbol{\iota}_{g_j}^{\prime} \Delta \tilde{L}_{it}\Delta \tilde{L}_{js}.
\end{equation*}
The $lw$th element of the matrices $\mathbf{V}_{3,nT}$ and $\widehat{\mathbf{V}}_{3,nT}$, corresponding to clusters $l$ and $w$ are
\begin{equation}\label{v3_lw}
v^{lw}_{3,nT} = \frac{1}{n^2 T} \sum_{i \in G_l} \sum_{j \in G_w} \sum_{t,s=1}^{T} (\boldsymbol{\lambda}'_i \boldsymbol{\Gamma}_{d_{ts}} \boldsymbol{\lambda}_j + \mathbf{r}_{i.}' \boldsymbol{\gamma}_{n,d_{ts}} \mathbf{r}_{j.}),
\end{equation}
and
\begin{equation*}
\hat{v}^{lw}_{3,nT} = \frac{1}{n^2 T} \sum_{i \in G_l} \sum_{j \in G_w} \sum_{t,s=1}^{T} k_{T}\left( \frac{d_{ts}}{d_{T}}\right)   \Delta \tilde{L}_{it}\Delta \tilde{L}_{js},
\end{equation*}
respectively. We will show that $\hat{v}^{lw}_{3,nT} - v^{lw}_{3,nT} = o_p (1)$ which gives the first result in the proposition. We have $\Delta \tilde{L}_{it} = \boldsymbol{\lambda}'_i \tilde{\mathbf{f}}_{t} + \mathbf{r}_{i.}' \tilde{\boldsymbol{\epsilon}}_{n,t}$ which gives
\begin{equation*}
\Delta \tilde{L}_{it} \Delta \tilde{L}_{js} = \boldsymbol{\lambda}'_i \tilde{\mathbf{f}}_{t} \tilde{\mathbf{f}}'_{s} \boldsymbol{\lambda}_j + \boldsymbol{\lambda}'_i \tilde{\mathbf{f}}_{t} \tilde{\boldsymbol{\epsilon}}'_{n,s} \mathbf{r}_{j.} + \mathbf{r}_{i.}' \tilde{\boldsymbol{\epsilon}}_{n,t} \tilde{\mathbf{f}}'_{s} \boldsymbol{\lambda}_j + \mathbf{r}_{i.}' \tilde{\boldsymbol{\epsilon}}_{n,t} \tilde{\boldsymbol{\epsilon}}'_{n,s} \mathbf{r}_{j.}.
\end{equation*}
Using the last three equations, we obtain
\begin{equation*}
\begin{split}
\hat{v}^{lw}_{3,nT} - v^{lw}_{3,nT} =& \frac{1}{n^2}  \sum_{i \in G_l} \sum_{j \in G_w} \boldsymbol{\lambda}'_i \left[ \frac{1}{T} \sum_{t,s=1}^{T} k_{T}\left( \frac{d_{ts}}{d_{T}}\right)  \tilde{\mathbf{f}}_{t} \tilde{\mathbf{f}}'_{s} - \bar{\boldsymbol{\Gamma}}_{T} \right] \boldsymbol{\lambda}_j \\
&+ \frac{1}{n^2}  \sum_{i \in G_l} \sum_{j \in G_w} \boldsymbol{\lambda}'_i \left[ \frac{1}{T} \sum_{t,s=1}^{T} k_{T}\left( \frac{d_{ts}}{d_{T}}\right)  \tilde{\mathbf{f}}_{t} \tilde{\boldsymbol{\epsilon}}'_{n,s} \right] \mathbf{r}_{j.} \\
&+ \frac{1}{n^2}  \sum_{i \in G_l} \sum_{j \in G_w} \mathbf{r}_{i.}' \left[ \frac{1}{T} \sum_{t,s=1}^{T} k_{T}\left( \frac{d_{ts}}{d_{T}}\right)  \tilde{\boldsymbol{\epsilon}}_{n,t} \tilde{\mathbf{f}}_{s} \right] \boldsymbol{\lambda}_j \\
&+ \frac{1}{n^2}  \sum_{i \in G_l} \sum_{j \in G_w} \mathbf{r}_{i.}' \left[ \frac{1}{T} \sum_{t,s=1}^{T} k_{T}\left( \frac{d_{ts}}{d_{T}}\right)  \tilde{\boldsymbol{\epsilon}}_{n,t} \tilde{\boldsymbol{\epsilon}}'_{n,t} - \bar{\boldsymbol{\gamma}}_{nT} \right] \mathbf{r}_{j.} \\
=& D_1 + D_2 + D_3 + D_4.
\end{split}
\end{equation*}
We will show that each of these four terms are $o_p(1)$. By Lemma \ref{lemma_2}, all expressions in square brackets are $o_p(1)$. The first term can be written as $D_1 = \frac{1}{n^2}  \sum_{i \in G_l} \sum_{j \in G_w} \boldsymbol{\lambda}'_i \boldsymbol{\lambda}_j o_p(1)$. By Hölder's inequality we have $|\boldsymbol{\lambda}'_i \boldsymbol{\lambda}_j| \leq ||\boldsymbol{\lambda}_i|| ||\boldsymbol{\lambda}_j||$ where the right hand side is bounded by Assumption \ref{ass_comcomp}\ref{ass_comcomp_loadingssecondorder}. This shows that $D_1 = o_p(1)$. Other terms can be shown to be $o_p(1)$ similarly which in turn gives $\widehat{\mathbf{V}}_{3,nT} - \mathbf{V}_{3,nT} = o_p(1)$ and hence $\widehat{\boldsymbol{\Omega}}_{3,nT} - \boldsymbol{\Omega}_{3,nT} = o_p(1)$. The consistency of $\widehat{\boldsymbol{\Omega}}_{3,nT}$ in turn implies the asymptotic null distribution which completes the proof of (i).

For the second result, we write $\underline{\widehat{\boldsymbol{\Omega}}}_{3,nT} = (\mathbf{D}_n/n)^{-1} \underline{\widehat{\mathbf{V}}}_{3,nT} (\mathbf{D}_n/n)^{-1}$, where
\begin{equation*}
\underline{\widehat{\mathbf{V}}}_{3,nT} = \frac{1}{n^2 T} \sum_{i,j=1}^{n} \sum_{t,s=1}^{T} k_{T} \left( \frac{d_{ts}}{d_{T}}\right) \boldsymbol{\iota}_{g_i}\boldsymbol{\iota}_{g_j}^{\prime} \widehat{\boldsymbol{\lambda}}'_i \widehat{\mathbf{f}}_{t} \widehat{\boldsymbol{\lambda}}'_j \widehat{\mathbf{f}}_{s} + \frac{1}{n^2 T}\sum_{i=1}^{n} \sum_{t,s=1}^{T} k_{T}\left( \frac{d_{ts}}{d_{T}}\right) \boldsymbol{\iota}_{g_i}\boldsymbol{\iota}_{g_i}^{\prime} \widehat{\varepsilon}_{it} \widehat{\varepsilon}_{is}.
\end{equation*}
The $lw$th element of this matrix is
\begin{equation*}
\underline{\hat{v}}^{lw}_{3,nT} = \frac{1}{n^2 T} \sum_{i \in G_l} \sum_{j \in G_w} \sum_{t,s=1}^{T} k_{T}\left( \frac{d_{ts}}{d_{T}}\right) \widehat{\boldsymbol{\lambda}}'_i \widehat{\mathbf{f}}_{t} \widehat{\boldsymbol{\lambda}}'_j \widehat{\mathbf{f}}_{s} + \frac{1}{n^2 T} \sum_{i \in G_l} \sum_{j \in G_w} \sum_{t,s=1}^{T} k_{T}\left( \frac{d_{ts}}{d_{T}}\right) \widehat{\varepsilon}_{it} \widehat{\varepsilon}_{is}.
\end{equation*}
Using this expression together with \eqref{v3_lw}, we find
\begin{equation*}
\begin{split}
\underline{\hat{v}}^{lw}_{3,nT} - v^{lw}_{3,nT} =& \frac{1}{n^2 } \sum_{i \in G_l} \sum_{j \in G_w} \left[ \frac{1}{T} \sum_{t,s=1}^{T} k_{T}\left( \frac{d_{ts}}{d_{T}}\right) \widehat{\boldsymbol{\lambda}}'_i \widehat{\mathbf{f}}_{t} \widehat{\boldsymbol{\lambda}}'_j \widehat{\mathbf{f}}_{s} - \boldsymbol{\lambda}'_i \bar{\boldsymbol{\Gamma}}_{T} \boldsymbol{\lambda}_j \right] \\
&+ \frac{1}{n^2 T} \sum_{i \in G_l} \sum_{t,s=1}^{T} k_{T}\left( \frac{d_{ts}}{d_{T}}\right) \widehat{\varepsilon}_{it} \widehat{\varepsilon}_{is} -  \frac{1}{n^2} \sum_{i \in G_l} \sum_{j \in G_w} \mathbf{r}_{i.}' \bar{\boldsymbol{\gamma}}_{nT} \mathbf{r}_{j.} \\
=& \frac{1}{n^2} \sum_{i \in G_l} \sum_{j \in G_w} \left[ \frac{1}{T} \sum_{t,s=1}^{T} k_{T}\left( \frac{d_{ts}}{d_{T}}\right) \left( \widehat{\boldsymbol{\lambda}}'_i \widehat{\mathbf{f}}_{t} \widehat{\boldsymbol{\lambda}}'_j \widehat{\mathbf{f}}_{s} - \boldsymbol{\lambda}'_i  {\mathbf{f}}_{t} {\mathbf{f}}'_{s} \boldsymbol{\lambda}_j   \right) \right] \\
&+ \frac{1}{n^2 T} \sum_{i \in G_l} \sum_{j \in G_w} \boldsymbol{\lambda}'_i  \left[\frac{1}{T}  \sum_{t,s=1}^{T} k_{T}\left( \frac{d_{ts}}{d_{T}}\right) {\mathbf{f}}_{t} {\mathbf{f}}'_{s} - \bar{\boldsymbol{\Gamma}}_{T} \right] \boldsymbol{\lambda}_j  \\
&+ \frac{1}{n^2 T} \sum_{i \in G_l} \sum_{t,s=1}^{T} k_{T}\left( \frac{d_{ts}}{d_{T}}\right) \widehat{\varepsilon}_{it} \widehat{\varepsilon}_{is} -  \frac{1}{n^2} \sum_{i \in G_l} \sum_{j \in G_w} \mathbf{r}_{i.}' \bar{\boldsymbol{\gamma}}_{nT} \mathbf{r}_{j.}
\end{split}
\end{equation*}
The desired result now follows from Lemma \ref{lemma_4}\ref{lemma_43}, Lemma \ref{lemma_4}\ref{lemma_44}, Lemma \ref{lemma_2}\ref{lemma_22} and by noting that the last term is $O(n^{-1})$. This completes the proof.

\section{Details on the Evaluation of the IMF Consumer Price Inflation Forecasts}\label{sec:app3}

In this appendix, we present the details on the evaluation of the IMF CPI forecasts which was summarized in Section \ref{sec:app2}.
As in the application on the comparison of the economic growth forecasts of the OECD and the IMF, the data for the IMF forecasts come from the Fund's Historical WEO Forecasts Database.
Once again we focus on their summer forecasts made for the following year, hence we are dealing with one year ahead forecasts.
Our data set contains 127 countries for which the forecasts are available from 1991 to 2019, i.e. the panel is balanced.
We exclude 5 countries from the original IMF data set as their loss differentials are very different from the rest of the sample.
These countries are Brazil, Democratic Republic of the Congo, Peru, Venezuela and Nicaragua.
Notice that all of these countries experienced hyperinflation, in late 2010's in the case of Venezuela, and in early to mid-90's for the rest.
For the first four countries there are very big drops or jumps in the CPI inflation, hence their RW forecasts are very poor for at least one year.
Whereas for the last country the situation is the contrary, that is IMF forecasts are much worse than the RW forecasts.
Our conclusions should be understood to apply to the 127 countries in our sample which includes the G7 countries, 22 OECD countries which are not part of G7 and 98 non-OECD countries.
As stated in the paper, in this application, in addition to the quadratic loss function, we use the absolute loss function.

\textbf{Cross-sectional and temporal dependence in loss differentials: CPI forecasts.} Before looking into the EPA test results we apply the methodology described in Section \ref{sec:imp} and check if we can find evidence for CD in our sample and identify its type.
To save space, we do not report the diagnostic results in tables in this subsection.
The two CD tests, namely BP and modified BP, provide $p$-values which are practically zero for both loss functions.
Hence, we conclude that the loss differentials contain CD.
Following the CD tests, we use the $IC_{p1}$ which indicate that there are 6 common factors in the loss differential series.
We therefore find out that both loss differential series display SCD and apply our tests robust to SCD.

\textbf{Panel tests for the EPA hypotheses: CPI forecasts.}
As before, we report the tests robust to SCD as well as the results for the non-robust tests as a benchmark.
We start the analysis with overall EPA tests and continue with the clustered EPA tests.
As stated in the main paper, we consider 2 different country clustering schemes: the first one divides the sample of countries into OECD and non-OECD countries whereas the second consists of G7 countries, non-G7 OECD countries and non-OECD countries.
We further split the sample into pre- and post-global financial crises periods and compute the average loss differentials.
The results for the average loss differentials of are given in Table \ref{tab:cpipanelavg}.
In the full sample with the absolute error loss, we see that the IMF does better than the RW for all clusters except the non-OECD countries.
The global average is found to be positive which shows that the IMF does worse than the RW overall.
The results are similar for the quadratic loss function except that the average loss is practically zero for G7 countries.
In the pre-crisis period, the overall differences are more pronounced between the IMF and the RW model.
In the post-crisis period however, the differences are very close to zero, especially for the quadratic loss.
In what follows we use our test to check the significance of these averages.

First, in Panel (a) of Table \ref{tab:cpipaneltests}, we see that all three overall EPA test statistics are statistically significant in 10\% level for the absolute loss function.
We also see here the effect of taking CD into account: with $S^{(1)}_{n,T}$, we can reject the overall EPA hypothesis at 5\% level but this is not the case for the tests $S^{(3)}_{n,T}$ and $\underline{S}_{n,T}^{(3)}$.
We remind that the average absolute loss differential of IMF and RW forecasts is 0.45 which is reported in the last row of Table \ref{tab:cpipanelavg}.
We conclude that, overall, there is a small but statistically significant difference between the average absolute loss differential of the IMF and RW forecasts in favor of the latter.

With the quadratic loss function, we cannot reject the overall EPA hypothesis in conventional levels using any test.
The question therefore is, if we can reject the clustered EPA hypothesis using the clusters under consideration.
In the second block of the table, we have the results for the clustered EPA tests using two clusters: OECD and non-OECD countries.
For these clusters, we can strongly reject the clustered EPA hypothesis with both loss functions using the $C^{(1)}_{n,T}$.
However, when we take into account the CD in the loss differentials, the magnitude of the test statistics decline dramatically and they are insignificant.
In the last block of the table, we have the results for the case of the three clusters, G7, non-G7 OECD, and non-OECD countries.
Here, a similar picture arises such that we can reject the clustered EPA hypothesis with $C^{(1)}_{n,T}$ but this is not the case for $C^{(3)}_{n,T}$ and $\underline{C}_{n,T}^{(3)}$.

In Panel (b), the results for the pre-crisis period are reported.
The results obtained from the overall tests are similar to those of the full sample, that is, we can reject the overall EPA hypothesis with the absolute loss function but this is not the case with the quadratic loss.
Similarly, with two country clusters (OECD and non-OECD countries), we can reject the clustered EPA hypothesis using only $C^{(1)}_{n,T}$ for both loss functions.
When we consider the three country clusters however, we can reject the clustered EPA hypothesis with the absolute loss function using any test, at least at 10\% level.
To conclude, a significant difference between the IMF and RW forecast accuracy exists in the pre-crisis period using the absolute loss function.

As can be seen in Panel (c), the results are slightly different for the post-crisis period.
First, the overall test statistics are negative for the absolute loss as IMF has less bias in this period.
However, these overall differences are not statistically significant.
When we look at the clustered EPA tests with two clusters, the differences are again statistically insignificant for both loss functions.
If we consider the case of three country clusters, similar to the pre-crisis period, we can reject the clustered EPA hypothesis with the absolute loss function using any test.
The statistics for the quadratic loss differentials are insignificant as before.
This is not surprising as we have found that the average loss differentials are very close to zero using the quadratic loss, as is reported in Table \ref{tab:cpipanelavg}.

\begin{sidewaystable}[htbp!]
  \centering
  \caption{Average Loss Differentials for the CPI Inflation Forecasts of Different Country Clusters (IMF \textit{vs.} Random Walk)}
    \begin{tabular}{ccccccc}
    \toprule
          & \multicolumn{2}{c}{Full Sample} & \multicolumn{2}{c}{1991-2006 (Pre-crisis)} & \multicolumn{2}{c}{2009-2019 (Post-crisis)} \\
    \midrule
    Cluster & Absolute Loss & Quadratic Loss & Absolute Loss & Quadratic Loss & Absolute Loss & Quadratic Loss \\
    \midrule
    G7    & -0.0004 & 0.0000 & 0.0006 & 0.0000 & -0.0019 & -0.0001 \\
    Non-G7 OECD & -0.0033 & -0.0397 & -0.0063 & -0.0719 & 0.0002 & 0.0000 \\
    OECD  & -0.0026 & -0.0301 & -0.0047 & -0.0546 & -0.0003 & 0.0000 \\
    Non-OECD & 0.0065 & 0.0134 & 0.0121 & 0.0243 & -0.0014 & -0.0002 \\
    All   & 0.0045 & 0.0035 & 0.0083 & 0.0063 & -0.0012 & -0.0002 \\
    \bottomrule
    \end{tabular}%

  \label{tab:cpipanelavg}%
\end{sidewaystable}

\begin{sidewaystable}[htbp!]
  \centering
  \caption{Panel Tests of EPA for the CPI Inflation Forecasts (IMF \textit{vs.} Random Walk)}
  \begin{threeparttable}
    \begin{tabular}{cccccccc}
    \toprule
          & \multicolumn{2}{c}{Overall EPA Tests} &       & \multicolumn{4}{c}{Clustered EPA Tests} \\
    \midrule
          &       &       &       & \multicolumn{2}{c}{Cluster 1: OECD} & \multicolumn{2}{c}{Cluster 1: G7} \\
          &       &       &       & \multicolumn{2}{c}{Cluster 2: Non-OECD} & \multicolumn{2}{c}{Cluster 2: Non-G7 OECD} \\
          &       &       &       & \multicolumn{2}{c}{} & \multicolumn{2}{c}{Cluster 3: Non-OECD} \\
    \midrule
    \multicolumn{8}{c}{Panel (a): Full Sample} \\
    \midrule
    Test  & Absolute Loss & Quadratic Loss & Test  & Absolute Loss & Quadratic Loss & Absolute Loss & Quadratic Loss \\
    \midrule
    $S^{(1)}_{n,T}$ & 2.04  & 0.42  & $C^{(1)}_{n,T}$ & 8.16  & 7.02  & 8.64  & 10.14 \\
          & (0.04) & (0.67) &       & (0.02) & (0.03) & (0.03) & (0.02) \\
    $S^{(3)}_{n,T}$ & 1.70  & 0.50  & $C^{(3)}_{n,T}$ & 3.23  & 3.65  & 3.62  & 4.49 \\
          & (0.09) & (0.62) &       & (0.20) & (0.16) & (0.31) & (0.21) \\
    $\underline{S}_{n,T}^{(3)}$ & 1.78  & 0.50  & $\underline{C}_{n,T}^{(3)}$ & 3.44  & 3.65  & 3.62  & 5.05 \\
          & (0.08) & (0.62) &       & (0.18) & (0.16) & (0.31) & (0.17) \\
    \midrule
    \multicolumn{8}{c}{Panel (b): 1991-2006 (Pre-crisis)} \\
    \midrule
    $S^{(1)}_{n,T}$ & 2.15  & 0.43  & $C^{(1)}_{n,T}$ & 9.08  & 7.29  & 9.78  & 7.53 \\
          & (0.03) & (0.67) &       & (0.01) & (0.03) & (0.02) & (0.06) \\
    $S^{(3)}_{n,T}$ & 2.02  & 0.51  & $C^{(3)}_{n,T}$ & 4.40  & 4.09  & 7.22  & 4.26 \\
          & (0.04) & (0.61) &       & (0.11) & (0.13) & (0.07) & (0.23) \\
    $\underline{S}_{n,T}^{(3)}$ & 1.99  & 0.51  & $\underline{C}_{n,T}^{(3)}$ & 4.28  & 4.08  & 8.20  & 4.23 \\
          & (0.05) & (0.61) &       & (0.12) & (0.13) & (0.04) & (0.24) \\
    \midrule
    \multicolumn{8}{c}{Panel (c): 2009-2019 (Post-crisis)} \\
    \midrule
    $S^{(1)}_{n,T}$ & -1.59 & -0.82 & $C^{(1)}_{n,T}$ & 2.66  & 2.25  & 8.29  & 6.31 \\
          & (0.11) & (0.41) &       & (0.26) & (0.33) & (0.04) & (0.10) \\
    $S^{(3)}_{n,T}$ & -0.45 & -0.40 & $C^{(3)}_{n,T}$ & 0.30  & 0.23  & 9.87  & 4.97 \\
          & (0.66) & (0.69) &       & (0.86) & (0.89) & (0.02) & (0.17) \\
    $\underline{S}_{n,T}^{(3)}$ & -0.45 & -0.40 & $\underline{C}_{n,T}^{(3)}$ & 0.41  & 0.24  & 8.27  & 6.34 \\
          & (0.65) & (0.69) &       & (0.82) & (0.89) & (0.04) & (0.10) \\
    \bottomrule
    \end{tabular}%
    \begin{tablenotes}
    	\item Note: The values shown in parentheses are $p$-values.
    \end{tablenotes}
  \end{threeparttable}
  \label{tab:cpipaneltests}%
\end{sidewaystable}

\restoregeometry

\end{appendices}

\end{document}